\documentclass{ws-ijmpa}

\let\draftnote\relax

\let\trimmarks\relax

\usepackage{graphicx,color}

\usepackage{graphicx}
\usepackage{wrapfig} 

\usepackage{mathptmx}
\usepackage[psamsfonts]{eucal}

\usepackage{latexsym}
\usepackage{amsmath,amssymb} 
\usepackage{bm} 
\usepackage{bbm} 

\usepackage{alphalph}

\numberwithin{equation}{section}

\allowdisplaybreaks

\begin{document}

\markboth{M.Yasuda}
{Generalized Analytical Mechanics in which Quantum Phenomena Appear}

%
%

\title{A GENERALIZED ANALYTICAL MECHANICS IN WHICH QUANTUM PHENOMENA APPEAR}

\author{Masao Yasuda \footnotemark \footnotetext{Current affiliation: None.  
 \   
 Previous affiliation, where most of the present paper was written: Department of Physics, Hokkaido University,  
Sapporo 060-0810, Japan}      }

\address{
myasuda@mbn.nifty.com
}

\maketitle


\begin{abstract}

\noindent 
Currently, dynamics of a massive macroparticle is given by classical analytical mechanics (CM), while that of a massive micro one is given by quantum mechanics (QM). 
We propose a mechanics effective for both: 
We transform, under coordinate transformation, the covariant tensor of order two underlying the kinetic energy term of the Hamilton-Jacobi (H-J) eq. of CM, not with the action of the diffeomorphism group, but with an action of an extended diffeomorphism group. 
Then, the H-J eq., a first-order partial differential eq., is modified to a third-order one. The Euler-Lagrange eq. of CM, a second-order ordinary differential eq., related to the H-J eq. through action integral of the least action principle is accordingly modified to a fourth-order one. 
We thus obtain a mechanics which, because of the higher-order eqs., accommodates phenomena corresponding to quantum phenomena. 
Energy of a particle in a confining potential is quantized to the same values as those of QM. 
Particle distribution in an ensemble disagrees with that of QM, but agrees with experiments within experimental uncertainty. 
The mechanics therefore is a testable alternative to QM.


\end{abstract}

\ccode{PACS numbers: 03.65.Ca}

\section{Introduction} \label{may0409-8}

We incorporate an extended diffeomorphism action, which is defined on p.\pageref{jul1920-1}, into classical analytical mechanics (CM) \cite{Landau(1976)} of nonrelativistic three-dimensional (3-D) systems comprised of a massive particle $m$ and a time-independent potential field $V$.

We first review CM of the system. 
We view the CM as being comprised of an Euler-Lagrange (E-L) eq. and a Hamilton-Jacobi (H-J) eq.  
The eqs. are obtained from the Hamilton's principle, a form of the principle of least action in which action $\mathcal{S} $ is given by an eq. $\mathcal{S} ( \textbf{x} , t) = \mathcal{S}_0 ( \textbf{x} ) - E t = \int ^  {\textbf{x} , \, t} L( \textbf{q} , \dot{\textbf{q}} ) d \tilde{t}$, \cite{Landau(1976)} \cite{Lanczos(1986)} where $L$ is a Lagrangian given as $L= m \dot{\textbf{q}}^2/2 - V(\textbf{q})$ and $E$ is total energy of the system. 
For variations $\delta \textbf{q}$ of a trajectory $\textbf{q}(t)$ and $\delta \textbf{x}$ of an endpoint $\textbf{x}$ of the $\textbf{q}(t)$ with $\delta t =0$, variation $\delta \mathcal{S}= \delta \mathcal{S}_0 $ of the action $ \mathcal{S}  ( \textbf{x} , t)$ is given as 
\begin{equation*}
 \delta \mathcal{S}_0 ( \textbf{x} ) 
=  \int ^{\textbf{x} , \, t} (  \delta \textbf{q}  \frac{ \partial L }{ \partial \textbf{q}} + \delta \dot{\textbf{q}} \frac{\partial L}{ \partial \dot{\textbf{q}}} ) d\tilde{t}  = \int ^{\textbf{x} , \, t} ( \frac{\partial L }{ \partial \textbf{q}} - \frac{d}{dt} \frac{\partial L }{ \partial \dot{\textbf{q}}}) \delta \textbf{q} d\tilde{t} + \frac{ \partial L }{ \partial \dot{\textbf{q}}} \delta \textbf{q} | ^ { \, t}  \ . 
\end{equation*} 
From the integrand, the E-L eq. $m \ddot{\textbf{q}} = - \partial V / \partial \textbf{q}$ is obtained. 
From the boundary term  $( \partial L / \partial \dot{\textbf{q}} ) \delta  \textbf{q} | ^ { \, t}$, $\nabla  \mathcal{S}_0( \textbf{x} ) = \lim _ { \delta \textbf{x} \to 0 } \delta \mathcal{S}_0 ( \textbf{x} ) / \delta \textbf{x} = \partial L / \partial \dot{\textbf{q}} |_ { \textbf{q} = \textbf{x}}$ is obtained for the integral $\mathcal{S} ( \textbf{x} , t) = \mathcal{S}_0 ( \textbf{x} ) - E t $ along a solution $\textbf{q}(t)$, along which $ \partial L / \partial \textbf{q} - (d / dt) (\partial L / \partial \dot{\textbf{q}}) =0$, of the E-L eq. 
Inserting $ \partial L / \partial \dot{\textbf{q}} = \nabla  \mathcal{S}_0$ into $E= \dot{\textbf{q}} \, \partial L / \partial \dot{\textbf{q}} -L = ( \partial L / \partial \dot{\textbf{q}} )^2 /2m + V( \textbf{q} )$, we have the H-J eq. $(\nabla \mathcal{S}_0 ( \textbf{x} ) )^2 /2m + V( \textbf{x} ) = E$. 
The E-L eq. is an ordinary differential eq. (ODE), which gives a trajectory $ \textbf{q}(t)$ of the particle $m$. 
While the H-J eq. is a partial differential eq. (PDE), which gives a momentum $ \nabla \mathcal{S}_0( \textbf{x} )$. 
The $\nabla \mathcal{S}_0( \textbf{x} )$ is a field in the sense that it is defined over whole domain. 
If the particle passes through $\textbf{x} _0$, an eq. 
 $ \nabla \mathcal{S}_0( \textbf{x} )  |_ { \textbf{x} = \textbf{x}_0} = \partial L / \partial \dot{\textbf{q}} |_ { \textbf{q} = \textbf{x}_0} = m \dot{\textbf{q}} |_ { \textbf{q} = \textbf{x}_0}$ holds true at $ \textbf{x}_0$.

The quantity $( \partial L / \partial \dot{\textbf{q}} )^2 /2m = m \dot{\textbf{q}} ^2 /2 $ represents kinetic energy in Lagrangian (Lag.) formalism, correspondingly $(\nabla \mathcal{S}_0 ( \textbf{x} ) )^2 /2m $ represents that in H-J formalism, where Lag. (H-J) formalism is a description of dynamics based on the E-L (H-J) eq. 
From the eq. $ \nabla \mathcal{S}_0 ( \textbf{x} ) |_ { \textbf{x} = \textbf{x} _0} = m \dot{\textbf{q}} |_ { \textbf{q} = \textbf{x}_0}$, an eq. $  (\nabla \mathcal{S}_0 ( \textbf{x} ) )^2 /2m |_ { \textbf{x} = \textbf{x}_0} = m \dot{\textbf{q}} ^2 / 2 |_ { \textbf{q} = \textbf{x}_0}$ follows.

The kinetic energy given in the H-J formalism of a 1-D system specified 
by a Lagrangian $L = m \dot{q} ^2 /2 - V(q)$ is $( \partial \mathcal{S}_0 (x) / \partial x )^2 / 2m$, which we write as $E_{\textrm{kin}}^{\textrm{HJ}} (x)$ for short. 
We regard the $E_{\textrm{kin}}^{\textrm{HJ}}(x )$ as a contraction 
\footnote{
Tensor contraction is defined as an operation to obtain a tensor $T _l^k $ of type $(k,l )$ from another tensor $T_{l+1}^{k+1}$ of type $(k+1 , l+1 )$. \cite{Lee(1997)} 
By the contraction of $\partial / \partial x \otimes \partial / \partial x$ and $dx \otimes dx $, we mean, by abuse of language, a double contraction over indexes of a tensor $ \partial / \partial x \otimes \partial / \partial x   \otimes dx \otimes dx \in T_2^2$. 
} 
of a contravariant metric tensor field \cite{Lee(2013)} $\partial / \partial x \otimes \partial / \partial x$ and a second-order covariant tensor field \cite{Lee(2013)} $(2m)^{-1}  ( \partial \mathcal{S}_0 / \partial x )^2 dx \otimes dx $, where $\partial / \partial x  \otimes \partial / \partial x $ and $dx \otimes dx$ are bases of the tensor fields. 
Under a coordinate transformation of a base space specified by a diffeomorphism \cite{Lee(2013)} $y = f(x )$, the contravariant field is transformed from a quantity on $x$- to that on $y$-coordinate system  by pushforward \cite{Lee(2013)}  $f_*$ by $f$ defined as $f_* \partial / \partial x \otimes \partial / \partial x  = ( \partial y/ \partial x )^2 \partial / \partial y \otimes \partial / \partial y $, 
while the covariant one is transformed by pullback \cite{Lee(2013)} ${f^{-1}}^*$ by $f^{-1}$ defined as ${f^{-1}}^* ( \partial \mathcal{S}_0 / \partial x )^2 dx \otimes dx   
= ( \partial \mathcal{S}_0 / \partial x )^2 ( \partial x / \partial y )^2 dy \otimes dy = ( \partial \mathcal{S} _0 / \partial y )^2 dy \otimes dy $. 
Accordingly, the contracted quantity is independent of coordinate systems on which they are contracted. 
Indeed, the contraction on $y$, $( \partial y / \partial x )^2 \partial / \partial y \otimes \partial / \partial y \cdot 
(2m)^{-1} ( \partial \mathcal{S} _0 / \partial y )^2 dy \otimes dy = ( \partial \mathcal{S}_0 / \partial x )^2 / 2m $, is identical to that on $x$. 
The coordinate transformation \label{mar3121-1} may be either passive or active, where passive coordinate transformation changes coordinate system of a base space while active one maps a point of a base space to another point. 
\cite{Rovelli(2004)} 
\footnote{
If, as in CM, the metric tensor field, $\partial / \partial x \otimes \partial / \partial x$ on $x$-coordinate system, is fixed on the base space, the active coordinate transformation for arbitrary diffeomorphisms does not make a contraction, say, $\partial \phi (x) / \partial x$, where $\phi (x) \in \mathbb{R}$ is a field on $x$, independent of coordinate systems, which we see as follows. 
The active one $x \mapsto f(x)$ moves a field value $\phi (x_0)$ ($\phi (x_0 + \delta x )$) from $x_0$ ($x_0 + \delta x $) to $f(x_0)$ ($f(x_0 + \delta x )$) on the $x$. 
Accordingly, $\Delta \phi = \phi ( x_0 + \delta x ) -  \phi ( x_0 )$ over $\delta x$ is transformed to the same $\Delta \phi$ over $f(x_0 + \delta x ) -f( x_0 ) \simeq (\partial f / \partial x ) \delta x $. 
Since $(\partial f / \partial x ) \delta x \, |_{f(x)} \neq \delta x \, |_x$ in general, $\partial \phi / \partial x$ is dependent on coordinate systems. 
If the diffeomorphisms are restricted to $f(x) = x + a $, where $a = const. \in \mathbb{R}$, on the $x$, the active one keeps the independence (cf. Ref. \refcite{Rovelli(2004)}). 
}

We write the pullback $f^*$ by $f$ as $\bar{T}_{f^{-1}}$ ($f^* \phi _y = \bar{T}_{f^{-1}} \phi _y$, where $\phi _y = ( \partial \mathcal{S}_0 / \partial y )^2 dy \otimes dy$). 
The pullback $\bar{T}$ is \label{jul2720-1} an example of diffeomorphism group action (diff. action)  $T$ defined as a transformation of $\phi _z , \phi _y , \cdots $ 
 satisfying i) $T_{{f^{-1}}{g^{-1}}} \phi _z = T_{f^{-1}} T_{g^{-1}} \phi _z$ and ii) $ T_1 \phi _z = \phi _z$, 
where $x = f^{-1}(y )$, $y = g^{-1}(z )$, $\phi _z = ( \partial \mathcal{S}_0 / \partial z )^2 dz \otimes dz$, and $ 1 $ is the identity $y = g^{-1}(z ) = z$ of the group. 
Indeed, the pullback $\bar{T}$ satisfies the i) 
\footnote{
Let $\phi _z = \phi (z) dz \otimes dz$, where $\phi (z) = ( \partial \mathcal{S} _0 / \partial z )^2$. Then, $ \bar{T}_{f^{-1}} (\bar{T}_{g^{-1}} \phi _z ) = \bar{T}_{f^{-1}} (( \partial g / \partial y )^2 \phi (z) dy \otimes dy ) $ \\ 
$= (\partial f/ \partial x )^2 ( \partial g/ \partial y)^2 \phi (z) dx \otimes dx = ( \partial g / \partial x )^2 \phi (z) dx \otimes dx = \bar{T}_{(gf)^{-1}} \phi _z $. 
} 
and ii). 
We note: 
the i) means that $T$ preserves group structure ($T_{{f^{-1}}{g^{-1}}} = T_{f^{-1}} T_{g^{-1}} $), and 
the ii) together with i) means that $T$ is one-to-one and onto ($T_{g^{-1}} T_g \phi _y = \phi _y$ and $ T_g T_{g^{-1}} \phi _z =  \phi _z$). \cite{Rotman(1995)} \cite{Kawakubo(1991)}

Another \label{jul1920-1} transformation rule $ \bar{\bar{T}}$, which we call an extended diff. action, defined as $ \bar{\bar{T}} _{ g^{-1}}( \partial \mathcal{S}_0 / \partial z )^2 dz \otimes dz = ( ( \partial \mathcal{S}_0 / \partial y )^2 + \alpha \{ g(y) ,y \} ) dy \otimes dy$, where $\alpha = const. ( \neq 0 ) \in \mathbb{R}$ and \label{jul1320-1} $\{ g(y) ,y \} := g'''/g' - (3/2){g''}^2 / {g'}^2$ ($g' =\partial g / \partial y$, etc.) is the Schwarzian derivative (SD), \cite{Ovsienko(2005)}---the $ \bar{\bar{T}} $ is an action generated by an extension \cite{Neeb(2004)} of the diffeomorphism group (see \ref{jun1418-1})---also is an example of the diff. action. 
Indeed, $ \bar{\bar{T}}$ satisfies i) and ii) above: 
\begin{align} 
\textrm{i)}& \ \bar{\bar{T}}_{f^{-1}}( \bar{\bar{T}}_{g^{-1}}  ( \partial \mathcal{S}_0 / \partial z )^2  dz \otimes dz  )  
= \bar{\bar{T}}_{f^{-1}} [ \, (   ( \partial \mathcal{S}_0 ( g(y) )/ \partial y )^2 + \alpha \{g (y) , y \}  ) dy \otimes dy \, ]   \notag   \\ 
&= ( \, ( \partial f/ \partial x )^2 (  ( \partial \mathcal{S}_0 ( g(y) ) / \partial y )^2  + \alpha \{g (y) , y \})  +    \alpha \{ f (x) , x \} \,  ) dx \otimes dx   \notag    \\  
&= ( ( \partial \mathcal{S}_0 (g(f(x))) / \partial x )^2 + \alpha \{g (f(x)) , x \} ) dx \otimes dx = ( \bar{\bar{T}}_{(gf) ^{-1}} )  ( \partial \mathcal{S}_0 / \partial z )^2 dz \otimes dz           \label{feb2519-1}  \\ 
\textrm{ii)}& \ \bar{\bar{T}}_1 ( \partial \mathcal{S}_0 / \partial z )^2 dz \otimes dz ) 
= ( ( \partial \mathcal{S}_0 / \partial z )^2 + \alpha \{ z , z \}) dz \otimes dz = ( \partial \mathcal{S}_0 / \partial z )^2 dz \otimes dz  \ ,  \notag   
\end{align} 
where we used the chain rule $\{ g (f(x)), x \} = ( \partial f / \partial x)^2 \{ g(y), y \} + \{ f(x),x \}$ 
\footnote{\label{dec0316-1}
The chain rule \label{sep1413-2} $\{ g (f(x)), x \} = ( \partial f / \partial x)^2 \{ g(y), y \} + \{ f(x),x \}$ is obtained as follows. \cite{Ovsienko(2005)} 
Writing 
$ \partial ^2 g / \partial y ^2 = g_{yy} $ and so on,  we have 
$   \partial   g(f(x)) / \partial x = g_y f_x$, 
$   \partial ^2   g(f(x)) / \partial x ^2 = g_{yy}f_x ^2  + g_y f_{xx}$, and 
$\partial ^3g(f(x)) / \partial x^3  = g_{yyy}  f_x^ 3  + 3 g_{yy} f_x f_{xx} + g_y f_{xxx}$. Thus,  \\ 
$\displaystyle{\{ g (f(x)), x \} 
 = \frac{ g_{yyy} f_x^ 3 + 3 g_{yy} f_x f_{xx} + g_y f_{xxx} }{g_yf_x}
-\frac{3}{2} ( \frac{  g_{yy}f_x^2 + g_y f_{xx} }{ g_yf_x } )^2 
= f_x^ 2  \big( \frac{g_{yyy}}{g_y} -\frac{3}{2} ( \frac{g_{yy}}{g_y} )^2 \big) + \big( \frac{f_{xxx}}{f_x} -\frac{3}{2} ( \frac{f_{xx}}{f_x} )^2 \big) \ . 
}$
} 
of the SD in (\ref{feb2519-1}). 
\footnote{
The $\bar{T}$ acts on 1-D second-order covariant tensor fields, which is written, say, on $y$ as $\phi (y) dy \otimes dy$ with a function $\phi (y)$ of $y$. 
The quantity $\{ g (y) , y \} dy \otimes dy$ is the tensor field. 
Therefore $\bar{T}$ acts on it. 
Action of $\bar{T} _{f^{-1}}$ on it is given as $\bar{T} _{f^{-1}}  \{g (y) , y \} dy \otimes dy = \{g (y) , y \} ( \partial y / \partial x )^2 dx \otimes dx = ( \{ g (f(x)), x \} - \{ f(x) , x \} ) dx \otimes dx$. 
} 
\footnote{\label{sep1220-1}
Repeated applications of $\bar{\bar{T}}$ on $( \partial \mathcal{S}_0 / \partial z )^2 dz \otimes dz $ give the same form $(( \partial \mathcal{S}_0 ( \tilde{g}(w) )/ \partial w )^2 + \alpha \{ \tilde{g} (w) , w \}) dw \otimes dw$, where $z = \tilde{g}(w)$, on any coordinate system $w$ according to the chain rule of the SD. 
If we set coordinate system $w$ equal to the coordinate system $z$, we have an SD-free form $( \partial \mathcal{S}_0 / \partial w )^2 dw \otimes dw$ since $\{ z , z \} =0$, which means that quantities transformed by the $\bar{\bar{T}}$ takes an SD-free form on some coordinate system. 
}

According to the definitions of $f_*$ and $\bar{\bar{T}}$, the contraction on $y$ of $\partial /\partial x \otimes \partial /\partial x $ and $( \partial \mathcal{S}_0 / \partial z )^2 dz \otimes dz / 2m$ is given as 
\begin{equation*}
( \frac{\partial f}{\partial x} )^2 ( \frac{\partial}{\partial y} \otimes \frac{\partial}{\partial y} ) \cdot 
 \frac{1}{2m} (( \frac{\partial \mathcal{S}_0}{\partial y} )^2 + \alpha \{g(y) , y \}) dy \otimes dy 
= ( \frac{\partial f}{ \partial x})^2 \frac{1}{2m} (( \frac{\partial \mathcal{S}_0}{ \partial y })^2 + \alpha \{ g (y) , y \}) \ , 
\end{equation*} 
which we rewrite, using the chain rule of SD, as a quantity on $x$ as
\begin{equation}
( \frac{\partial f}{ \partial x})^2 \frac{1}{2m} ( ( \frac{\partial \mathcal{S}_0}{ \partial y })^2 + \cdots ) 
= \frac{1}{2m} \big( ( \frac{\partial \mathcal{S}_0 ( g(f(x)) ) }{ \partial x })^2  + \alpha \{ g(f(x)) , x \} - \alpha \{ f(x) , x \} \big)   \ .  \label{oct1619-3} 
\end{equation} 
We write the quantity on $x$, the right hand side (RHS) of (\ref{oct1619-3}), as $\bar{\bar{E}}_{\textrm{kin}}^{\textrm{HJ}} (x) $. 
The $\bar{\bar{E}}_{\textrm{kin}}^{\textrm{HJ}} (x) $ explicitly depends on coordinate systems $y=f(x)$ on which they are contracted and $z = g(f(x))$ on which the covariant one is SD-free (such $z$ exists, see footnote \ref{sep1220-1}). 
However, i) if coordinate systems on which they are contracted are restricted to $f(x) =(Ax + B) / (Cx +D)$, where $A , \cdots , D = const. \in \mathbb{C}$ and $AD - BC \neq 0$, we have $\{ f(x) , x \} = 0$ since $ \{ (A x + B ) / (C x +D ) , x \} = \{ x , x \} $ and $\{ x,x \} =0$ \cite{Faraggi(2000)} 
(indeed, for a diffeomorphism $y=f(x)$ of $\mathbb{RP}^1$, 
$\{ f(x), x \} =const. \in \mathbb{R} \leftrightarrow f(x) =(Ax + B)/(Cx+D)$), 
\footnote{
If $\{ f(x) , x \}$ is independent of coordinate systems, then $\{ f(x) , x \} = const.$ 
The last eq. is satisfied by a diffeomorphism of real projective line $\mathbb{RP}^1$ if and only if $\{ f(x) , x \} = 0$ with $f(x) =(Ax + B)/(Cx+D)$, which we see as follows. 
Solutions of a differential eq. $\{ f(x) , x \} = c$, where $x , f(x) \in \mathbb{RP}^1$ and $ c = const. \in \mathbb{R}$, are,  
if $c < 0$, $f (x) = (A e^{ x \surd (-2c)} + B )/ ( C  e^{ x \surd (-2c)}  +D)$, if $c >0$, $f (x) = (A \tan (x \surd (c/2))  + B )/ ( C \tan (x \surd (c/2))  +D)$, and
if $c =0$, $f (x) = (A x + B )/ ( C x +D)$, where $A , \cdots , D = const. \in \mathbb{C}$ and $AD-BC \neq 0$ (Mathematica solves the eq.). 
For $c \neq 0$, $f(x)$ is not a diffeomorphism \cite{Lee(2013)} of  $\mathbb{RP}^1$ because it is not bijective. 
For $c=0$, $f (x)$ is that of $\mathbb{RP}^1$. 
} 
and ii) if $g$ is a diffeomorphism from the base space to some other space (for example, base space $\to \mathcal{S}_0$-space; $y \mapsto \mathcal{S}_0 (y)$), 
the $\{ g(f(x)) , x \} $ does not explicitly depend on coordinate systems of the base space. 
According to i) and ii), $\bar{\bar{E}}_{\textrm{kin}}^{\textrm{HJ}} (x) $ with such $f$ and $g$ is independent of coordinate systems of the base space.

A note on topology and geometry of the base space: 
According to the appearance in (\ref{oct1619-3}) of the SD defined on real projective line $\mathbb{RP}^1 $, which we regard as one-point compactification \cite{Bredon(1993)} of real line $\mathbb{R}^1$, $\mathbb{RP}^1 = \mathbb{R}^1 \cup \{ \pm \infty \}$, the base space on which coordinate systems are defined is $\mathbb{RP}^1$. 
The $\mathbb{RP}^1 $, as a base space on which a coordinate-independent quantity $( \partial \mathcal{S}_0 / \partial x )^2 $ is defined, accompanies metric. 
The metric is $\partial / \partial x \otimes \partial / \partial x $ on $- \infty < x < + \infty$. 
We do not define it at $x = \{ \pm \infty \}$, according to which $( \partial \mathcal{S}_0 / \partial x )^2 $ is undefined there. 
The $y$ ($z \, , \cdots$) -coordinate of a point $p$ on $- \infty < x < + \infty$ is allowed to be $\{ \pm \infty \}$---a coordinate value $x = x_0$, where $|x_0 | < + \infty$, of the point $p$ may be transformed to $y = \{ \pm \infty \}$ by a passive coordinate transformation (see p.\pageref{mar3121-1}), say, $y= (Ax +B)/(Cx +D)$ of $\mathbb{RP}^1$---because values at $p$ of coordinate-independent quantities are irrelevant to coordinate-values of $p$. 
The active coordinate transformation (see p.\pageref{mar3121-1}) is meaningless under $\bar{\bar{T}}$ because it may move $( \partial \mathcal{S}_0 / \partial x )^2 $ at $x = x_0$ to $x = \{ \pm \infty \}$ at which $( \partial \mathcal{S}_0 / \partial x )^2 $ is undefined.

Under \label{may0121-1} $\bar{\bar{T}}$, a H-J eq. of CM $ ( \mathcal{S}_0 (x) / \partial x )^2 /2m =E$ is modified, according to (\ref{oct1619-3}), to 
\begin{equation} 
( ( \partial \mathcal{S}_0 ( g (f(x)) ) / \partial x )^2  + \alpha \{ g (f(x)) , x \} - \alpha \{ f(x) , x \}) /2m =E  \ ,    \label{may0321-1} 
\end{equation} 
where $E (>0)$ is kept intact because $\bar{\bar{T}}$ is irrelevant to $E$. 

If (\ref{may0321-1}) is independent of coordinate systems of the base space, we have $f(x) =(Ax + B)/(Cx+D)$ as we saw in the third last paragraph.


Renaming $g(f(x))$ as $q^ \star (x)$ and $\mathcal{S}_0 ( g (f(x)) ) = \mathcal{S}_0 ( q^ \star (x) ) $ as $\tilde{ \mathcal{S}}_0 (x)$, we rewrite (\ref{may0321-1}) with the $f$ as $( \tilde{ \mathcal{S}}_0 ' (x) )^2 + \alpha \{ q^ \star (x) , x \} = 2mE$. 
The last eq. determines a solution $\tilde{ \mathcal{S}}_0 ' (x)$ by itself if 
$q^ \star (x) = G ( \tilde{ \mathcal{S}}_0 (x) , \tilde{ \mathcal{S}}_0 ' (x) , \cdots ,  \tilde{ \mathcal{S}}_0 ^{(n-3 )} (x) )$, where $G$ is a function of $(  \cdots )$ and $3 \leq n < + \infty $. 
The solution $\tilde{ \mathcal{S}}_0 ' (x)$ is however unstable ($ \tilde{\mathcal{S}}_0 ' \to  \pm \infty$ for $x \to - \infty$ or $x \to + \infty$) unless $G ( \cdots ) = \tilde{ \mathcal{S}}_0 (x)$, which follows from stability analysis \cite{Perko(2001)} of the solution; see \ref{may1618-1}. 
We thus have an eq. $( \tilde{ \mathcal{S}}_0 ' (x) )^2 + \alpha \{ \tilde{ \mathcal{S}}_0 (x) , x \} = 2 m E$ which may have stable solutions $\tilde{ \mathcal{S}}_0 ' (x)$\,s. 
\footnote{\label{may2121-1} 
The eq. $( \tilde{ \mathcal{S}}_0 ' (x) )^2 + \alpha \{  \tilde{ \mathcal{S}}_0 (x) , x \} = 2 m E $ is written with original variables as $(\partial \mathcal{S}_0 (f(x)) / \partial x ) ^2  + \alpha \{ \mathcal{S}_0 (f(x)),x \}   = 2mE$. 
Note that from $q^ \star (x) = g(f(x))$ and $q^ \star (x)= \tilde{ \mathcal{S}}_0 (x) = \mathcal{S}_0 ( g (f(x)) )$, $g =  \mathcal{S}_0$ follows because $g$ and $\mathcal{S}_0$ behaves as a coordinate function \cite{Lee(2013)} giving coordinate values of points $p$'s of a base space. 
} 
We rename $\tilde{ \mathcal{S}}_0 (x)$ as $\mathcal{S}_0 (x)$ for readability, which is allowed because $\tilde{ \mathcal{S}}_0 (x)$ is the unknown of the differential     eq. 
We fix $\alpha$ to $\hbar ^2 /2$, where $\hbar$ is a reduced Planck constant $h / 2 \pi$, because eventually it is determined to that by atomic-scale experiments. 
We thus have
\begin{equation}
 (\partial \mathcal{S}_0 (x) / \partial x ) ^2 /2m  + \hbar ^2 \{ \mathcal{S}_0 (x),x \} /4m  =E  \ .     \label{apr2421-oct0116-1}
\end{equation}

We solve (\ref{apr2421-oct0116-1}) for $E > 0$. 
We rewrite (\ref{apr2421-oct0116-1}), 
using a formula $(\partial \mathcal{S}_0 / \partial x ) ^2 + \hbar ^2 \{ \mathcal{S}_0 (x),x \} /2 = \hbar ^2 \{ e^{ 2i \mathcal{S}_0 / \hbar } ,x \} /2$, 
\footnote{
Use the chain rule (footnote \ref{dec0316-1}): 
$\{ g (f(x)), x \} = ( \partial f / \partial x)^2  \{ g(y), y \} + \{ f(x),x \}$ with $g(f(x)) =  e^{ 2i f (x) / \hbar }$ and $y = f(x) = \mathcal{S}_0 (x)$. 
The mapping $x \mapsto g(f(x))$ is not a diffeomorphism of $\mathbb{RP}^1$. 
Indeed, it is not bijective. 
It however is a local diffeomorphism \cite{Lee(2013)} on $\mathbb{RP}^1$. 
} 
as $\hbar ^2 \{ e^{ 2i \mathcal{S}_0 / \hbar } ,x \} / 4m =E \ \cdots (*1)$. 
The $(*1)$ has a solution $ \mathcal{S} _0 (x) = \hbar k x$, where $k= \hbar ^{-1} \surd (2mE)$. 
Because $\{ f(x) , x \} = \{ (Af(x) +B)/( Cf(x) +D ) , x \}$, \cite{Faraggi(2000)} $\bar{\bar{\mathcal{S}}}_0 (x)$ constructed from $e^{2 i \hbar k x / \hbar} = e^{2 i k x}$ as 
\begin{equation*}
 e^{ 2i \bar{\bar{\mathcal{S}}}_0 / \hbar } 
= \frac{\tilde{A} e^{ 2i kx } + \tilde{B} }{ \tilde{C} e^{ 2ikx } +\tilde{D} }
= \frac{\tilde{A} e^{ i kx  } + \tilde{B}  e^{ - i kx  } }{ \tilde{C} e^{ i kx  } + \tilde{D} e^{- i kx  } }  
= \frac{A \sin kx +B \cos kx }{ C \sin kx +D \cos kx } 
\end{equation*} 
also satisfies $(*1)$, though $\bar{\bar{\mathcal{S}}}_0(x)$ may not stay in $\mathbb{R}$. 
Differentiating both sides, we have 
$\partial \bar{\bar{\mathcal{S}}}_0 / \partial x 
=   \frac{- i \hbar k (AD - BC )}{2AC} / ( ( \sin kx + \frac{AD+BC}{2AC} \cos kx )^2 - ( \frac{AD - BC}{2AC})^2 \cos ^2 kx   ) $, which stays in $ \mathbb{R} ( \neq 0 ) $ if $l_1 := i(AD-BC)/2AC = const. \in \mathbb{R} \backslash \{ 0 \}$ and $l_2 := (AD+BC)/2AC =  const. \in \mathbb{R}$. 
Thus 
\begin{equation}
\partial \bar{\bar{\mathcal{S}_0}} / \partial x = - \hbar k l_1 / ( \ ( \sin kx + l_2 \cos kx )^2 + l_1^2 \cos ^2 kx \ )  \in \mathbb{R} \backslash \{ 0 \}  \ ,    \label{oct0116-2} 
\end{equation}
where $l_1 \in \mathbb{R} \backslash \{ 0 \}$ and $l_2 \in \mathbb{R}$, is a real ($\partial \bar{\bar{\mathcal{S}_0}} / \partial x  \in \mathbb{R}$) solution of $(*1)$. 

The (\ref{oct0116-2}) is a general solution of (\ref{apr2421-oct0116-1}), a 2nd-order ODE for $\partial \mathcal{S}_0 / \partial x$, because it has two parameters, $ l_1$ and $l_2 $. 
The $( l_1 , l_2 )$ encodes amplitude and phase of the periodic $\partial \bar{\bar{\mathcal{S}}}_0 / \partial x$. 
\footnote{
We have 
$( \sin kx + l_2 \cos kx )^2 + l_1^2 \cos ^2 kx=  2^{-1}( l_1^2+l_2^2 +1) + l_2 \sin 2kx + 2^{-1}(l_1^2+l_2^2 -1)\cos 2kx = R_1 + R_2 \cos (2k(x - \theta ))$, where $R_1= 2^{-1}( l_1^2+l_2^2 +1)$, $R_2 = \surd (l_2 ^2 + 4^{-1}(l_1^2+l_2^2 -1)^2 )$, and $2 k \theta = \tan ^{-1} (2l_2 / (l_1^2+l_2^2 -1))$. 
The phase is specified by the $\theta$. 
The peak-to-peak amplitude is given as $| \hbar k l_1 (R_1+R_2 )^{-1} - \hbar k l_1 (R_1 - R_2 )^{-1} | = | \hbar k l_1^{-1} \surd ( 4 l_2^2 + (l_1^2+l_2^2 -1)^2  )|$. 
} 
Because $l_1 \in \mathbb{R} \backslash \{ 0 \}$ and $l_2 \in \mathbb{R}$ corresponds to $\mathcal{S}_0 '(x) |_{x = x_1} \in \mathbb{R} \backslash \{ 0 \}$ and $\mathcal{S}_0 '' (x) |_{x = x_1} \in \mathbb{R}$,where $x_1 \in \mathbb{R}$, respectively, 
\footnote{
From (\ref{oct0116-2}), we have 
$\partial ^2 \bar{\bar{\mathcal{S}_0}} / \partial x ^2 =  \hbar k^2 l_1 ( ( 1 -l_1^2 -l_2^2 ) \sin 2kx + 2 l_2 \cos 2kx ) / (( \sin kx + l_2 \cos kx )^2 + l_1^2 \cos ^2 kx )^2$. 
From (\ref{oct0116-2}) and the last eq., we have $\bar{\bar{\mathcal{S}_0}} ' = - \hbar k l_1$ and $\bar{\bar{\mathcal{S}_0}} '' = -2 \hbar k^2 l_1 l_2$ at $x  = x_0 = \pi /2k$. 
The $\bar{\bar{\mathcal{S}_0}} ' ( \neq 0 )$ and $\bar{\bar{\mathcal{S}_0}} '' $ take arbitrary values at $x_0$ since $l_1 (\neq 0)$ and $l_2 $ are arbitrary. 
At $x = x_1 \neq \pi /2k$, $\bar{\bar{\mathcal{S}_0}} ' ( \neq 0 )$ and $\bar{\bar{\mathcal{S}_0}} '' $ also take arbitrary values because i) we can, keeping the amplitude of $\bar{\bar{\mathcal{S}_0}} '$ invariant, change the value of the phase $\theta$ from $\theta _ { \, 0 }$ to $\theta _1 $ adjusting $l_1$ and $l_2$, and ii) $(\bar{\bar{\mathcal{S}_0}} ' , \bar{\bar{\mathcal{S}_0}} '' )| _ {x = x_0} = (\bar{\bar{\mathcal{S}_0}} ' , \bar{\bar{\mathcal{S}_0}} '' )| _ {x = x_1}$ if 
the amplitude at $x_0$ and $x_1$ of $\bar{\bar{\mathcal{S}_0}} '$  are equal and $(x_0 - \theta _{ \, 0 }) =  (x_1 - \theta _1 )$. 
} 
solutions of (\ref{apr2421-oct0116-1}) are stable for any initial condition $\mathcal{S}_0 '(x) ( \neq 0 )  , \mathcal{S}_0 '' (x) |_{x = x_1} \in \mathbb{R}$. 

We thus have an eq. (\ref{apr2421-oct0116-1}) being independent of coordinate systems, determining ${\mathcal{S}_0} '$ by itself, and having stable solutions. 
If the $f$ and $g$ introduced by the $\bar{\bar{T}}$ are not those given above, the three properties are unsatisfied. 
An eq. not satisfying the three properties is unqualified as a H-J eq.

If $l_1 = -(1+ \epsilon _1 )$ and $l_2 = \epsilon _2$ for $\epsilon _1 , \epsilon _2 \ll 1$, we have a solution $\partial \bar{\bar{\mathcal{S}}}_0 / \partial x = \hbar k ( 1+ \epsilon \cos ( 2kx + \theta ) + \mathcal{O}( \epsilon ^2 )) $, where $\epsilon = \surd ( \epsilon _1 ^2 + \epsilon _2 ^2 )$. 
If $l_1 = \mp 1$ and $l_2 =0$, we have a solution $\partial \bar{\bar{\mathcal{S}}}_0  / \partial x = \pm \surd (2mE )$. 
We call a solution, a system, and so on having $\epsilon _1 , \epsilon _2 \ll 1$ as a semiclassical solution, a semiclassical system, and so on.

We show that a 3-D eq. reducing to (\ref{apr2421-oct0116-1}) exists for 3-D systems comprised of a massive particle $m$ and a potential field $V(\textbf{x})$. 
\footnote{
We do not construct 3-D H-J eq. of GCM applying 3-D extended diff. action on the 3-D second-order covariant tensor in the classical 3-D H-J eq. as we did in 1-D space because we do not know the formula of the 3-D extended diff. action on the 3-D tensor. The 3-D extended diff. action of $\textbf{y} = \textbf{g}^{-1} ( \textbf{z})$, where  $ \textbf{g}( \textbf{y})  = (g^1 ( y^1,y^2 , y^3 ) , \cdots ,  g^3 ( y^1,y^2 , y^3 ) )$, on a 3-D second-order covariant tensor field $\phi _{ij} ( \textbf{z}) dz^i \otimes dz^j$ should be given as 
$\bar{\bar{T}}_ {\textbf{g}^{-1}} ( \phi _{ij} ( \textbf{z}) dz^i \otimes dz^j )  
 = (( \partial z^p / \partial y ^i)( \partial z^q / \partial y ^i)  \phi _{pq }( \textbf{z}) 
 + \beta \textsf{S} _{ij}( \textbf{g}( \textbf{y}) ,   \textbf{y}   )) dy^i \otimes dy^j $, 
where  $\beta ( \in \mathbb{R}) \neq 0$ and $ \textsf{S} _{ij}$ satisfies a chain rule  
$ \textsf{S} _{ij}( \textbf{g}( \textbf{f} ( \textbf{x})) ,  \textbf{x}  ) 
 = ( \partial y ^p / \partial  x^i )( \partial y ^q / \partial  x^j ) 
 \textsf{S} _{pq}( \textbf{g}( \textbf{y}) , \textbf{y} ) +  \textsf{S} _{ij}( \textbf{f}( \textbf{x}) ,  \textbf{x}     ) $, where $\textbf{y} = f ( \textbf{x})$. 
The explicit form of $ \textsf{S} _{ij}$, a 3-D version of the 1-D Schwarzian derivative, is not reported in the literature as far as the present author knows (cf., e.g., Ref. \refcite{Bouarroudj(2000)}). 
} 
In the case that variables separate on a Cartesian coordinate system (in the separable case), the 3-D eq. reducing to (\ref{apr2421-oct0116-1}) is given as a sum of eqs. in $x^1 (=x)$-, $x^2 (=y)$-, and $x^3 (=z)$-direction as 
\begin{equation*} \textstyle 
 \sum _{i=1} ^3  (\partial \mathcal{S}_{0 \, i} (x ^i) / \partial x^i ) ^2 /2m  + \sum _{i=1} ^3 \hbar ^2 \{ \mathcal{S}_{0 \, i}  (x^i ),x ^i \} /4m + \sum _{i=1} ^3 V_i(x ^i) = \sum _{i=1} ^3 E_i \ . 
\end{equation*} 
Note that $\bar{\bar{T}}$ is irrelevant to $V$ as well as $E$. 
The above forms of the terms are effective in the separable case. 
Forms effective in separable and inseparable cases and reduce to $\sum (\partial \mathcal{S}_{0 \, i } (x^i) / \partial x^i ) ^2 $, $\sum V_i$, and $\sum E_i$ in the separable case are $ (\nabla  \mathcal{S}_0 (\textbf{x} )) ^2 $, $V (\textbf{x} ) $, and $E$ respectively, which we see from CM. 
A form effective in both cases and reduces to $\sum \{ \mathcal{S}_{0 \, i} (x ^i ),x ^i \}$ in the separable case is given as follows. 
Using an expression \cite{Osgood(1998)} 
\begin{equation*}
 \{ \mathcal{S}_{0 \, i} (x^i),x^i \} = -2 ( \partial \mathcal{S}_{0 \, i} / \partial x^i )^{ 1/2 } (\partial ^2 /  \partial { x^i }^2 ) ( \partial \mathcal{S}_{0 \, i } / \partial x ^i )^{- 1/2}  = -2 R_i ^{- 1} \partial ^2  R_i / \partial {x^i} ^2   \ , 
\end{equation*}
where $R_i = ( \partial \mathcal{S}_{0 \, i } / \partial x^i )^{-1/2}$, 
\footnote{
Actually, $R_i = const. ( \partial \mathcal{S}_{0 \, i } / \partial x^i )^{-1/2}$, where $const. \neq 0$. 
We omit $const.$ for simplicity without losing generality. 
} 
we write $\sum \{ \mathcal{S}_{0 \, i } (x ^i),x ^i \}$ as $\sum \{ \mathcal{S}_{0 \, i } (x ^i),x ^i \} = - 2 \nabla ^2 R / R$, where 
$R = R( \textbf{x} )  = R_x (x) R_y (y) R_z (z) = ( ( \partial \mathcal{S}_{0 \, x } / \partial x ) ( \cdots  )( \partial \mathcal{S}_{0 \, z } / \partial z )) ^{-1/2} $. 
The last eq. is equivalent to $\nabla (R^2 \nabla \mathcal{S}_0 ) =0$ 
\footnote{
We show $R_x (x) R_y (y) R_z (z) = ( ( \partial \mathcal{S}_{0 \, x } / \partial x ) ( \partial \mathcal{S}_{0 \, y } / \partial y  )( \partial \mathcal{S}_{0 \, z } / \partial z )) ^{-1/2} \Leftrightarrow  \nabla (R^2 \nabla \mathcal{S}_0 ) =0  $ for $\mathcal{S}_0 ( \textbf{x} ) =  \mathcal{S}_{0 \, x}(x) + \mathcal{S}_{0 \, y}(y) + \mathcal{S}_{0 \, z}(z)$. 
We have 
$\nabla (R^2 \nabla \mathcal{S}_0 ) =0 
\Leftrightarrow 
2 \nabla R \nabla \mathcal{S}_0 + R \nabla ^2 \mathcal{S}_0 =0 
\Leftrightarrow 
\sum _ { i =1}^3 ( 2 R_i ' \mathcal{S}_{0 \, i} ' + R_i \mathcal{S}_{0 \, i} '' ) = 0 
 \Leftrightarrow  
2 R_i ' \mathcal{S}_{0 \, i} ' + R_i \mathcal{S}_{0 \, i} '' ) = 0 
 \Leftrightarrow  
R_i = ( \mathcal{S}_{0 \, i} ' ) ^{-1/2} $, where 
$R_i ' = \partial R_i /\partial x^i$ and so on. 
We have $\sum _ { i =1}^3 ( 2 R_i ' \mathcal{S}_{0 \, i} ' + R_i \mathcal{S}_{0 \, i} '' ) = 0 
 \Rightarrow  
2 R_i ' \mathcal{S}_{0 \, i} ' + R_i \mathcal{S}_{0 \, i} ''  = 0 $ 
because we have i) $2 R_i ' \mathcal{S}_{0 \, i} ' + R_i \mathcal{S}_{0 \, i} '' = c_i$, where $c_i$ is a const. satisfying $c_1 + c_2 +c_3 =0$, because $\mathcal{S}_{0 \, i}$ is an arbitrary diffeomorphism, and ii) $c_1 = c_2 = c_3 $ because each direction is identical. 
} 
meaningful even in inseparable---$\mathcal{S}_0 ( \textbf{x} ) \neq \mathcal{S}_{0 \, x }(x) + \mathcal{S}_{0 \, y }(y) + \mathcal{S}_{0 \, z } (z)$---case. 
Thus, the form reducing to $\sum \{ \mathcal{S}_{0 \, i} (x ^i ),x ^i \}$ is $- 2 \nabla ^2 R / R$, where $R = R( \textbf{x} )$ is a function satisfying $\nabla (R^2 \nabla \mathcal{S}_0 ) =0$. Thus the 3-D eq. reducing to (\ref{apr2421-oct0116-1}) is 
\begin{equation} 
 - \hbar ^2 \nabla ^2 R / 2mR + ( \nabla \mathcal{S}_0 )^2 /2m + V(\textbf{x}) = E \ , \ \ \ \ \nabla ( R^2 \nabla \mathcal{S}_0 ) = 0 \ . \label{jul1510-1} 
\end{equation}

We define a generalized CM (GCM) as a mechanics comprised of a H-J eq. given by (\ref{jul1510-1}) and an eq. of motion (EoM) related through an action           integral. 
The EoM need not be derived from Lagrangian (cf. p.\pageref{dec1017-1}). 
The `related through' means that a solution of the H-J eq. is equal to an action integral along a solution $\textbf{q}(t)$ of the EoM. 
The equality is equivalent to 
$\nabla \mathcal{S}_0 ^{HJ} ( \textbf{x} ) = \nabla \mathcal{S}_0 ^{NW} ( \textbf{x} )$, where $\nabla \mathcal{S}_0 ^{HJ} ( \textbf{x} )$ is a solution of the H-J eq. and  $\nabla \mathcal{S}_0 ^{NW} ( \textbf{x} )$ is a momentum constructed from a solution $\textbf{q}(t)$ of the EoM (NW for Newton), which follows from expressions of boundary terms of the action integral (see (\ref{apr2619-1-051606-k}) and (\ref{sep1016-2})).

Assume that the GCM exists. 
Then, $\nabla \mathcal{S}_0 ^{HJ} ( \textbf{x} ) = \nabla \mathcal{S}_0 ^{NW} ( \textbf{x} )$. 
Since $\nabla \mathcal{S}_0 ^{HJ} ( \textbf{x} )$ for $V=0$ is fixed by four parameters $( \textbf{x} , \bm{l} _1 , \bm{l} _2 , \textbf{k} )$---cf. (\ref{oct0116-2})---, the $\nabla \mathcal{S}_0 ^{NW} ( \textbf{x} )$ for $V=0$ also is fixed by four parameters. 
Since the $\nabla \mathcal{S}_0 ^{NW} ( \textbf{x} )$ is fixed by EoM's solution determined by EoM's initial conditions, the EoM is a 4th-order ODE having four initial conditions $( \textbf{q}(t) , \dot{\textbf{q} }(t) , \ddot{ \textbf{q}}(t) , \dot{\ddot{ \textbf{q}}}(t))|_{ \, t = t_0 } $. 
Note that $\nabla \mathcal{S}_0 ^{NW} ( \textbf{x} )$ is fixed by $( \textbf{q}(t) , \dot{\textbf{q} }(t) , \ddot{ \textbf{q}}(t) , \dot{\ddot{ \textbf{q}}}(t))|_{ \textbf{q}(t) = \textbf{x} } $, and that $\nabla \mathcal{S}_0^{HJ} ( \textbf{x}; \bm{l} _1 , \bm{l} _2 , \textbf{k}  ) = \nabla \mathcal{S}_0^{NW} ( \dot{\textbf{q} } , \ddot{ \textbf{q}} , \dot{\ddot{ \textbf{q}}} )|_{ \textbf{q}(t) = \textbf{x}}$ along a solution $\textbf{q}(t)$.

The EoM is subordinate to the H-J eq. in the following sense. 
The value of $\mathcal{S}_0 ^{NW}$ along a solution $ \textbf{q}(t)$ of the EoM does not depend on $V(\textbf{x} )$ away from the $\textbf{q}(t) $, while that of $\mathcal{S}_0 ^{HJ}$ along the $ \textbf{q}(t)$ does since the EoM (the H-J eq.) is an ODE (a PDE). 
Nevertheless, $ \mathcal{S}_0^{NW} = \mathcal{S}_0^{HJ}$ has to hold true for GCM to be consistent. 
To hold the eq. true, we introduce into the EoM a force term $-\partial Q / \partial \textbf{q}$ generated by a pseudo-potential $ Q ( \textbf{q})$. 
We determine the $Q(\textbf{q})$ as follows. 
We first determine an expression of $\nabla \mathcal{S}_0 ^{NW} ( \dot{\textbf{q}}, \ddot{\textbf{q}}, \dot{\ddot{\textbf{q}}}) $ for free system (see p.\pageref{may1021-1} and p.\pageref{dec1017-1}). 
The expression is effective also for systems having $V(\textbf{x} ) \neq 0$. 
Using the expression, we next determine a trajectory $\textbf{q}(t)$ under $V(\textbf{x} ) \neq 0$ independently from $Q$ solving an ODE for $\textbf{q}(t)$: $\nabla \mathcal{S}_0 ^{NW} ( \dot{\textbf{q}}, \ddot{\textbf{q}}, \dot{\ddot{\textbf{q}}}) = \nabla \mathcal{S}_0 ^{HJ} (\textbf{x}) |_{\textbf{x}= \textbf{q}} \ \ \cdots (*2)$ for a given solution  $\nabla \mathcal{S}_0 ^{HJ}$ of (\ref{jul1510-1}) and a set of EoM's initial conditions  $(\textbf{q} , \dot{\textbf{q}}, \ddot{\textbf{q}}, \dot{\ddot{\textbf{q}}})|_{t=t_0}$ compatible with the $\nabla \mathcal{S}_0 ^{HJ}$. 
We lastly determine a force term $-\partial Q / \partial \textbf{q}$ in the EoM so as to make the solution of the $(*2)$ a solution of the EoM. 
The $Q$ approaches zero for $ \hbar ^2 /m \to 0$ or, because of translational invariance, for $V \to const.$ 
Thus, the EoM contains a quantity $Q$ depending on the solution of the H-J eq. and initial conditions of the EoM.

We show that there exists a GCM of which EoM---it is effective for classical and semiclassical systems---is derived from a second-order Lagrangian \cite{Leon(1985)} $L = L(\textbf{q} ,  \dot{ \textbf{q}}, \ddot{ \textbf{q}} )$. 

A Lagrangian $L$ for $V(\textbf{x}) = 0$ giving the fourth-order EoM is restricted by dimensional analysis to a third-order one \label{feb1618-1} $L = m \dot{\textbf{q}} ^2 /2  + c_1 \hbar \ddot{\textbf{q}} \dot{\textbf{q}} / \dot{\textbf{q}} ^2  +   \hbar ^2 ( c_{2a} \dot{\ddot{\textbf{q}}}\dot{\textbf{q}} / m \dot{\textbf{q}}^4 + c_{2b} \ddot{\textbf{q}}^2 / m \dot{\textbf{q}}^4 ) + c_3 \hbar ^3  \ddot{\textbf{q}}^2 \ddot{\textbf{q}}  \dot{\textbf{q}} / m^2 \dot{\textbf{q}}^8 + \cdots $, 
where $c_1 , c_{2a}, \cdots = const. \in \mathbb{R}$ are coefficients to be determined, $\dot{\textbf{q}} ^2 =  \dot{q}_1^2 + \dot{q}_2^2 + \dot{q}_3 ^2 $, $\dot{\textbf{q}} ^4 =  \dot{\textbf{q}} ^2 \cdot \dot{\textbf{q}} ^2$, and so on. 
For the $L$, the variation $\delta \mathcal{S}_0$ is given as 
\begin{align}
&\delta \mathcal{S} _0 (\textbf{x} , \dot{\textbf{x}} ,  \ddot{\textbf{x}}      )
= \int _{ \textbf{x}_0 , \dot{\textbf{x}}_0 , \ddot{\textbf{x}}_0 , \, t_0 }  ^{ \textbf{x} , \dot{\textbf{x}} , \ddot{\textbf{x}} , \, t} \big( \ L(  \textbf{q}  +  \delta   \textbf{q}  , \dot{ \textbf{q} }+  \delta  \dot{ \textbf{q} },\ddot{ \textbf{q} }+  \delta  \ddot{ \textbf{ \textbf{q} } } , \dot{\ddot{ \textbf{q} }}+  \delta  \dot{\ddot{ \textbf{ \textbf{q} } }} ) -  L(  \textbf{q} ,\dot{ \textbf{q} },\ddot{ \textbf{q} }  , \dot{ \ddot{ \textbf{q} }}) \ \big) d \tilde{t}       \notag \\ 
&= \int ( \frac{ \partial L}{\partial  \textbf{q} }  \delta   \textbf{q}  + \frac{ \partial L}{\partial \dot{ \textbf{q} }}  \delta  \dot{ \textbf{q} } + \frac{ \partial L}{\partial \ddot{ \textbf{q} }}  \delta  \ddot{ \textbf{q} } + \frac{ \partial L}{\partial \dot{\ddot{ \textbf{q}} }}  \delta  \dot{\ddot{ \textbf{q} }}  ) d \tilde{t}    
= \int ( \frac{ \partial L}{\partial  \textbf{q} }  - \frac{d}{dt} \frac{ \partial L}{\partial \dot{ \textbf{q} }}  + \frac{d^2}{dt^2} \frac{ \partial L}{\partial \ddot{ \textbf{q} }} - \frac{d^3}{dt^3} \frac{ \partial L}{\partial \dot{\ddot{ \textbf{q} }}}  )  \delta   \textbf{q}  d \tilde{t}   \notag \\  
& \ \ \ \  
+ (\frac{ \partial L}{\partial \dot{ \textbf{q} }}  - \frac{d}{dt} \frac{ \partial L}{\partial \ddot{ \textbf{q} }}  + \frac{d^2}{dt^2} \frac{ \partial L}{\partial \dot{\ddot{ \textbf{q} }}}    )  \delta   \textbf{q}  | _{t_0} ^t + (\frac{ \partial L}{\partial \ddot{ \textbf{q} }}  - \frac{d}{dt} \frac{ \partial L}{\partial \dot{\ddot{ \textbf{q} }}})  \delta  \dot{ \textbf{q} } | _{t_0}  ^t + (\frac{ \partial L}{\partial \dot{\ddot{ \textbf{q} }}}   )  \delta  \ddot{ \textbf{q} } |_{t_0}  ^t  \ .
  \label{apr2619-1-051606-k}    
\end{align} 
From the integrand of the RHS, we have an E-L eq. 
$ \partial L / \partial \textbf{q} -  \cdots  - ( d^3 / dt^3 )  ( \partial L / \partial \dot{\ddot{ \textbf{q} }} ) =0$. 
We call boundary terms, $ \partial L / \partial \dot{ \textbf{q} } - \cdots + ( d ^2 / dt^2 )  \partial L / \partial \dot{\ddot{ \textbf{q} }} \, $, 
$ \partial L/ \partial \ddot{ \textbf{q} } - (d / dt)  \partial L / \partial \dot{\ddot{ \textbf{q} }}$, and $ \partial L / \partial \dot{\ddot{ \textbf{q} }}$ as $\textbf{p}_{(1)}$, $\textbf{p}_{(2)}$, and $\textbf{p}_{(3)}$ respectively.

We show that $c_{2a} = 0$, that is, the Lagrangian is of second-order. 
The $\textbf{p}_{(3)}$ is defined as $\textbf{p}_{(3)} = \lim _{ \delta \ddot{\textbf{q}} \to 0 } \delta \mathcal{S}_0 / \delta  \ddot{\textbf{q}} |_t$, 
where $\delta \mathcal{S}_0 = \mathcal{S}_0  ( \ddot{\textbf{q}} + \delta \ddot{\textbf{q}} |_t) - \mathcal{S}_0 ( \ddot{\textbf{q}}|_t ) $, 
where $\mathcal{S}_0 ( \ddot{\textbf{q}} + \delta \ddot{\textbf{q}}|_t )$ is a definite integral along a solution $\textbf{q}(t)$ of the E-L eq. from $( \textbf{q} , \dot{\textbf{q}} , \ddot{\textbf{q}} ) | _ {t_0}$ to $( \textbf{q}  , \dot{\textbf{q}} , \ddot{\textbf{q}} + \delta \ddot{\textbf{q}} )| _ t$. 
Likewise for $\mathcal{S}_0 ( \ddot{\textbf{q}} |_t ) $. 
When we evaluate $ \lim _{ \delta \ddot{\textbf{q}} \to 0 } \delta \mathcal{S}_0 / \delta \ddot{\textbf{q}} |_t$, 
the three lower limits $( \textbf{q} , \dot{\textbf{q}}  , \ddot{\textbf{q}}  ) | _ {t_0}$ and two $( \textbf{q} , \dot{\textbf{q}} ) | _ t$ of the three upper limits of the integral have to be the same for both $\mathcal{S}_0 ( \ddot{\textbf{q}} + \delta \ddot{\textbf{q}} |_t)$ and $\mathcal{S}_0 ( \ddot{\textbf{q}}|_t )$. 
A solution of the fourth-order E-L eq. passing through the $( \textbf{q} , \dot{\textbf{q}} , \ddot{\textbf{q}} ) | _ {t_0}$ and the $( \textbf{q} , \dot{\textbf{q}} ) | _ t$ however cannot pass through both $\ddot{\textbf{q}} |_t$ and $( \ddot{\textbf{q}} + \delta  \ddot{\textbf{q}} )|_t$ because the solution is completely fixed by four initial conditions, say, by $( \textbf{q}  , \dot{\textbf{q}} ) | _ {t_0}$ and $( \textbf{q} , \dot{\textbf{q}} ) | _ t$. 
Accordingly, $\textbf{p}_{(3)}$ is indefinable. 
Therefore $c_{2a} = 0$.

From here on, we use $\textbf{p}$ and $\nabla \mathcal{S}_0$ interchangeably.

We determine $c_1 $. \label{may1021-1} 
The Lagrangian with $c_1 \neq  0$ does not make the system time-reversal invariant because, under time reversal $t \to -t$, $ c_1 \hbar \ddot{\textbf{q}} \dot{\textbf{q}} / \dot{\textbf{q}} ^2$ changes sign unlike $m \dot{\textbf{q}} ^2 /2 $ and $\hbar ^2 ( \cdots ) $. 
Indeed, an eq. $ \textbf{p} ^ {EL} (-t) = - \textbf{p} ^ {EL} (t)$ does not hold true, where $\textbf{p} ^ {EL} = ( \textbf{p}_{(1)}\dot{\textbf{q}} + \textbf{p}_{(2)} \ddot{\textbf{q}}) \dot{\textbf{q}} / \dot{\textbf{q}} ^2$ (see \S \ref{jun1609-1}). 
The system however is time-reversal invariant in QM. \cite{Messiah(1961)} \cite{Sozzi(2008)} 
Then, it is natural to consider that the system in GCM also is. 
We accordingly set $c_1 =0$.


We determine $c_{2b}$. 
The 1-D free E-L eq. 
$m \ddot{q} - 2 c_{2b} \hbar ^2 \ddot{\ddot{q}} / m \dot{q}^4 + \cdots =0$ 
derived from 1-D version of the candidate 3-D free Lagrangian with $c_1=c_{2a} =0$ has a solution $ q(t) = vt + A \cos \omega t + \cdots $ for $ c_{2b} <0$, where $v, A , \omega = const. \in \mathbb{R}$, 
$ \omega A /v \ll 1$, and $\omega = mv^2 / \hbar \surd ( - 2 c_{2b} ) $. 
For the solution, we have $ p ^{EL} = p_{(1)} + p_{(2)} \ddot{q} / \dot{q}  = const. + c_{2b} \hbar ^2 A^2 \omega ^4 \cos 2\omega  t /mv^5 + \cdots $. 
It is equal to the $p ^{HJ} = \hbar k ( 1 + \epsilon \cos 2 k x + \cdots )$ with $2 \omega = 2 kv$ for $c_{2b} =-1/2$. 
Thus $c_{2b} =-1/2$. 
Note \label{jun1619-1} that the $\omega $ is not the $\omega ^{QM}$ of de Broglie relation $E  = \hbar \omega ^{QM}= ( \hbar \textbf{k} )^2/2m $. 
De Broglie relation relates energy $E$ and momentum $\textbf{p}$ to $\omega$ and $\textbf{k}$ of matter wave. 
\footnote{
Removing $\textbf{v}$ from $E = mc^2 / \surd ( 1- \textbf{v}^2 / c^2 )$ and $\textbf{p} = m \textbf{v} / \surd ( 1- \textbf{v}^2 / c^2 )$, we have 
$E/c = \surd ( \textbf{p} ^2 + m^2c^2 ) $. 
Inserting $E= \hbar \omega '$ and $\textbf{p} = \hbar \textbf{k}$, which is true for light quanta, to the eq., we have $\omega '/c = \surd (  \textbf{k} ^2 + m^2c^2 / \hbar ^2 )$. 
From the last eq., $\omega ' = mc^2 / \hbar + \hbar \textbf{k} ^2 / 2m + \cdots $ follows. 
Defining $\omega ^{QM} = \omega ' - mc^2 / \hbar$, we have $\hbar \omega ^{QM} \simeq ( \hbar \textbf{k}) ^2 / 2m$. \cite{Pauli(1973)}
} 
It however does not relate the $\omega$ and $\textbf{k}$ of matter wave to those of oscillatory linear motion of matter. 
Indeed, we have $\hbar \omega = \hbar kv = v \surd (2mE) \simeq mv^2$.

From $c_1 = c_{2a} = 0$ and $c_{2b} =-1/2$, a 3-D free Lagrangian is determined up to $\hbar ^2$ as $L= m  \dot{\textbf{q} } ^2 / 2 -  \hbar ^2  \ddot{ \textbf{q}}^2 / ( 2m  \dot{ \textbf{q}}^4 ) $. 
Incorporating $V ( \textbf{q})$ and $Q ( \textbf{q})$, we have a 3-D Lagrangian
\begin{equation}
  L( \textbf{q} , \dot{\textbf{q} }  , \ddot{ \textbf{q}} ) = m  \dot{\textbf{q} } ^2 / 2 -  \hbar ^2  \ddot{ \textbf{q}}^2 / ( 2m  \dot{ \textbf{q}}^4 ) -V ( \textbf{q}) - Q ( \textbf{q})  \ ,  \label{feb1414-1} 
\end{equation}
which gives the EoM and $ \nabla \mathcal{S}_0^{EL} $ as 
\begin{align}
m \ddot{q}_i &= - \hbar ^2  ( \ddot{ \ddot{q}}_i / \dot{\textbf{q} }^4 + \cdots ) / m -  \partial V(\textbf{q}) / \partial q_i - \partial Q(\textbf{q}) / \partial q_i  \ ,   \label{aug0316-1}   \\ 
\partial \mathcal{S}_0^{EL} /\partial q_i &= m \dot{q}_i + \hbar ^2 ( \dot{ \textbf{q}} \dot{\ddot{\textbf{q}}} / \dot{\textbf{q}}^6 + \ddot{ \textbf{q}} ^2 / \dot{ \textbf{q}}^6 - 4 ( \dot{ \textbf{q}} \ddot{ \textbf{q}})^2 / \dot{ \textbf{q}}^8 ) \dot{q}_i /m       \ .  \label{jul0819-2}    
\end{align}
For full form of (\ref{aug0316-1}), see (\ref{june1308-2apr15}).

Two comments: 
i) The formalism breaks down at $\dot{\textbf{q}} =0$ because of $\dot{\textbf{q}} $ in the denominator of the Lagrangian (\ref{feb1414-1}). 
No solution $\textbf{q}(t)$ of the E-L eq. however takes value $\dot{\textbf{q}} =0$ because $\textbf{p}_{(1)}= const.$ and $E =const. $ satisfied by a free particle (see \ref{0428ab}) are incompatible with $\dot{\textbf{q}} \to 0$; see footnote \ref{may1716-1}. 
The $\dot{\textbf{q}}$ in the denominator therefore is allowed. 
ii) The velocity $\dot{\textbf{q}}$ of a solution $\textbf{q}(t)$ of the 3-D free E-L eq. is written as a Fourier series (see p.\pageref{may2516-1}). No solution of the free E-L eq., a nonlinear fourth-order ODE, therefore is chaotic \cite{Strogatz(1994)} or runs away ($ \dot{\textbf{q}}  \to \pm \infty$ for $t \to +\infty$) \cite{Jackson(1999)}.

We thus see that there exists a GCM comprised of the H-J eq. (\ref{jul1510-1}),  and the E-L eq. (\ref{aug0316-1}) effective for classical and semiclassical systems. 
\footnote{
The E-L eq. is transformable to Hamilton's canonical eqs. by higher-order Legendre transformation. See \ref{0428ab} and Ref. \refcite{Leon(1985)}. 
We do not use them in the present paper. 
} 
Next three paragraphs further clarifies the structure of GCM.

In general, \label{feb0221-1} energy of a single particle system specified by a higher-order Lagrangian is unbounded from below. \cite{Woodard(2015)} 
\footnote{
As an example, consider a system specified by a Lagrangian $L= -m \ddot{q} ^2 /2 \omega _{ \, Q} ^2 + m \dot{q}^2 /2 - m \omega _ { \, C} ^2 q^2 /2$, where $4 \omega _{ \, C } ^2 <  \omega _{ \, Q} ^2 $. 
The E-L eq. derived from the $L$ is $m \ddot{\ddot{q}} / \omega _{ \, Q}^2 + m \ddot{q} + m \omega _{ \, C} ^2 q =(m / \omega _{ \, Q} ^2) ( d^2/dt^2 + k_+ ^2 )(d^2/dt^2 + k_- ^2 ) q(t) =0$, where $k_{\pm }^2  = \omega _{ \, Q}^2 ( 1 \mp \surd ( 1- 4 \omega _{ \, C} ^2 / \omega _{ \, Q} ^2 ))/ 2$. 
The general solution $q(t)$  of the eq. is $q(t)=C_+ \cos k_+ t + S_+ \sin k_+ t + C_- \cos k_- t + S_- \sin k_- t$, where $\{ C_+ , S_+ , C_- , S_- \} \in \mathbb{R}$ are arbitrary consts. 
The energy of the solution  $q(t)$ is 
\begin{equation*} 
 E =  \dot{q} \big( \frac{\partial L}{ \partial \dot{q}} - \frac{d}{dt} \frac {\partial L }{ \partial \ddot{q}} \big) + \ddot{q} \frac{\partial L }{ \partial \ddot{q}} -L = \sqrt{ 1 - \frac{4 \omega _{ \, C } ^2 }{ \omega _{ \, Q} ^2} } \ \big(  \frac{m  k_+ ^2  \ (C_+ ^2 + S_+ ^2 ) }{2} - \frac{ m k_- ^2 \ (C_- ^2 + S_- ^2 ) }{2} \big) \ . 
\end{equation*} 
It is unbounded from below (cf. Ref. \refcite{Woodard(2015)}).  
For semiclassical systems, $ k_- (C_- ^2 + S_- ^2 ) ^{1/2} / k_+ (C_+ ^2 + S_+ ^2 ) ^{1/2} \ll 1$ by definition. 
Accordingly $E> 0$. 
} 
Accordingly, a many-particle system specified by the one is unstable. 
\footnote{
Consider a system comprised of statistical mechanically many particles in an isolated box. 
If energy $e_i$ of each constituent particle is unbounded from below, the $e_i$ of some particles limitlessly increases and that of some others limitlessly decreases with energy of the system kept intact by, say, three body collisions regardless of nature of interaction among particles because it is entropically favored. \cite{Woodard(2015)} 
} 
It is known as the Ostrogradsky instability. \cite{Woodard(2015)} 
It however does not dismiss the Lag. formalism of GCM based on (\ref{feb1414-1}) because (\ref{feb1414-1}) is effective only for classical and semiclassical systems, for which the formalism gives energy bounded from below.  
Indeed, a discrepancy between $ \nabla \mathcal{S}_0^{HJ} ( \textbf{x})$ and $ \nabla \mathcal{S}_0^{EL} ( \textbf{x})$, 
 $ \nabla \mathcal{S}_0^{HJ} ( \textbf{x}) - \nabla \mathcal{S}_0^{EL} ( \textbf{x}) = \mathcal{O}(\epsilon ^{3/2})$ does not decrease beyond $\mathcal{O} ( \epsilon ^{3/2} ) $ even if $\hbar ^3$, $\hbar ^4$, $\cdots$ -terms are taken into account in the $L$, which means that a force generated by any Lagrangian does not completely reproduce the one generated by the extended diff. action.

To \label{dec1017-1} determine a free EoM giving a solution making $ \nabla \mathcal{S}_0^{HJ} ( \textbf{x}) - \nabla \mathcal{S}_0^{NW} ( \textbf{x} ) = \mathcal{O}( \epsilon ^n)$ true, where $n > 3/2$, is beyond the scope of the present paper.      
However, a theoretical framework in which the EoM would be determined is worth mentioning. 
It is a higher-order version of the integral form $\int ( m \ddot{\textbf{q}} - \textbf{F} ) \delta \textbf{q} dt =0$ \cite{Lanczos(1986)} of the classical d'Alembert's principle $ m \ddot{\textbf{q}} - \textbf{F} =0$, a form of the principle of least action more basic than the Hamilton's principle (see \ref{sep2212-730}). 
In the principle, variation $\delta \mathcal{S}_0$ is given as 
\begin{align} 
&\delta \mathcal{S}_0 
= \textstyle\int  ^ t ( \textbf{G}_0\delta \textbf{q} + \textbf{G}_1 \delta \dot{\textbf{q}} + \textbf{G}_2 \delta \ddot{\textbf{q}} + \cdots ) d \tilde{t} = \textstyle\int ( \textbf{G}_0 - \dot{\textbf{G}}_1 + \ddot{\textbf{G}}_2 -   \cdots ) \delta \textbf{q} d\tilde{t}  \notag  \\ 
&+ ( \textbf{G}_1 - \dot{\textbf{G}}_2  + \ddot{\textbf{G}}_3 - \cdots ) \delta \textbf{q} |^t 
 + ( \textbf{G}_2 - \dot{\textbf{G}}_3 + \cdots ) \delta \dot{\textbf{q}}|^t  
+ ( \textbf{G}_3 - \cdots ) \delta \ddot{\textbf{q}}|^t + \cdots      \ , \label{sep1016-2} 
\end{align}
where $\textbf{G}_i$ ($i = 0,1, \cdots $) is a function of $\textbf{q}, \dot{\textbf{q}}, \cdots$, 
$\textbf{G}_i = \textbf{G}_i ( \textbf{q} , \dot{\textbf{q} } , \cdots )$. 
We consider a free system. 
Then, we have $\textbf{G}_0 =0$ according to the translational invariance of the system (see \ref{sep2212-730}). 
We assume that each of $\textbf{G}_1 , \textbf{G}_2 , \cdots$ is given as a series in powers of $\hbar$. 
The series is determined by dimensional analysis up to coefficients of terms in the series---for example, for 1-D systems, $ G_2 = \sum _{k=1}^{+\infty } c_{2k} \hbar ^k \ddot{q}^{\, k-1} / m^{\, k-1} \dot{q}^{\, 3k -2} $, where $ c_{2k} = const. \in \mathbb{R}$ are coefficients to be determined. 
We have $\textbf{G}_3 = \textbf{G}_4 = \cdots = 0$ as in the Lag. formalism (see p.\pageref{feb1618-1}). 
We have $\textbf{p} ^{NW} = \textbf{p}_{(1)} + \textbf{p}_{(2)} \ddot{\textbf{q}} / | \dot{\textbf{q}} |$ with $ \textbf{p}_{(1)} = \textbf{G}_1 - \dot{\textbf{G}}_2 $ and $ \textbf{p}_{(2)} = \textbf{G}_2$ like the $\textbf{p}^{EL}$. 
The $\textbf{p} ^{NW}$ has more adjustable coefficients than the $\textbf{p} ^{EL} $ constructed from the candidate Lagrangian (see p.\pageref{feb1618-1}) has because $ \textbf{G}_1 $ and $ \textbf{G}_2 $ need not be derived from a single Lagrangian. 
Indeed, the number of adjustable parameters in $\hbar ^4$-term of $\textbf{p}^{NW}$ is four, while that of $\textbf{p}^{EL}$ is one. 
Accordingly, $\textbf{p} ^{NW}$ reproduces $\textbf{p} ^{HJ}$ more accurately than $\textbf{p} ^{EL}$ does. 
Note that (\ref{sep1016-2}) is equal to (\ref{apr2619-1-051606-k}) if $ \textbf{G}_0 = \partial L / \partial \textbf{q},  \textbf{G}_1 = \partial L / \partial \dot{\textbf{q}} $, and so on.

The \label{aug1419-1} H-J eq. (\ref{jul1510-1}) of GCM is transformed to (from) the Schr\"{o}dinger eq. $\hbar ^2 \nabla ^2 \psi = 2m (V-E) \psi$ of QM with  $\psi ( \textbf{x} )  = R ( \textbf{x} ) e^{i \mathcal{S}_0 ( \textbf{x} ) / \hbar }$, where $R ( \textbf{x} ), \mathcal{S}_0 ( \textbf{x} ) \in \mathbb{R}$. \cite{Holland(1993)} 
The GCM however is not a particle picture (the E-L eq.) incorporated interpretation of QM because $| \psi |^2 $ constructed from a solution $R$ of the H-J eq. as $| \psi |^2 = R^2$ does not represent, unlike QM, particle density in an ensemble. 
The $R^2 ( = | \psi |^2 )$ is irrelevant to the particle density because GCM determines a trajectory $\textbf{q}(t)$ of a single particle. 
Even in an ensemble of systems, $R^2$ is irrelevant to particle density: 
Assume that particle density at time $t_0$ in an ensemble comprised of systems having identical $R$ and $\mathcal{S}_0$ is equal to $R^2 ( \textbf{x} )$---we can make particle density equal to the $R^2 ( \textbf{x} )$ because we can put a particle at any point $\textbf{x} = \textbf{q} (t_0)$ as far as $\textbf{q}(t_0)$ satisfies $\nabla \mathcal{S}_0 ^{NW} ( \dot{\textbf{q}}, \ddot{\textbf{q}}, \dot{\ddot{\textbf{q}}}) |_{t = t_0} = \nabla \mathcal{S}_0 ^{HJ} (\textbf{x}) |_{\textbf{x}= \textbf{q}} $. 
If $R^2$ represents particle density for $- \infty < t < + \infty $, then $\nabla \mathcal{S}_0 = m \dot{\textbf{q}}$ from the second eq. $\nabla ( R^2 \nabla \mathcal{S}_0 ) =0$ of (\ref{jul1510-1}). 
However $\nabla \mathcal{S}_0 \neq m \dot{\textbf{q}}$ from (\ref{jul0819-2}). 
The $R^2$ therefore does not represent particle density in the ensemble. 
Thus, GCM is not a QM.

\label{dec0616-1} 
References \refcite{Faraggi(2000)}, \refcite{Bertoldi(2000)}, and \refcite{Bouda(2003)} precede the present paper. 
Faraggi et al.\,\cite{Faraggi(2000)} \cite{Bertoldi(2000)} 
obtained (\ref{jul1510-1}), 
which we sometimes call the 3-D Quantum Stationary H-J eq. (QSHJE) after Ref. \refcite{Faraggi(2000)}, 
 from a requirement that two 1-D systems classically described as $-(\partial \mathcal{S}_0 / \partial x )^2 /2m = V(x)-E \neq 0$ and $(\partial \tilde{\mathcal{S}}_0  / \partial y )^2 /2m =  0$ be  transformed to each other by a time-independent coordinate transformation $y|_p=y(x|_p)$ satisfying $\mathcal{S}_0 (x)|_p = \tilde{\mathcal{S}}_0 (y)|_p$, which cannot be implemented in CM. 
They \cite{Faraggi(2000)} showed that 1-D energy is quantized by a requirement that $e^ { 2i \mathcal{S}_0 (x) / \hbar }$, where $\mathcal{S}_0 (x)$ is a solution of (\ref{apr2421-oct0116-1}) of which $E$ replaced with $E-V$, be continuous on the compactified real line $\mathbb{RP}^1$ even at $x = \{ \pm \infty \}$. 
They  \cite{Faraggi(2000)} showed that tunneling---a momentum $ \partial \mathcal{S}_0 ^{HJ} / \partial x$ and a particle velocity $ \dot{x}$ are real ($ \partial \mathcal{S}_0 ^{HJ} / \partial x , \,\, \dot{x} \in \mathbb{R}$) in the region where $V(x) - E > 0$---occurs because 
i) in the 1-D QSHJE written as 
$ \partial \mathcal{S}_0 / \partial x = \pm \sqrt{ 2m(E- V -Q^{HJ} )}$, where $Q^{HJ}(x) =  \hbar ^2 \{ \mathcal{S}_0 (x) , x \} /4m  $, $ \partial \mathcal{S}_0 / \partial x$ can be real in the region because of the $Q^{HJ}$, and 
ii) in an eq. $\dot{x} = ( \partial \mathcal{S}_0 / \partial x ) / m ( 1- \partial Q^{HJ} /\partial E)$ \cite{Floyd(1982)} 
obtained from 
$ \partial \mathcal{S}_0 / \partial x = \pm \sqrt{ \cdots }$ given above, 
$\dot{x} = ( dt/dx)^{-1}$, and 
$t-t_0 = \partial \mathcal{S}_0 / \partial E = ( \partial / \partial E )\int dx  \partial \mathcal{S}_0 / \partial x $ resulting from Jacobi's theorem \cite{Lurie(2002)} 
\footnote{\label{sep0614-1}Jacobi's theorem says that  partial derivatives $\partial \mathcal{S} / \partial \bm{\alpha }$ 
of a complete integral---a solution containing $n$ non-additive consts.---$\mathcal{S}(t , \textbf{x} , \bm{\alpha})$, 
where $ \bm{\alpha} = ( \alpha _1 , \cdots , \alpha _n ) = const.$, of $n$-D H-J eq.: $H( \textbf{x} , \partial \mathcal{S} / \partial \textbf{x}) + \partial \mathcal{S} / \partial t =0$ are consts. of motion. 
From a 1-D complete integral  $ \mathcal{S}(t , x ,  E) =  \mathcal{S}_0 (x , E) -Et$ constructed from  a solution $\mathcal{S}_0 (x ; E)$ of the 1-D QSHJE written as $ (\partial \mathcal{S}_0 / \partial x)^2 /2m + V(x) + Q^{HJ} (x) =E$, $\partial  \mathcal{S}_0 (x ; E) / \partial E  - t = const.$ follows, where $E$ works as the $\alpha _1$. 
} 
in CM, $\dot{x}$ is real since $ \partial \mathcal{S}_0 / \partial x $ and $ \partial Q^{HJ} /\partial E$ are. 
Bouda  \cite{Bouda(2003)} constructed a higher-order 1-D Lagrangian compatible with (\ref{apr2421-oct0116-1}) of which $E$ replaced with $E-V$.

We i) construct the 3-D QSHJE (\ref{jul1510-1}) using the extended diff. action.  The present method has wider applicability: the extended diff. action is  applicable also on the tensor in theories other than CM.   
We ii) determine, using higher-order Lag. mechanics, \cite{Leon(1985)} \cite{Miron(1997)} a 3-D Lagrangian (\ref{feb1414-1}), which 
 is given for the first time. 
The (\ref{feb1414-1}) leads to a 3-D trajectory-determining E-L eq. 
 The above method making use of Jacobi's theorem to determine trajectories works, set aside exceptions, only in 1-D space. 
\footnote{
In order to use the method in 3-D space,  
we need to construct a complete integral $ \mathcal{S}( t, \textbf{x} , \bm{\alpha}  )$, where $ \bm{\alpha} = ( \alpha _1 , \alpha _2 , \alpha _3 ) = const.$ (see footnote \ref{sep0614-1}), from  (\ref{jul1510-1}) regarding $\hbar ^2  \nabla ^2 R / 2mR $ as an effective potential $Q^{HJ} ( \textbf{x} )$, which however is impossible in general.  \cite{Lurie(2002)} 
}  
We note that 1-D form of (\ref{feb1414-1}) differs from the 1-D Lagrangian constructed in Ref. \refcite{Bouda(2003)}; see footnote \ref{sep1314-1}. 
We iii) show that,  maintaining trajectory concept at    all times, following quantum phenomena appear: 
1) 1-D and 2) 3-D  energy quantization, 
3) angular momentum quantization, 
4) uncertainty relation between position and momentum, 
5) tunneling, and 
6) interference. 
The reasoning for 1) does not refer to $x= \{ \pm \infty \}$ unlike that of Ref. \refcite{Faraggi(2000)}. That for  5) with the higher-order Lag. formalism sharpens the one in Ref. \refcite{Faraggi(2000)}. 
Reasonings for 2), 3), 4), and 6) with GCM is given for the first time. 
Lastly, we iv) show that GCM is testable with present-day technology because particle distribution in an ensemble it gives differs from that of QM. 
An alternative to QM testable with present-day technology is given for the first time. 
\footnote{Dynamical reduction models \cite{Bassi(2003)} are testable, but are not with present-day technology.
}

Phenomena corresponding to quantum phenomena appear as follows.

$\bullet$ Energy quantization: 
For concreteness, we consider a particle in a 3-D potential well $V (\textbf{x}) = V_x(x) + V_y(y) + V_z(z) $, where $ V_x(x)=  0$ for $ | x | \leq L_x $ and $V_x(x) = + \infty$ for $ |x | > L_x $, likewise for $V_y(y)$ and $V_z(z)$. 
The quantization follows from i)--iii) below. 
i) The value of $ \textbf{p} ^{NW} $ averaged over macroscopic length along trajectory $\textbf{q}(t)$ is independent from the parameters---for example $l_1$ and $l_2$  of (\ref{oct0116-2})---introduced by the extended diff. action, 
which follows from 1) the value of $ \textbf{p} ^{NW}$ along curved trajectory is equal to that along straightened trajectory because $ \textbf{p} ^{NW}( \dot{\textbf{q}} ,  \ddot{\textbf{q}} , \dot{\ddot{\textbf{q}}} ) |_{\textbf{q}(t) = \textbf{x}} =   \textbf{p} ^{HJ}(\textbf{x})   $ and  $ \textbf{p} ^{HJ}(\textbf{x}) $ is fixed solely by particle position $\textbf{x}$, and 
2) the value of $ \textbf{p} ^{NW}$ of the straightened trajectory averaged over macroscopic length is given by CM free from the parameters. 
ii) The value of $x$-component $p_x^{NW}$ of the $ \textbf{p} ^{NW}$ averaged over one-cycle-run length in $x$-direction---a length along $\textbf{q}(t)$ from $\textbf{q} _0(t _0)$ of which $x$-coordinate is $x_0$ to  $\textbf{q} _1(t _1)$ of which $x$-coordinate is the same $x_0$---is equal to double the value of $x$-component $p_x^{HJ}$ of $ \textbf{p} ^{HJ}(x)$ averaged over $ -L_x \leq x \leq L_x$ regardless of the shape of trajectory (see \S \ref{nov1615-1}). 
and iii) Macroscopic length is a multiple of the one-cycle-run length. 
From i)--iii), we see that the value of $p_x^{HJ}$ averaged over $ -L_x \leq x \leq L_x$, which is given by (\ref{oct0116-2}), has to be independent from the parameters. 
It is so iff $k_x= \surd (2mE_x) / \hbar = n\pi / 2L_x$, where $n= 1,2, \cdots$ (see \S \ref{oct0114-1}). 
The $x$-direction energy $E_x$ therefore is quantized to $E_x = (n \pi \hbar /2L_x )^2 / 2m $. 
  Likewise for $E_y$ and $E_z$.

$\bullet$ Angular momentum quantization: 
Angular momentum of a 3-D system separable in spherical polar coordinates $(\varphi , \theta ,r)$ are quantized, like the 3-D energy, by a requirement that the value of $|\textbf{p}^{HJ}|$ averaged along trajectory is independent, in $\varphi$- and $\theta$-direction, from parameters corresponding to the $(l_1 , l_2 )$.

\begin{wrapfigure}{r}{30 ex} 
\vspace*{- 0.5 \intextsep} 
\includegraphics[width = 30 ex, clip]{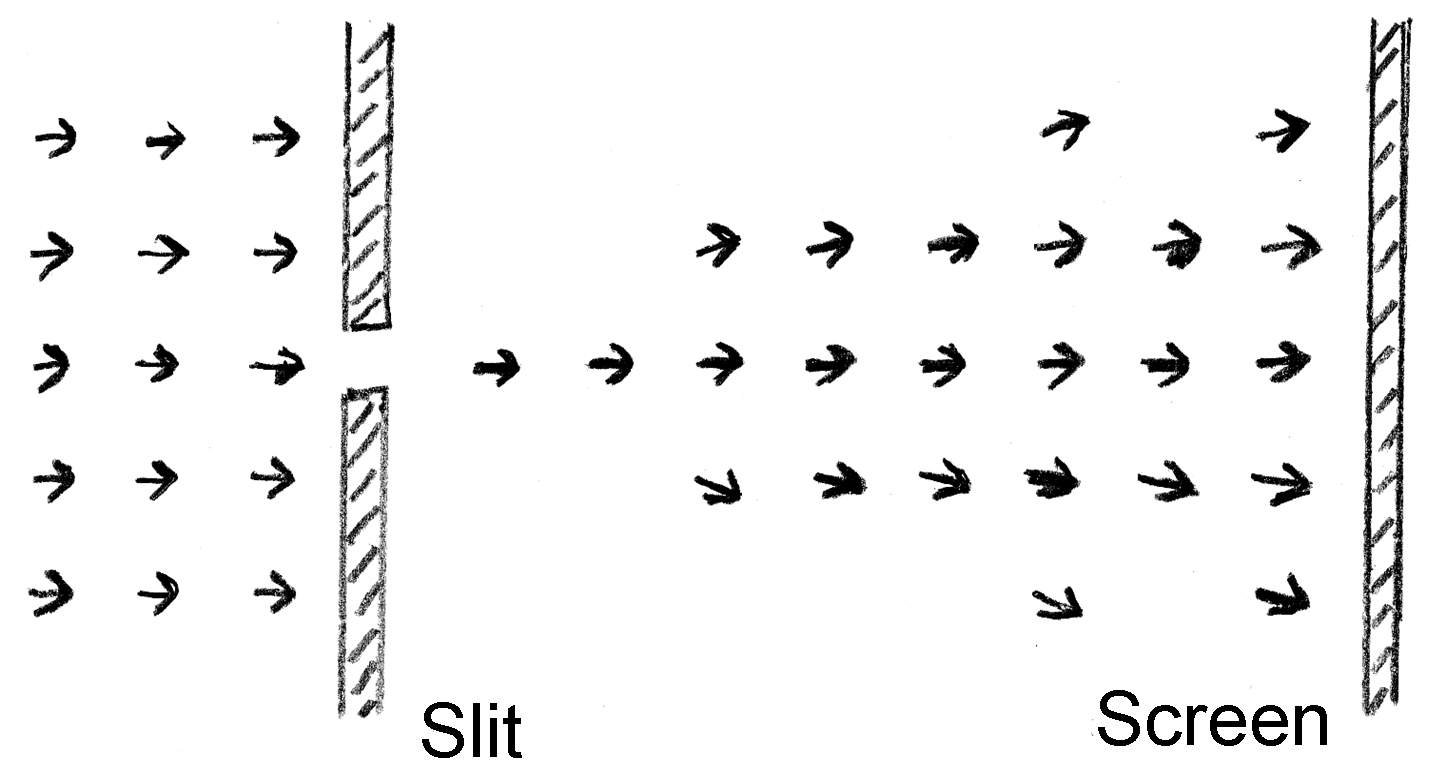} 
\caption{Diffracted $\textbf{p}^{HJ}$} 
\label{jan1918-1} 
\vspace*{- 0.5 \intextsep} 
\end{wrapfigure}

$\bullet$ 
Uncertainty: In GCM, position $\textbf{x}$ and momentum $\textbf{p}^{NW}$ of a particle always have definite values since the particle has a trajectory $\textbf{q}(t)$.  
Nevertheless the particle exhibits phenomena corresponding to the uncertainty relation between $\textbf{x}$ and $\textbf{p}$ of QM. 
We show that i) the $| \textbf{p} |$ fluctuates in a box, and 
ii) we cannot simultaneously measure $\textbf{x}$ and $\textbf{p}^{NW}$ of a particle beyond some precision:

i) The $x$-direction component $ p _x ^{HJ}$ of $\textbf{p}^{HJ}$ of a particle in the above given 3-D potential well is given by (\ref{oct0116-2}). 
When $k_x= \pi / 2L_x$, which corresponds to the lowest allowed $E_x$, the formula gives $\Delta p_x := \textrm{max.} p_x^{HJ} - \textrm{min.} p_x^{HJ} \simeq 0.37 \hbar k_x $ for, say, $l_1 \simeq \pm 1.2$ and $l_2 =0$. 
We therefore have, corresponding to QM, $\Delta x \Delta p_x \simeq \hbar$ for $\Delta x := 2L_x$. 
Allowed values of $(l_1 , l_2 )$ would be determined by an eq. 
$ \textbf{p} ^{NW} ( \dot{\textbf{q} } , \ddot{ \textbf{q}} , \dot{\ddot{ \textbf{q}}} )|_{ \textbf{q} = \textbf{x}} = \textbf{p}^{HJ} ( \textbf{x}; \bm{l} _1 , \bm{l} _2 ,E) $ for $ \dot{\textbf{q}} (t)$ under a given  $ \textbf{p}^{HJ} ( \cdots )$, where $\bm{l}_1 = ( l_{1x} , \cdots , l_{1z})$ and $\bm{l}_2 = ( l_{2x} , \cdots , l_{2z})$. 
That is, the $( \bm{l}_1 , \bm{l}_2 )$ are allowed if the eq. has a solution $\dot{\textbf{q}}(t)$ satisfying $0 < | \dot{\textbf{q}}| < + \infty$ over $-\infty < t < + \infty $. 
For the eq. to have such a solution, the narrower the well is, the larger $( \Delta p_x ^{HJ} , \cdots , \Delta p_z ^{HJ} )$ the  $( \bm{l}_1 , \bm{l}_2 )$ have to give because the narrower the well is, the larger the fluctuations of $( \ddot{\textbf{q}} , \dot{\ddot{\textbf{q}}} )$ are. 
We cannot determine $ (\bm{l}_1 , \bm{l}_2 ) $ because we do not have a formula of $\textbf{p}^{NW}$ precise enough in atomic scale confining potential.

ii) We consider a measurement of $x$-direction components of $\textbf{x}$ and $\textbf{p}^{NW}$ of a particle $m$ running from $z$-minus direction to $z$-plus direction. \cite{Feynman(1965)} 
We measure $x$-coordinate of the $m$ with a slit at $z=0$ along $y$-axis on  $xy$-plane. 
We measure $p_x^{NW} \simeq m \dot{q} _x$ with a hitting position on the screen behind the slit. 
See Fig.\ref{jan1918-1}. 
The $ \textbf{p} ^{HJ}$ is diffracted by the slit. 
Since $\textbf{p} ^{HJ} = \textbf{p} ^{NW} \simeq m \dot{ \textbf{q} } $, the $m$ follows an integral curve of $ \textbf{p} ^{HJ}$ behind the slit. 
The curve the $m$ follows is determined by the position of $m$ in the slit. 
Since the position is unknowable, the curve the $m$ follows is unknowable. 
Accordingly, we cannot determine the original value of $p_x^{NW}$ beyond the uncertainty by diffraction. 
The uncertainty is larger when certainty in the measured value of $x$-coordinate is larger because the diffraction is larger when the slit is narrower.

$\bullet$ 
Tunneling:  
 In the Lag. formalism of GCM, energy $E$ of the system is conserved, and  kinetic energy $E _{\textrm{kin}}$ of a 1-D  particle is written as $E _{\textrm{kin}} = m \dot{q} ^2/2 + ( \hbar^2 /m )( \dot{\ddot{ q }} / \dot{q} ^3 - 5 \ddot{q} ^2 / 2 \dot{q} ^4 )$. 
Assume a 1-D particle having  $ \dot{q} _0 >0$ and $ \ddot{q}_0 = \dot{ \ddot{q}}_0 =0$  runs towards a bell-shaped potential hill $V$ of which height is $m \dot{q}_0 ^2/ 2 + \delta ^2$, where $0 < \delta ^2 \ll m \dot{q} _0 ^2/ 2 $. 
Then, near the top of the hill, $ 0 \approx | \dot{q}| \ll | \dot{q} _0 |$  and  $\ddot{q}  \neq 0$. 
Accordingly, $- 5 \hbar ^2 \ddot{q} ^2 / 2 m \dot{q} ^4 $ is dominant in the $E _{\textrm{kin}}$ formula. 
That is, $E _{\textrm{kin}}$ is negative ($E _{\textrm{kin}} < 0$). 
Thus,  even at the top of the hill, $\dot{q}$ can be real ($\dot{q} \in \mathbb{R}$) with $E = m \dot{q} _0 ^2/ 2 = E _{\textrm{kin}} +  m \dot{q} _0 ^2/ 2 + \delta ^2 = const.$ 
We note that this reasoning is qualitative because the Lagrangian (\ref{feb1414-1}) is inaccurate for small $\dot{q}$.

$\bullet$ Interference: 
In a double-slit experiment of a massive particle, \cite{Holland(1993)} a solutions $\textbf{p}^{HJ} $ of the QSHJE undulates behind slits. 
Accordingly, trajectories $\textbf{q} (t)$'s oscillate there because $\textbf{p}^{HJ} (\textbf{x} ) = \textbf{p}^{NW} (\textbf{x} ) \simeq  m \dot{ \textbf{q} } |_  { \textbf{q} = \textbf{x} }$. 
 Thus, we have shot marks which look like an interference pattern of QM on the screen.


Predictions by GCM on particle distribution in an ensemble sometimes differ from that by QM as was mentioned on p.\pageref{aug1419-1}. 
Indeed, a distribution of electrons Fresnel-diffracted by an electron biprism calculated with GCM and that calculated with QM differ by a few percent (\S \ref{jan2107-6}). 
Whereas experimental result and QM's calculation on the distribution differ by more than several percent. \cite{Komrska(1967)} \cite{Yamamoto(2000)} 
In addition, to the best of the author's knowledge, no experiments to date quantitatively confirm QM's prediction on the distribution of massive particles. 
\footnote{
The best report to date to quantitatively test  QM's prediction on massive particle distribution, to the best of the author's knowledge, is  Ref.\refcite{Tschernitz(1992)}. 
In it, an experimental data and a calculation result of the distribution of cold neutrons Fraunhofer-diffracted by a single slit are reported. 
The experiment is precise enough to detect difference of the slit width by a few tenths of a percent, while the calculation requires slit width $1.5\%$ narrower than that measured with an optical microscope to reproduce the observed distribution. 
Thus, on massive particle distribution, experiment and theory (QM) seem to disagree beyond experimental uncertainty. 
} 
The GCM therefore is not rejected by experiments to date.


\newpage

\section{Formalisms} \label{jun3009-1}

\subsection{Generalized Hamilton-Jacobi formalism} \label{oct0114-1} 

We construct a generalized H-J eq. of  nonrelativistic 3-D systems comprised of a massive particle $m$ and a time-independent potential field $V$. 
We first construct a generalized relativistic H-J eq. 
We then take a nonrelativistic limit to obtain the generalized H-J eq. 
This approach gives an example of application of the extended diff. action to theories other than nonrelativistic CM. 
We also give a generalized 1-D H-J eq.

$\bullet$ \textit{Construction of the generalized H-J eq.} \ \ 
We first construct a relativistic H-J eq. of GCM from the relativistic H-J eq. \cite{Landau(1975)} of CM---we include the relativistic eq. too under labels of CM and GCM. 
Our Minkowski metric is $ \textrm{diag}(1,-1,-1,-1)$. 
We use the Einstein summation convention. 
We work with the Gaussian system of electromagnetic units. \cite{Landau(1975)} 
The relativistic H-J eq. of CM is 
\begin{equation}
( \frac{\partial \mathcal{S}}{ \partial  \textbf{x}} - \frac{e}{c} \textbf{A} )^2 - \frac{1}{c^2}(   \frac{\partial \mathcal{S}}{ \partial t} + e A^0 )^2 + m^2c^2 =0 \ ,  \label{oct1609-a1}
\end{equation}
where $\mathcal{S}$ is the relativistic action, $e$ is the electric charge of the particle, $m$ is the mass of the particle, $c$ is the velocity of light, and $(A^0, \textbf{A})$ \label{jun2212-1} is an electromagnetic potential. 
In the case that $(A^0, \textbf{A})$ in (\ref{oct1609-a1}) is separable $(A^0, \textbf{A}) = (A^0 (x^0) , \cdots , A^3(x^3)) $, 
the eq. separates to four 1-D eqs. 
To each separated eq., we apply the 1-D extended diff. action $\bar{\bar{T}}$. 
We then combine $\bar{\bar{T}}$-applied eqs. to a single 4-D eq. 
Lastly, we rewrite the eq. to a form effective for both separable- and inseparable-$(A^0, \textbf{A})$ to have
\begin{subequations}
\begin{align}
( \partial _{\mu}  \mathcal{S}    + \frac{e}{c} A_{\mu}) ( \partial ^{\mu}  \mathcal{S}   + \frac{e}{c} A^{\mu}) 
&= m^2 ( c^2 + \tilde{Q} ) ,     \ \ \ 
  \tilde{Q} = \frac{ \hbar ^2  \partial _{\mu} \partial ^{\mu}   R}{R m^2}  \label{aug0608-1}       \\
\partial _{\mu} j^{\mu} 
&= 0 ,    \ \ \ \ \ \ \ \   j^{\mu} = -\frac{R^2}{m} ( \partial ^{\mu}  \mathcal{S} + \frac{e}{c} A^{\mu})  \ ,  \label{aug0608-2} 
\end{align}  \label{aug0608-3}
\end{subequations}
where $\mu$ runs over 0--3, and $R$ is a function of $\mathcal{S}$. 
For details of the derivation, see \ref{may0309-a3}.

We introduce the extended one because 
i) it still qualified as a diff. action since it satisfies axioms for a group action (see p.\pageref{jul2720-1}), 
and ii) a term introduced by the one, $\tilde{Q}$ in (\ref{aug0608-1}), leads to quantum phenomena. 

We then take non-relativistic limit of (\ref{aug0608-3}). 
For simplicity, we consider only the case that $A^1 = A^2 = A^3 =0$ and $A^0$ is time-independent. 
We set, in (\ref{aug0608-3}), $A^1 = A^2 = A^3 =0$, $A^0 = A^0 ( \textbf{x} )$, and $\mathcal{S} = \mathcal{S} _{\textrm{nr}} -mc^2t$, \cite{Landau(1975)} where $\mathcal{S} _{\textrm{nr}} $ is non-relativistic action. 
We then ignore $ \bullet /c^2 $-terms to have 
\begin{equation}
 - \frac{\hbar ^2}{2m} \frac{ \partial _i \partial ^i R}{R}- \frac{1}{ 2m} ( \partial _i \mathcal{S}_ {\textrm{nr}} )^2  - (  e  A^0 + \frac{ \partial  \mathcal{S} _{\textrm{nr}} }{ \partial  t }) =0   \ , \ \ \ \ 
\frac{ \partial R^2}{ \partial t } + 
\frac{ \partial }{ \partial x^i} ( \frac{R^2}{m} \partial _i \mathcal{S}_{\textrm{nr}} ) = 0  \ ,   \label{jan1712-1}
\end{equation}
where $i$ runs over 1--3.  
In eqs. of (\ref{jan1712-1}), variables, $t$ and $ \textbf{x}$, separate with $R(t , \textbf{x}) = R(t) R( \textbf{x}) $ and $ \mathcal{S}_{\textrm{nr}} (t , \textbf{x}) = \mathcal{S}_{\textrm{nr}} (t ) + \mathcal{S}_{\textrm{nr}} ( \textbf{x}) $. 
From the former eq. of (\ref{jan1712-1}), we have $ \hbar ^2 \partial _i \partial ^i R ( \textbf{x})/ 2mR( \textbf{x}) + ( \partial _i \mathcal{S}_ {\textrm{nr}}( \textbf{x}) )^2 / 2m + e A^0 ( \textbf{x}) = - \partial  \mathcal{S} _{\textrm{nr}}(t) / \partial t = k_1$, where $k_1 = const. \in \mathbb{R}$. 
Because $- \partial \mathcal{S} _{\textrm{nr}}(t) / \partial t$ represents      energy $E$ of the system in CM and the $E$ is kept invariant under change of diff. action from $\bar{T}$ to $\bar{\bar{T}}$, we have $k_1 = E$. 
From the latter eq. of (\ref{jan1712-1}), we have 
$( 1/ R^2 ( \textbf{x} ) ) ( \partial / \partial x^i )( m^{-1} R^2 ( \textbf{x} ) \partial _i \mathcal{S}_ {\textrm{nr}}( \textbf{x}) ) = (1/ R^2 (t) ) ( \partial R^2 (t)  / \partial t ) = k_2 $, where $k_2 = const. \in \mathbb{R}$, from which we have $ R^2 (t) = k_3 e^{k_2 t}$, where $k_3 = const. (\neq 0) \in \mathbb{R}$. 
However, $R^2 (t)$ satisfies $R^2 (t) ( \partial _0 \mathcal{S} + eA^0 / c ) = const.$ (see \ref{may0309-a3}), according to which $k_2 = 0 $. 
We thus have 
\begin{equation}
 \frac{\hbar ^2}{2m} \frac{\nabla ^2 R( \textbf{x})}{R( \textbf{x})}- \frac{1}{ 2m} ( \partial _i \mathcal{S}_ 0( \textbf{x}) )^2  - V ( \textbf{x}) + E =0   \ , \ \ \ \ 
\frac{ \partial }{ \partial x^i} ( \frac{R^2( \textbf{x})}{m} \partial _i \mathcal{S}_0 ( \textbf{x}) ) =   0  \ ,     \label{june1308-1}
\end{equation}
where $V = eA^0$ and $\mathcal{S}_0 ( \textbf{x})  =  \mathcal{S}_{\textrm{nr}} ( \textbf{x})$.

On 1-D space, the second eq. of (\ref{june1308-1}) is equivalent to $R= const. /  ( \partial _x \mathcal{S}_0 ) ^ {1/2}$. 
Inserting it into the first eq., we have (cf. footnote \ref{jan1009-ss}) 
\begin{equation}
 \frac{1}{2m} (\frac{\partial \mathcal{S}_0}{\partial x}  ) ^2  + \frac{ \hbar ^2 }{4m} \{ \mathcal{S}_0,x \}  +V(x) -E = 0  \ ,  \label{may0409-2}
\end{equation}
where $\{ \mathcal{S}_0,x \}$ is the Schwarzian derivative defined on  p.\pageref{jul1920-1}.

As was mentioned on p.\pageref{dec0616-1}, we call (\ref{june1308-1}) and (\ref{may0409-2}) 3-D and 1-D QSHJE \cite{Faraggi(2000)} respectively. 


$\bullet$ \textit{Solving 1-D QSHJE} \ \ We solve 1-D QSHJE (\ref{may0409-2}). 
We transform (\ref{may0409-2}) to
\begin{equation}
 V(x) - E = - \hbar ^2  \{ e^{ \frac{2i}{\hbar} \mathcal{S}_0} ,x \} / 4m \ ,   \label{sep0308-1} 
\end{equation}
with a replacement: 
 $q^a \to x ,  \ q^b \to  \mathcal{S}_0    ,  \   q^c \to e^{ 2i  \mathcal{S}_0 / \hbar}$ in the chain rule (p.\pageref{sep1413-2}) of the Schwarzian derivative: $\{ q^c,q^a \} = ( \partial q^b / \partial q^a)^2  \{ q^c,q^b  \} +  \{ q^b,q^a \}$. 
Whereas using a theorem: `Let two linearly independent solutions of  
$
 d^2 z / d x ^2  + u( x ) z =0   
$
 be $ z_1, z_2$. Then,  $ \{ z_2/z_1 , x \} = \{ (Az_1 + Bz_2)/(Cz_1 +Dz_2) ,x \} = 2u(x) $, where $A, \cdots , D = const. \in \mathbb{C} $ and $AD-BC = const. \in \mathbb{C} \backslash \{ 0 \}$.' (e.g., Ref. \refcite{Ovsienko(2005)}), we have 
\begin{equation}
V(x) -E =-  \frac{\hbar ^2}{4m} \{ \frac{A\Psi ^D + B \Psi }{ C \Psi ^D + D \Psi } , x \} \ , \label{apr0915-1}
\end{equation} 
where $\Psi , \Psi ^D \in \mathbb{R}$ are two linearly independent solutions of the corresponding, having the same $V(x) $ and $E$, second-order linear differential eq.: 
$ d^2 z / d x ^2  - (2m/ \hbar ^2 )( V(x) - E ) z =0  $. 
Since it  is the same as the 1-D Schr\"{o}dinger eq. (SE) of QM, we henceforth call it 1-D SE. 
\footnote{
Keep in mind that in the present formulation  SE is used as a mathematical tool to solve QSHJE. 
We do not give any physical meaning  such as 
`$|\Psi  (x) |^2$ represents the probability density of finding a particle at $x$\,' to the $\Psi$ or $\Psi ^D$. 
Physics is discussed with $\mathcal{S}_0$. 
} 
We set $\Psi$ and $\Psi ^D$ to be real without losing generality. 
We define $\check{\mathcal{S}}_0 \in \mathbb{C}$ as 
\begin{equation}
 e^{ \frac{2i}{\hbar}  \check{\mathcal{S}}_0} :=  ( A \Psi ^D + B \Psi ) / ( C \Psi ^D + D \Psi )    \ ,  \label{apr0915-3}
\end{equation} 
from which 
\begin{equation}
 \frac{ \partial \check{\mathcal{S}}_0}{ \partial x} = \frac{i}{2} \frac{\hbar (AD-BC)W}{( A \Psi^{D} + B \Psi )(C \Psi^{D}  + D \Psi )} 
 =  \frac{ \frac{i \hbar (AD - BC )W }{2AC}}{ ( \Psi^D + \frac{AD+BC}{2AC} \Psi)^2 - ( \frac{AD - BC}{2AC})^2 \Psi ^2  }   \ , \label{may0109-1}
\end{equation} 
where $W= \Psi ' \Psi^{D} - {\Psi^{D}}' \Psi = const. \neq 0$ is the Wronskian,  \cite{Messiah(1961)} follows. 
Comparing (\ref{sep0308-1}), (\ref{apr0915-1}), and (\ref{apr0915-3}), we see that the $ \check{\mathcal{S}}_0$ of (\ref{apr0915-3}) satisfies  (\ref{sep0308-1}). 
For $\{ e^{ 2i \mathcal{S}_0 / \hbar } ,x \} $ in (\ref{sep0308-1}) to be definable ($  e^{ 2i \mathcal{S}_0 / \hbar } \neq const. $), $ \partial \check{\mathcal{S}}_0 / \partial x  \neq 0$  in (\ref{may0109-1}) is required. In addition, for $ \partial \check{\mathcal{S}}_0 / \partial x  $ in (\ref{may0109-1}) to be a solution of (\ref{may0409-2}),  $ \partial \check{\mathcal{S}}_0 / \partial x \in \mathbb{R}$ is required. 
Therefore $l_1 := i (AD -BC)/2AC  \in \mathbb{R} \backslash \{ 0 \}$, and $ \ l_2 := ( AD +BC)/2AC \in \mathbb{R} $. 
We thus have 
\begin{equation}
 \frac{\partial \check{\mathcal{S}}_0}{\partial x} 
=  \frac { \hbar   W l_1   }{(\Psi^{D} + l_2 \Psi )^2 + l_1^2 \Psi ^2 }  \in \mathbb{R} \backslash \{ 0 \} \ . \label{may0409-s1}
\end{equation} 
The (\ref{may0409-s1}) is at least a solution of (\ref{may0409-2}). 
Actually, (\ref{may0409-s1}) is a general solution of (\ref{may0409-2}) because arbitrary initial values $\{ \partial \mathcal{S}_0 / \partial x , \partial ^2 \mathcal{S}_0 / \partial x^2 \} |_{ x=x_0}$ for the unknown $\partial \mathcal{S}_0 / \partial x $ of (\ref{may0409-2}) are realized as those for the $\partial \check{\mathcal{S}}_0 /\partial x$ of (\ref{may0409-s1}) with suitable values of $\{l_1 , l_2 \}$.

We solve 1-D free QSHJE, (\ref{may0409-2}) of which $V(x) = 0$, for $E \geq 0$.  
If $E>0$, we take $\Psi = \cos kx$ and $\Psi ^D = \sin kx$, where $k = \surd (2mE)/\hbar $, as two linearly independent solutions of the corresponding SE. The momentum $p := \partial \mathcal{S}_0 / \partial x$ is given from (\ref{may0409-s1}) as 
\begin{equation}
p =  \frac { - \hbar k l_1 }{(\sin kx + l_2 \cos kx )^2 + l_1^2 \cos ^2 kx } 
= \frac{ - 2 \hbar k l_1}{ (l_1^2+l_2^2 +1)+ (l_1^2+l_2^2 -1) \cos 2 k x + 2 l_2  \sin 2k x } \ . \label{may0409-j2}
\end{equation} 
If $l_1 = \mp 1$ and $l_2 =0$, we have the classical value: $p = \pm \sqrt{2mE}$. \label{may0315-2} 
If $l_1    = \pm ( 1 + \epsilon_1 )$ and $l_2 =  \epsilon_2$ with $|\epsilon_1 | , |\epsilon_2 | \ll 1$, 
we describe  $p$ of (\ref{may0409-j2}) as semiclassical and rewrite (\ref{may0409-j2}), for $p>0$, as
\begin{equation}
p =  \frac {  \hbar k (1 + \epsilon_1 ) }{(\sin kx + \epsilon_2 \cos kx )^2 + (1 + \epsilon_1 )^2 \cos ^2 kx } \ . \label{may2009-pp5}
\end{equation}

Regardless of values of $l_1 $ and $l_2 $, 
the value of $p$ averaged over an interval $x_0 \leq x \leq x_0 + \pi /k$ takes the classical value 
 $\pm \sqrt{2mE}$, which is seen from 
$\int _0 ^{\pi /k} p dx  = \pm \pi \hbar $ 
\footnote{
Let $l_1^2+l_2^2 +1 =a$, $l_1^2+l_2^2 - 1 = b$, and $2 l_2 = c$. 
Then $\sqrt{a^2 -b^2-c^2} = 2| l_1 |$ and 
\begin{equation*}
\int _0 ^{\pi /k} p dx  = \int _0 ^{\pi /k} \frac{ - 2 \hbar k l_1 dx}{ a + b \cos 2kx + c \sin 2kx }
=  -\hbar l_1 \int _0 ^{2 \pi } \frac{d ( 2 k x ) }{ a + b \cos  2 k x + c \sin  2 k x } 
=  -\hbar l_1  \frac{2 \pi}{ \surd ( a^2 -b^2-c^2 ) } = \pm \pi \hbar \ . 
\end{equation*}
} 
obtained from a formula \cite{Gradshteyn(1994)} 
\begin{equation*}
\int \frac{dy}{(a + b \cos y + c \sin y ) }= \frac{2}{ \sqrt{ a^2 -b^2-c^2  }} \arctan \big( \frac{ (a-b) \tan (y/2) +c }{\sqrt{ a^2 -b^2 -c^2 }} \big) 
\end{equation*}
for $a^2 > b^2 + c^2$.

If $E = 0$, we take $\Psi = \textbf{1} $ and $\Psi ^D = x$, where \textbf{1} is a constant having the same dimension as $x$. 
  The momentum $p$ is given as 
\begin{equation}
 p =  \frac{-  \hbar l_1 \textbf{1}  }{ (x + l_2 \textbf{1} )^2 + (l_1 \textbf{1}  ) ^2 } \ .               \label{nov0916-1} 
\end{equation}

As is seen from (\ref{may0409-j2}) and (\ref{nov0916-1}),  solutions $\mathcal{S}_0 (x)$'s of the 1-D QSHJE (\ref{may0409-2}) have, even if $E-V(x) \equiv 0$,  a property $\partial \mathcal{S}_0 / \partial x \neq 0$ over $-\infty < x < + \infty$ in accordance with the existence of  $\partial \mathcal{S}_0 / \partial x $ in the denominators of the Schwarzian derivative in (\ref{may0409-2}). 

$\bullet$ \textit{Solving 3-D QSHJE} \ \ 
We solve 3-D QSHJE (\ref{june1308-1}). 
A solution set ($R$, $\mathcal{S}_0$) of  (\ref{june1308-1}) is constructed from a solution $\psi$ of the corresponding 3-D SE 
$- \hbar ^2 \nabla ^2 \psi /2m + (V-E) \psi =0 $ 
as $\psi = R e ^ {i \mathcal{S}_0 / \hbar }$ since (\ref{june1308-1}) is equivalent to the SE with $\psi = R e ^ {i \mathcal{S}_0 / \hbar }$. \cite{Holland(1993)} 
Accordingly, the momentum $\textbf{p} := \nabla \mathcal{S}_0$ is given as 
\begin{equation}
\textbf{p} = \nabla \mathcal{S}_0 = \frac{\hbar }{i} \frac{ \bar{ \psi } \nabla  \psi - \psi \nabla \bar{ \psi } }{2 \psi \bar{\psi}} \ , \label{may0609-1}
\end{equation}
where $\bar{ \psi }$ is the complex conjugate of $\psi$. 
In 1-D space, the $p$ obtained from (\ref{may0609-1}) with $\psi = c_1 \Psi + c_2 \Psi ^D $ is equal to the $p$ obtained from (\ref{may0109-1}) with $A = c_1 , B = c_2 , C = \bar{c}_1  $ and $D = \bar{c}_2$.


\subsection{Generalized Lagrangian formalism} \label{jun1609-1}

We construct a generalized Lag. formalism compatible with the generalized H-J formalism. 
That is, we construct a Lagrangian a solution of the E-L eq. derived from which makes $\nabla \mathcal{S}_0 ^{HJ}( \textbf{x})  = \nabla \mathcal{S}_0 ^{EL}( \textbf{x})$ true, where $\nabla \mathcal{S}_0 ^{HJ}( \textbf{x})$ is a solution of the 3-D QSHJE and $\nabla \mathcal{S}_0 ^{EL}( \textbf{x})$ is a momentum constructed from a solution of the E-L eq., though we will see that no solution of the E-L eq. makes $\nabla \mathcal{S}_0 ^{HJ} = \nabla \mathcal{S}_0 ^{EL}$ true beyond $\nabla \mathcal{S}_0 ^{HJ} - \nabla \mathcal{S}_0 ^{EL}= \mathcal{O} ( \epsilon ^{3/2} )$ because the eq. of motion making $\nabla \mathcal{S}_0 ^{HJ} = \nabla \mathcal{S}_0 ^{EL}$ exact is of a kind not derivable from Lagrangian.  Indeed, not all ordinary differential eqs. are derived from Lagrangians. \cite{Anderson(1992)}

\sloppy

$\bullet$ \textit{Construction of 1-D free Lagrangian} \ \ 
The E-L eq. to be determined is of fourth-order because four parameters $( x , l_1 , l_2 , k )$ to specify the 1-D free momentum $p^{HJ}$ of (\ref{may0409-j2}) correspond to four initial conditions of the E-L eq.

We expand a 1-D free Lagrangian $L = L( \dot{q} , \ddot{q} , \dot{ \ddot{q}} , \cdots )$ to be determined in powers of $\hbar$ as 
$ L=  m \dot{q}^2 /2 + \hbar f_1( \dot{q} , \ddot{q} , \dot{ \ddot{q}} ,  \cdots ) + \hbar ^2 f_2( \dot{q} , \ddot{q} , \dot{ \ddot{q}}   , \cdots ) +  \cdots  $. 
The parameter in $f_1 , f_2 , \cdots $ having dimensions is only mass $m$ because the parameter introduced to the H-J eq. of CM by the extended diff. action is only $\hbar$.  
We assume that $f_i = m^ \alpha   \dot{q} ^\beta  \ddot{q} ^\gamma \dot{\ddot{q}} ^\delta \cdots $, where $\alpha ,\beta , \cdots \in \mathbb{Z}$---since  $\ddot{q},  \dot{\ddot{q}} , \cdots $ may take negative values, to keep $f_i$ real ($f_i \in \mathbb{R}$), $\gamma , \delta , \cdots $ have to be integers. 
We observe that 
i) the $L$ does not contain $\ddot{\ddot{q}}$ and higher-order derivatives, otherwise the resulting E-L eq. cannot be of fourth order, and 
ii) the $\dot{\ddot{q}}$-containing term, if any, appears only in the form: $\hbar ^{1- \alpha } m^\alpha \dot{q}^\beta \dot{\ddot{q}}$, otherwise the order of the resulting E-L eq. exceeds four (\ref{0428ab}). 
Dimensional analysis together with the observations above restricts $f_k$'s to 
\begin{equation}
f_1 = \frac{c_1  \ddot{q}}{ \dot{q}}  \ , \ \ 
f_2 = \frac{c_{2a} }{m} \frac{ \dot{\ddot{q}}}{ \dot{q}^3} + \frac{c_{2b} }{m} \frac{ \ddot{q}^2}{ \dot{q}^4}  \ ,  \ \  
f_k = \frac{c_k} {m^{k-1}} \frac{ \ddot{q}^k}{ \dot{q}^{3k-2}} \ \ (k = 3,4, \cdots) \ ,    \label{jul0211-1}
\end{equation}
where $c_1 , c_{2a}, \cdots \in \mathbb{R}$ are dimensionless consts.  
Thus, we have
\begin{equation}
L = \frac{1}{2} m \dot{q} ^2 + c_1 \hbar  \frac{\ddot{q}}{ \dot{q}} + \frac{c_{2a} \hbar ^2} {m}  \frac{ \dot{\ddot{q}}}{ \dot{q}^3} +  \frac{c_{2b} \hbar ^2} {m}  \frac{ \ddot{q}^2}{ \dot{q}^4} 
+ \sum_{k=3}^{ \infty } \frac{c_k \hbar ^k } {m^{k-1}} \frac{ \ddot{q}^k}{ \dot{q}^{3k-2}}   \ . \label{jun1211-1}   
\end{equation}

\fussy

We obtain the E-L eq., Jacobi-Ostrogradsky (J-O) momenta $( p_{(1)}, p_{(2)} , p_{(3)} )$, and energy $E$ within higher-order Lagrangian formalism (\ref{0428ab}) as 
\begin{multline}
0 = m\ddot{q} 
 + \frac{ 2(3 c_{2a} +  c_{2b}) \hbar ^2 }{ m}   \Big( - \frac{  \ddot{\ddot{q}} }{ \dot{q}^4}     
  +  \frac{ 8 \ddot{q}  \dot{\ddot{q}} }{\dot{q} ^5}
 -  \frac{10  \ddot{q}^3 }{\dot{q} ^6} \Big)   
+  \sum_{k=3}^{ \infty } \frac{ (k-1) c_k \hbar ^k}{m^{k-1}}    \\  
  \times \Big(  - \frac{ (k-2)k \ddot{q} ^{k-3} \dot{ \ddot{q}} ^2 + k \ddot{q}^{k-2} \ddot{ \ddot{q}}}{ \dot{q}^{3k-2}} 
   + \frac{ 2(3k-2)  k \ddot{q}^{k-1}  \dot{ \ddot{q}}}{\dot{q}^{3k-1}} 
   -   \frac{(3k-2) (3k-1)  \ddot{q}^{k+1}}{ \dot{q}^{3k}}   \Big) 
   \ ,  \label{jun1311-1}
\end{multline} 
\begin{align}
p_{(1)} &=  \frac{ \partial L}{ \partial \dot{q}} -  \frac{d}{dt} \frac{ \partial L}{ \partial \ddot{q}}     + \frac{d^2}{dt^2} \frac{ \partial L}{\partial \dot{\ddot{ q }}}   
=   m\dot{q} - \frac{ 2( 3 c_{2a} +  c_{2b}) \hbar ^2 }{m} \frac{ \dot{\ddot{q}}}{ \dot{q}^4}   +\frac{ 4(3 c_{2a} +  c_{2b} ) \hbar ^2} {m} \frac{ \ddot{q}^2}{ \dot{q}^5}      \notag \\ 
& \ \  \ \  + \sum _{k=3}^{\infty }\frac{ c_k \hbar ^k}{m^{k-1}} \Big(  - \frac{k(k-1) \ddot{q}^{k-2} \dot{\ddot{q}}}{ \dot{q}^{3k-2}} + \frac{ (3k-2)(k-1) \ddot{q} ^k }{ \dot{q}^{3k-1}} \Big)  \ ,  \label{jun2711-1}  \\ 
p_{(2)} &= \frac{ \partial L}{ \partial \ddot{q}} - \frac{d}{dt} \frac{ \partial L}{\partial \dot{\ddot{ q }}}      
= \frac{c_1 \hbar }{ \dot{q}} +  \frac{( 3 c_{2a} + 2 c_{2b}) \hbar ^2} {m}  \frac{ \ddot{q}}{ \dot{q}^4} 
+ \sum _{k=3}^{\infty }\frac{ k c_k \hbar ^k}{m^{k-1}} \frac{\ddot{q}^{k-1}}{ \dot{q}^{3k-2}}    , \ 
  p_{(3)} = \frac{ \partial L}{\partial \dot{\ddot{ q }}} 
= \frac{ c_{2a} \hbar ^2}{m \dot{q}^3  }  ,  \notag 
\end{align} 
\begin{align}
E  &= p_{(1)} \dot{q} + p_{(2)} \ddot{q}  +p_{(3)}  \dot{\ddot{q}}   -L   
=  \frac{1}{2} m\dot{q}^2   -\frac{ 2( 3 c_{2a} + c_{2b}) \hbar ^2}{m} \frac{ \dot{\ddot{q}}}{ \dot{q}^3}      + \frac{5 ( 3 c_{2a} + c_{2b} ) \hbar ^2}{m} \frac{ \ddot{q}^2}{ \dot{q}^4}   \notag \\  
& \ \ \ \  +  \sum _{k=3}^{\infty }\frac{ c_k \hbar ^k}{m^{k-1}} \Big(  - \frac{k(k-1) \ddot{q}^{k-2} \dot{\ddot{q}}}{ \dot{q}^{3k-3}} + \frac{ (3k-1)(k-1) \ddot{q} ^k }{ \dot{q}^{3k-2}} \Big)    \ .   \label{jun2711-2}  
\end{align}

We determine $c_1 , c_{2a} , \cdots $ in the Lagrangian (\ref{jun1211-1}) step by step for a momentum $\partial \mathcal{S}_0 ^{EL} / \partial x$ constructed from a solution of the E-L eq. (\ref{jun1311-1}) to satisfy $\partial \mathcal{S}_0 ^{HJ} / \partial x = \partial \mathcal{S}_0 ^{EL} / \partial x$.

The E-L eq. (\ref{jun1311-1}) has to have an oscillatory linear solution $q(t) = vt + f(t) + q_0$, where $v =const. \neq 0$, $q_0 = const.$, and $f(t)$ is a function which satisfies $| \dot{f} / v |< 1$ and oscillates around $f=0$, for $p^{EL}( \dot{q} , \ddot{q} , \dot{\ddot{q}}) |_{q = x}$ defined below to be equal to the $p^{HJ} |_x $ of (\ref{may2009-pp5}). 
The E-L eq. has the oscillatory one only if $c_2 := 3 c_{2a} + c_{2b} < 0$; see footnote. 
\footnote{
For free particle systems, $p_{(1)} = const. $ (\ref{0428ab}). 
Assume that (\ref{jun1311-1}) has the oscillatory linear solution. Then, at time $t_0$ at which the velocity takes a local minimum value $\dot{q} = v - \delta \dot{q} $, $p_{(1)} $ of (\ref{jun2711-1}) is written as 
$p_{(1)}^- =  m  (v - \delta \dot{q} ) -  2 c_2 \hbar ^2  \dot{\ddot{q}} ^- / m  (v - \delta \dot{q} ) ^4$, where $\dot{\ddot{q}} ^- > 0$, since $\ddot{q} =0$ at $t_0$. 
Likewise, at time $t_1$ at which the velocity has a local maximum value $\dot{q} = v + \delta \dot{q} $, $p_{(1)} $ is written as 
$p_{(1)}^+ =  m ( v + \delta \dot{q} ) -  2 c_2 \hbar ^2  \dot{\ddot{q}}^+  / m  ( v + \delta \dot{q} )^4$, where $\dot{\ddot{q}}^+ < 0$. 
We have $p_{(1)}^+ - p_{(1)}^- \simeq 2 m \delta \dot{q} + 2 c_2 \hbar ^2 ( \dot{\ddot{q}} ^- - \dot{\ddot{q}} ^+ ) / mv^4 =0 $ only if $c_2 <0$.  
}

The $f(t)$ of the oscillatory linear solution is written as a Fourier series because, \label{aug2114-1} at any $t_0$ at which $\dot{q}$ takes a local maximum (minimum) value, at which $\ddot{q} (t_0) =0$, values of $\dot{q}(t_0) $ and $\dot{\ddot{q}}(t_0) < 0$ ($\dot{\ddot{q}}(t_0) > 0$) are the same according to $p_{(1)} ( \dot{q} , \ddot{q} , \dot{ \ddot{q}} )= const. $ and $E( \dot{q} , \ddot{q} , \dot{ \ddot{q}} )= const.$ 
Accordingly, the solution $q(t)$ is written as 
\begin{align}
q(t) &= vt +f + q_0  \notag \\ 
     &= v t + \sum _{ n=1}^{\infty} ( a_n \cos n \omega t + b_n \sin n \omega t ) + q_0 = v t + \sum _{ n=1}^{\infty} A_n \cos ( n \omega t + \theta _n ) + q_0                 \ ,   \label{may0409-cv3}
\end{align}
and the interval $\delta t$ between the local velocity maximums is determined as $\delta t = 2 \pi / \omega $ by $A_1$-term; 
the amplitudes $A_2 , A_3 , \cdots $ are not large enough to generate velocity maximum. 
If $| \dot{f}/v | \ll 1$, we call the $q(t)$ a semiclassical solution.

In the solution (\ref{may0409-cv3}), a relation $\hbar \omega = mv^2 / \surd ( - 2c_2 )$ between $\omega$ and $v$ holds true regardless of the magnitude of the oscillation because the $\omega$-term  $(\omega A_1 / v )^1 \cos ( \omega t + \theta _1 )$ having infinitesimal (inf.) order $(\omega A_1 / v )^1$ sums up to zero in 
$m \ddot{q} - 2 c_2 \hbar ^2 \ddot{ \ddot{q}}/ m \dot{q} ^4$ of (\ref{jun1311-1}) regardless of the one.

\sloppy

In \label{feb1521-1} the solution (\ref{may0409-cv3}), inf. order of $( n \omega A_n / v) $ is equal to that of $(\omega A_1 / v ) ^n $, $\mathcal{O}( n \omega A_n /v ) = \mathcal{O} ( ( \omega A_1 /v ) ^n )$, as $A_1 \to 0$, which we show. 
From (\ref{may0409-cv3}), we have 
\begin{equation}
\dot{q}(t) = v ( 1 - \sum \nolimits _{ n=1}^{\infty} ( n \omega A_n / v ) \sin ( n \omega t + \theta _n ) ) \ .   \label{feb1221-2}
\end{equation} 
We insert (\ref{feb1221-2}) into the E-L eq. (\ref{jun1311-1}). 
We then expand and recombine to have 
\begin{equation}
\sum \nolimits _{n=1}^{\infty} -( m n\omega + \frac{2c_2 \hbar ^2 (n \omega )^3 }{mv^4} ) \frac{n \omega  A_n}{v} \cos ( n \omega t + \theta _n ) + \sum \nolimits _{n=1}^{\infty} d_n \cos ( n \omega t + \eta _n )=0 \ .  \label{feb1921-1}
\end{equation} 
The first sum comes from $m\ddot{q} -  2 c_2 \hbar ^2  \ddot{\ddot{q}} / m \dot{q}^4$ of (\ref{jun1311-1}).  
The $d_n \cos ( n \omega t + \eta _n )$ is a term constructed from a sum of products of Fourier components in the expansion of (\ref{jun1311-1}). 
For example, $d_3 \cos ( 3 \omega t + \eta _3 ) $  is constructed from a sum of $3\omega $-terms resulting from 
$ \cos ( \omega t + \theta _1 ) \cos ( 2 \omega t + \theta _2 )$, $\cos ^3 ( \omega t + \theta _1 )$, $ \cos ( \omega t + \theta _1 )  \cos ( 4 \omega t + \theta _4)$, and so on  
(remind $\cos a \cos b =  \cos ( a+b)/2 + \cos (a-b)/2$ and $\cos ( \omega t + \alpha ) + \cos ( \omega t + \beta ) = \surd ( 2 + 2 \cos ( \alpha - \beta ) ) \cos ( \omega t + \gamma )$, where $\gamma$ is an angle satisfying $\cos \gamma = \cos \alpha + \cos \beta$ and $\sin \gamma = \sin \alpha + \sin \beta$). 
We have $\theta _n = \eta _n$ and $d_n = ( m n\omega + 2c_2 \hbar ^2 (n \omega )^3  / mv^4 ) ( n \omega  A_n / v )$ since $q(t)$ is a solution. 
Given an $\omega$-term $ ( \omega A_1 /v )  \sin ( \omega t + \theta _1 )$ of (\ref{feb1221-2}), we construct $(A_2 , \theta _2 ) , (A_3 , \theta _3 ) , \cdots $ step by step as follows. 
 The $2 \omega $-term having the lowest inf. order in the second sum of (\ref{feb1921-1}) results from $( \omega A_1 / v )^2  \sin ^2 ( \omega t + \theta _1 )$. 
It sums up to zero with the $2 \omega $-term in the first sum, from which $(A_2 , \theta _2 )$ is determined. 
 The $3 \omega $-term having the lowest inf. order in the second sum of (\ref{feb1921-1}) results from 
 $ ( \omega A_1 / v ) \sin ( \omega t + \theta _1 ) \cdot ( 2 \omega A_2 / v )  \sin ( 2 \omega t + \theta _2 ) $ and $(( \omega A_1 / v ) \sin ( \omega t + \theta _1 ))^3$. 
It sums up to zero with the $3 \omega $-term in the first sum, from which $(A_3 , \theta _3 )$  is determined. 
\footnote{
From the given $\omega$-term and the determined $2 \omega$- and $3\omega$-term, an $\omega$-term having inf. order $(\omega A_1 / v )^3$ and $(\omega A_1 / v )^5$ is obtained from $( \omega A_1 / v ) \cos ( \omega t + \theta _1 ) \cdot   ( 2 \omega A_2 / v ) \cos ( 2\omega t + \theta _2 ) $ and $ 2 \omega A_2 / v ) \cos ( 2 \omega t + \theta _2 ) \cdot ( 3 \omega A_3 / v ) \cos ( 3\omega t + \theta _3 ) $ respectively. 
Similarly, a $2 \omega$-term having inf. order $(\omega A_1 / v )^4$ 
 is obtained from $ ( \omega A_1 / v ) \cos ( \omega t + \theta _1 ) \cdot ( 3 \omega A_3 / v ) \cos ( 3\omega t + \theta _3 ) $. 
The terms thus obtained is irrelevant to the present discussion aiming at $\mathcal{O}( n \omega A_n /v ) = \mathcal{O} ( ( \omega A_1 /v ) ^n )$. 
} 
In the same manner, we determine $(A_4 , \theta _4 )$ and higher. 
In the step by step process, we see that inf. order of $n\omega A_n /v$ is equal to that of $(\omega A_1 / v )^n$.

\fussy

We see that the Lagrangian (\ref{jun1211-1}) does not contain $\dot{\ddot{q}}$, that is $c_{2a} =0$, as follows. 
The Hamilton's principle for (\ref{jun1211-1}) is 
\begin{align}
&\delta \mathcal{S} _0 (x , \dot{x} ,  \ddot{x}      )
= \int _{ x_0 , \dot{x}_0 , \ddot{x}_0 , \, t_0 }  ^{ x , \dot{x} , \ddot{x} , \, t} \big( \ L(  q  +  \delta   q  , \dot{ q }+  \delta  \dot{ q },\ddot{ q }+  \delta  \ddot{q } , \dot{\ddot{ q }}+  \delta  \dot{\ddot{q }} ) -  L(  q ,\dot{ q },\ddot{ q }  , \dot{ \ddot{ q }}) \ \big) d \tilde{t}       \notag \\ 
&= \int ( \frac{ \partial L}{\partial  q }  \delta   q  + \frac{ \partial L}{\partial \dot{ q }}  \delta  \dot{ q } + \frac{ \partial L}{\partial \ddot{ q }}  \delta  \ddot{ q } + \frac{ \partial L}{\partial \dot{\ddot{ q} }}  \delta  \dot{\ddot{ q }}  ) d \tilde{t}    
= \int ( \frac{ \partial L}{\partial  q }  - \frac{d}{dt} \frac{ \partial L}{\partial \dot{ q }}  + \frac{d^2}{dt^2} \frac{ \partial L}{\partial \ddot{ q }} - \frac{d^3}{dt^3} \frac{ \partial L}{\partial \dot{\ddot{ q }}}  )  \delta   q  d \tilde{t}   \notag \\  
& \ \ \ \  
+ (\frac{ \partial L}{\partial \dot{ q }}  - \frac{d}{dt} \frac{ \partial L}{\partial \ddot{ q }}  + \frac{d^2}{dt^2} \frac{ \partial L}{\partial \dot{\ddot{ q }}}    )  \delta   q  | _{t_0} ^t + (\frac{ \partial L}{\partial \ddot{ q }}  - \frac{d}{dt} \frac{ \partial L}{\partial \dot{\ddot{ q }}})  \delta  \dot{ q } | _{t_0}  ^t + (\frac{ \partial L}{\partial \dot{\ddot{ q }}}   )  \delta  \ddot{ q } |_{t_0}  ^t  \ .         \label{may1019-1}
 \end{align} 
We write boundary terms, $ \partial L / \partial \dot{ q } - \cdots + ( d ^2 / dt^2 )  \partial L / \partial \dot{\ddot{ q }} \, $, 
$ \partial L/ \partial \ddot{ q }  - (d / dt)  \partial L / \partial \dot{\ddot{ q }}$, and $ \partial L / \partial \dot{\ddot{ q }}$ as $p_{(1)}$,  $p_{(2)}$, and $p_{(3)}$ respectively. 
The $p_{(3)} $ should be evaluated by a formula $p_{(3)} |_{ \, t} = \lim _{\delta \ddot{q} \to 0 } \delta \mathcal{S}_0 / \delta \ddot{q} | _{ \, t }$, where  $\delta \ddot{q} |_{ \, t} $ should be moved on a solution of the E-L eq. passing through fixed  $( q (t_0) , \dot{q} (t_0), \ddot{q}(t_0) )$ and $( q (t),\dot{q}(t) )$.  
We however cannot move $\delta \ddot{q} |_{ \, t} $ with the five fixed---a solution of the fourth-order E-L eq. is made immobile with four initial conditions, say, $ q (t_0) , \dot{q} (t_0), \ddot{q}(t_0) $ and $ q (t) $. 
Therefore $\partial L / \partial \dot{\ddot{ q }} =0$ in  (\ref{may1019-1}).

We determine $c_1$ of the Lagrangian (\ref{jun1211-1}). 
Let $c_{2a} =0$. 
Then, from $\delta \mathcal{S}_0 = p_{(1)} \delta q + p_{(2)} \delta \dot{q} $ and $\delta  \dot{q} =  ( \ddot{q} / \dot{q}) \delta q$ (cf. Ref. \refcite{Bouda(2003)}), we have $p^{EL} =  \lim _{ \delta q \to 0} \delta  \mathcal{S}_0 / \delta q = p_{(1)} + p_{(2)}( \ddot{q} / \dot{q}) $. 
If $c_1 \neq 0$, $p^{EL} (-t) \neq - p^{EL} (t)$ because of $c_1$ in $p_{(2)}$, which means that a free particle system of GCM is not time-reversal invariant. 
It is natural to consider that the system is time-reversal invariant because the system is in QM. \cite{Messiah(1961)} \cite{Sozzi(2008)} 
Accordingly, we set $c_1 =0$. 
\footnote{
For a solution $q(t) = vt + A_1 \cos ( \omega t + \theta _1 ) + \cdots $ of the E-L eq., we have 
$p^{EL} = p_{(1)} +  c_{2b} \hbar ^2 \omega ^4 A_1 ^2 /mv^5 + ( c_{2b} \hbar ^2 \omega ^4 A_1 ^2 /mv^5 ) \cos (2 \omega t + 2 \theta _1 ) + \cdots $ if $c_1 = 0$, and $p^{EL} = p_{(1)} - ( c_1 \hbar \omega ^2 A_1 / v^2 ) \cos( \omega t + \theta _1 ) + \cdots$ if $c_1 \neq 0$. 
Comparing each $p^{EL}$ with $p^{HJ} = \hbar k ( 1+ \epsilon \cos ( 2kx + \theta ) +\cdots )$ (cf. (\ref{may2009-pp5})), we see $\omega = kv$ if $c_1 =0$ and $\omega = 2 kv $ if $c_1 \neq 0 $, where $k = \surd (2mE ) / \hbar$. 
}

For $c_1 = c_{2a} =0$, $p^{EL} $ is explicitly given as 
\begin{equation}
 p^{EL} (x) 
=   p_{(1)} + p_{(2)} \frac{ \ddot{q}}{\dot{q}}  \ |_{q(t) = x}   
=  p_{(1)}    +  \frac{2 c_{2b} \hbar ^2} {m}  \frac{ \ddot{q}^2}{ \dot{q}^5}
+ \sum _{k=3}^{\infty }\frac{ k c_k \hbar ^k}{m^{k-1}} \frac{ \ddot{q}^k}{\dot{q}^{3k-1}} \ |_{q(t) = x}  \ , \label{jun1211-3}
\end{equation}
where $p_{(1)} = const.$ (\ref{0428ab}) is left intact for later convenience.



We determine $ c_{2b}$ of (\ref{jun1211-1}) with $c_1 = c_{2a} =0$ (see \ref{mar2009-01} for detail). 
We obtain the $\omega ^{EL}$ inserting $q(t)$ of (\ref{may0409-cv3}) into the E-L eq. (\ref{jun1311-1}) as $\omega ^{EL} = mv^2 / \hbar \sqrt{-2 c_2 } +  \mathcal{O} ( ( \dot{f}  /v) ^2 ) $. 
We obtain $\omega ^{HJ} := kv$ inserting $E$ of (\ref{jun2711-2}) into $\omega ^{HJ} = kv = v \sqrt{2mE} / \hbar $ as $\omega ^{HJ} = mv^2 / \hbar +  \mathcal{O} ( ( \dot{f}  /v) ^2 )$. 
Comparing both, we see that $c_{2b} = -1/2$. 
The 1-D free Lagrangian thus is determined up to $\hbar ^2$-term as 
\footnote{\label{sep1314-1}
The 1-D Lagrangian (\ref{may0109-7}) 
 differs from 1-D Lagrangian 
$L = m \dot{q}^2 /2 + ( \hbar ^2 / 4m) ( 5 \ddot{q} ^2 / 2\dot{q}^4 -  \dot{\ddot{q}} / \dot{q}^3 ) \ \ \cdots (*1)$ constructed in Ref. \refcite{Bouda(2003)}. 

We outline the method used to derive $(*1)$. 
  Bouda \cite{Bouda(2003)} derives $\partial \mathcal{S} _0 / \partial x =  p_{(1)} +  \ddot{q} p_{(2)} / \dot{q} + \dot{\ddot{q}} p_{(3)} / \dot{q} \ |_{q=x} $ and $E= \dot{x} \partial \mathcal{S} _0 / \partial x -L \ \  \cdots (*2)$ from $d \mathcal{S} = d \mathcal{S}_0 -Edt = Ldt$ and $E = \dot{q} p_{(1)} + \ddot{q} p_{(2)} + \dot{\ddot{q}} p_{(3)} - L = const.$, where 
$L = L (q ,\dot{q} ,\ddot{q} , \dot{\ddot{q}} )$,  
$p_{(1)} =   \partial L /  \partial \dot{ q }  -  ( d / dt ) (\partial L / \partial \ddot{ q })  +(d^2 / dt^2 ) ( \partial L / \partial \dot{\ddot{ q }})$, 
$p_{(2)} =  \partial L / \partial \ddot{ q }  - (d / dt)  \partial L / \partial \dot{\ddot{ q }}$, and 
$ p_{(3)} =  \partial L / \partial \dot{\ddot{ q }} $. 
Using dimensional analysis, he expands the Lagrangian to be determined in powers of $\hbar$ as $L= \sum _{n=0}^{\infty } ( \hbar ^n / m^{n-1} ) ( \alpha _n \ddot{q}^n / \dot{q} ^{3 n-2} + \beta _n \ddot{q} ^{n - 2} \dot{\ddot{q}} / \dot{q} ^{3n-3} )$, where $\alpha _n$ and $\beta _n$ are consts. to be determined. 
(We simplified the original form \cite{Bouda(2003)} to the one appropriate to the present context.) 
He writes $ \partial _x \mathcal{S} _0  $ as a power series of $\hbar$ using the Lagrangian. 
Making use of  $\partial /\partial x = \dot{x} ^{-1} \partial / \partial t$, 
he also writes $ \partial _x ^2 \mathcal{S} _0  $ and  $ \partial _x ^3 \mathcal{S} _0  $  as power series of $\hbar$. 
Inserting them, together with $(*2)$, into the 1-D QSHJE written as 
$ ( \partial _x \mathcal{S} _0 )^4 /2m  + (\hbar ^2 / 4m )( \partial _x \mathcal{S} _0    \cdot \partial _x ^3 \mathcal{S} _0  -  (3/2)( \partial _x ^2 \mathcal{S} _0  )^2 )  -  E ( \partial _x \mathcal{S} _0 )^2 =0$, he has an eq. 
$\hbar ^0 (\cdots ) + \hbar ^1 ( \cdots ) + \hbar ^2 ( \cdots ) + \cdots =0$
 for $\alpha _n$ and $\beta _n$.   
He obtains $\alpha _2 = 5/8$ and  $ \beta _2 = - 1/4 $ leading to $(*1)$ from  the eq. at $\hbar ^2$-level:  
$m \hbar ^2 ( ( \alpha _2 - 5/8 ) \ddot{q} ^2 / \dot{q} ^2 + ( \beta _2 + 1/4 ) \dot{\ddot{q}} / \dot{q} ) =0 $ regarding $\ddot{q}$ and  $\dot{\ddot{q}} $ in it as independent.

Whereas we consider that $\ddot{q}$ and  $\dot{\ddot{q}} $ are dependent as quantities derived from a solution $q(t)$ of the E-L eq. 
} 
\begin{equation}
L  =  m \dot{q} ^2 / 2 -  \hbar ^2 \ddot{q}^2 /  2m \dot{q}^4   \ .                     \label{may0109-7}    
\end{equation}

Given an $E^{HJ} >0$ and the $p^{HJ}$ of (\ref{may2009-pp5}), a solution $q(t) = vt + a_1 \cos \omega t + b_1 \sin \omega t + \mathcal{O}((\dot{f}/v)^2)$ of the E-L eq. derived from (\ref{may0109-7}) 
makes $E^{EL}$ of (\ref{jun2711-2}) and $p^{EL} $ of (\ref{jun1211-3}) agree with the $E^{HJ}$ and $p^{HJ}$ up to $\mathcal{O}(( \dot{f} / v )^2 ) $, where $v = \sqrt{2mE} ( 1 - \epsilon + \mathcal{O}( \epsilon ^2) )/m$, $\omega =kv = 2E ( 1 -  \epsilon + \mathcal{O}( \epsilon ^2) ) / \hbar$,  and $( a_1, b_1 )$ are given in (\ref{oct1410-7}) and (\ref{062906ab}); 
the $\epsilon$ in the above $v = \cdots $ and $\omega = \cdots$ is given as $ \epsilon  = \surd ( \epsilon _1 ^2 + \epsilon _2^2 )$ with $\epsilon _1$ and $\epsilon _2$ in (\ref{may2009-pp5}). 
The $p^{HJ}= p^{EL} + \mathcal{O}((\dot{f}/v)^3)$ is given in \ref{oct0512-5}. 
The $E^{HJ} = E^{EL} + \mathcal{O}((\dot{f}/v)^3)$ is seen from (\ref{mar2116-1}) with $v$, $\omega$, and $A_1 = \surd ( a_1^2 + b_1^2 )$ in it replaced with those given above.

\sloppy

\label{jun3011-1} 
No value of $c_3 $ makes $ p^{EL}$ and $p^{HJ} $ agree at $\mathcal{O}(( \dot{f} / v )^3 ) $ because the Lagrangian (\ref{jun1211-1}) does not have enough adjustable parameters to make them so (see \ref{jun2511-1}). 
As was mentioned in p.\pageref{dec1017-1}, a higher-order version of the integral form of the classical d'Alembert's principle would make $ p^{NW}$ and $p^{HJ} $  agree at $\mathcal{O}(( \dot{f} / v )^3 ) $ and higher (see \ref{sep2212-730}).


\fussy

$\bullet$ \textit{Construction of 3-D  Lagrangian} \ \ 
A 3-D free Lagrangian is obtained from (\ref{may0109-7}) as 
\begin{equation}
L =  m  \dot{\textbf{q} } ^2 /2 -   \hbar ^2 \ddot{ \textbf{q} } ^2 / 2 m   \dot{ \textbf{q}}^4       \ , \label{jan3112-1} 
\end{equation}
where $ \dot{\textbf{q} } = ( \,  \dot{q}_1(t)$,  $\dot{q}_2(t), \dot{q}_3(t) \,  )$, $ \dot{\textbf{q} } ^2 =  \dot{\textbf{q} } \cdot  \dot{\textbf{q} }  = \dot{q}_1\dot{q}_1 + \dot{q}_2 \dot{q}_2 + \dot{q}_3 \dot{q}_3 $, and $ \dot{\textbf{q} } ^4 = ( \dot{\textbf{q} } \cdot  \dot{\textbf{q} } )^2 $; likewise for  $\ddot{ \textbf{q} } ^2 $. 
It is meaningless to consider $\hbar ^3$ and higher terms; see the second last paragraph.

A \label{apr1912-1} 3-D Lagrangian of a system having a potential field $V(\textbf{q})$, which is assumed to be almost const. over $1/ | \textbf{k} | = \hbar / \surd ( 2mE)$ is $L =  m  \dot{\textbf{q} } ^2 /2 -  V(\textbf{q}) - \hbar ^2 \ddot{ \textbf{q} } ^2 / 2 m   \dot{ \textbf{q}}^4$. 
Whereas in an interference region, even if $  V(\textbf{q}) =0$, streamlines of $\textbf{p}^{HJ}  $ of (\ref{may0609-1}) may macroscopically curve (an example in \S \ref{jan2107-6}). 
Correspondingly trajectories, too, may because $\textbf{p}^{HJ} = \textbf{p}^{EL}$ has to hold true for GCM to be consistent. 
That is, the momentum field  $\textbf{p}^{HJ} ( \textbf{x} )$, in addition to $V(\textbf{x})$, affect the motion of the particle in GCM. 
To describe the curving, we introduce a pseudo-potential $Q(\textbf{q})$. 
Thus, we have the 3-D Lagrangian
\begin{equation}
L = m   \dot{\textbf{q} } ^2 /2 - V(\textbf{q}) 
-  Q(\textbf{q})  -  \hbar ^2   \ddot{ \textbf{q}}^2 / 2m \dot{ \textbf{q}}^4    \ . \label{may0909-ee1}
\end{equation}

We characterize Lagrangians (\ref{may0109-7}), (\ref{jan3112-1}), and (\ref{may0909-ee1}) as semiclassical in the sense that no $\hbar ^3$-term makes $p^{HJ} =p^{EL}$ at $\mathcal{O}(( \dot{f} / v )^3 ) $.

$\bullet$ \textit{Dynamical eqs.} \ \  
From (\ref{may0909-ee1}),  we obtain (see \ref{0428ab}) 
\footnote{
One-dimensional forms of (\ref{june1308-2apr15}), (\ref{may0209-a2apr15}), and $\textbf{p}_{(1)}$ of (\ref{jan2012-1apr15}) are (cf. (\ref{jun1311-1}), (\ref{jun2711-1}), and (\ref{jun2711-2})) 
\begin{equation*}
\frac{\partial V}{ \partial q} + \frac{\partial Q}{ \partial q} +m \ddot{q} 
+ \frac{ \hbar ^2}{m} (  \frac{ \ddot{\ddot{q}}}{ \dot{q} ^4} - \frac{ 8 \ddot{q}\dot{\ddot{q}}}{\dot{q}^5} + \frac{ 10 \ddot{q}^3}{ \dot{q}^6} ) =0 , 
E = \frac{m}{2} \dot{q}^2 + V+Q 
+ \frac{ \hbar ^2}{m} ( \frac{ \dot{\ddot{q}}}{ \dot{q}^3} - \frac{ 5 \ddot{q}^2 }{ 2 \dot{q}^4}) , 
p_{(1)} = m \dot{q} +  \frac{ \hbar ^2}{m} ( \frac{ \dot{\ddot{q}}}{ \dot{q}^4} - \frac{ 2 \ddot{q}^2 }{ \dot{q}^5}) . 
\end{equation*}
} 
\footnote{
The \label{may1716-1} velocity $\dot{q} (t)$ of solutions of the E-L eq. derived from (\ref{may0109-7}) with initial conditions $( q(t_0) , \cdots , \dot{\ddot{q}}(t_0) )$ satisfying  $E =const. >0$ never approach zero, which we see as follows. 
We have 
$   m \dot{q}^2 /2 - p_{(1)} \dot{q} +E = -  \hbar ^2 \ddot{q} ^2 /  2 m \dot{q} ^4 $ removing $\dot{\ddot{q}}$ from 
$p_{(1)} = m \dot{q} + ( \hbar ^2/m ) ( \dot{\ddot{q}}/\dot{q}^4 -2\ddot{q}^2/ \dot{q}^5 ) = const.$ and $E = m \dot{q}^2/2 + ( \hbar ^2/m ) (  \dot{\ddot{q}}/\dot{q}^3 -5\ddot{q}^2/ 2\dot{q}^4 ) = const. $  
Assume $\dot{q} (t) \to 0$. Then, we have a contradiction between $ m \dot{q}^2 /2 - p_{(1)} \dot{q} +E \to E >0$ and $ -  \hbar ^2 \ddot{q} ^2 /  2 m \dot{q} ^4 \leq  0 $. Therefore $\dot{q}(t) \not\to 0$.
} 
\begin{align}
 &\textrm{(E-L eq.)} \ \  0= -  \partial V(\textbf{q}) / \partial q_i - \partial Q(\textbf{q}) / \partial q_i 
- m \ddot{q}_i       \notag \\  
&   + \frac{ \hbar ^2  } {m}  
 \big(  - \frac{ \ddot{ \ddot{q}}_i }{ \dot{\textbf{q} }^4 }  - \frac{ 4 \dot{q}_i \dot{ \ddot{\textbf{q}}} \ddot{\textbf{q}}}{ \dot{\textbf{q}}^6} +  \frac{ 8 \dot{ \ddot{q}}_i \dot{\textbf{q}} \ddot{\textbf{q}} }{ \dot{\textbf{q}} ^6} 
+ \frac{ 4 \ddot{q}_i  \dot{\textbf{q}} \dot{\ddot{\textbf{q}} }}{ \dot{\textbf{q}} ^6}  
+ \frac{ 2 \ddot{q}_i  \ddot{\textbf{q}}^2  }{ \dot{\textbf{q}} ^6}  + \frac{12 \dot{q}_i  \ddot{\textbf{q}} ^2   ( \dot{\textbf{q}} \ddot{\textbf{q}} )}{ \dot{\textbf{q}} ^8} - \frac{ 24 \ddot{q}_i (\dot{\textbf{q}} \ddot{\textbf{q}} )^2  }{ \dot{\textbf{q}}^8 }  \big)    \ ,       \label{june1308-2apr15}   
\end{align} 
\begin{align}
E&= \frac{m  \dot{\textbf{q}} ^2  }{2}     +   V( \textbf{q}) + Q(\textbf{q})  + \frac{ \hbar ^2  }{ m}  \big( \frac{  \dot{ \textbf{q}}   \dot{ \ddot{\textbf{q}}} }{ \dot{ \textbf{q}}^4}   + \frac{3}{2}  \frac{          \ddot{ \textbf{q}}^2   }{ \dot{ \textbf{q}}^4}   -  \frac{ 4 (\dot{ \textbf{q}} \ddot{ \textbf{q}} )^2   }{ \dot{ \textbf{q}}^6} \big)    \ ,  \label{may0209-a2apr15}  \\ 
 p_{(1)i} &= m  \dot{q}_i  + \frac{  \hbar ^2 }{ m} \big( \frac{ \dot{ \ddot{q}} _i}{ \dot{ \textbf{q}}^4} +        \frac{ 2 \ddot{ \textbf{q}} ^2  \dot{q}_i }{ \dot{ \textbf{q}}^6} -  \frac{ 4 \dot{ \textbf{q}} \ddot{ \textbf{q}}   \ddot{q}_i }{ \dot{ \textbf{q}}^6} \big)    \  ,  \ \ \ \   \ \ \ \  p _{(2)i} = - \frac{  \hbar ^2  }{ m } \frac{ \ddot{ q} _i}{\dot{ \textbf{q}}^4}  \ .    \label{jan2012-1apr15}  
\end{align} 
If $ V(\textbf{q}) + Q(\textbf{q}) $ is separable on a Cartesian coordinate system, the conserved energy $E$ is decomposed to three conserved components $E_i = m\dot{q}_i ^2 /2 +V_i + Q_i + \hbar ^2 ( \cdots )/m $ though the $\hbar ^2$-term is inseparable, which follows from (\ref{42606va}) and (\ref{may1311-1}). 
\footnote{
We give a formula of $E_i$ of a free particle. 
We have from (\ref{may1311-1})
\begin{align}
\frac{dE_i}{dt} 
&=  m \dot{q} _i \ddot{q}_i  + \frac{ \hbar ^2  } {m}  
 \big(   \frac{\dot{q} _i \ddot{ \ddot{q}}_i }{ \dot{\textbf{q} }^4 }  + \frac{ 4 \dot{q}_i ^2 \dot{ \ddot{\textbf{q}}} \ddot{\textbf{q}}}{ \dot{\textbf{q}}^6} -  \frac{ 8 \dot{q} _i \dot{ \ddot{q}}_i \dot{\textbf{q}} \ddot{\textbf{q}} }{ \dot{\textbf{q}} ^6} 
- \frac{ 4 \dot{q} _i \ddot{q}_i  \dot{\textbf{q}} \dot{\ddot{\textbf{q}} }}{ \dot{\textbf{q}} ^6}  
- \frac{ 2 \dot{q} _i \ddot{q}_i  \ddot{\textbf{q}}^2  }{ \dot{\textbf{q}} ^6}  - \frac{12 \dot{q}_i ^2 \ddot{\textbf{q}} ^2   ( \dot{\textbf{q}} \ddot{\textbf{q}} )}{ \dot{\textbf{q}} ^8} 
+ \frac{ 24 \dot{q} _i \ddot{q}_i (\dot{\textbf{q}} \ddot{\textbf{q}} )^2  }{ \dot{\textbf{q}}^8 }  \big)   \notag \\ 
&= \frac{d}{dt} \Big(  \frac{ m\dot{q} _i ^2 }{2} + \frac{ \hbar ^2  } {m}  
 \big(  \frac{ \dot{q}_i \dot{\ddot{q}}_i }{ \dot{\textbf{q}} ^4}- \frac{ \ddot{q}_i ^2 }{ 2 \dot{\textbf{q}} ^4} +   \frac{ 2 \ddot{\textbf{q}} ^2 \dot{q}_i^2 }{ \dot{\textbf{q}} ^6} -   \frac{ 4 \dot{\textbf{q}}\ddot{\textbf{q}}  \dot{q}_i \ddot{q}_i  }{ \dot{\textbf{q}} ^6}   \big)  \Big) 
+ \frac{ \hbar ^2  } {m} \frac{ 2 ( \dot{\textbf{q}} \ddot{\textbf{q}} \ddot{q}_i^2 - \ddot{\textbf{q}}^2 \dot{q}_i \ddot{q}_i ) }{  \dot{\textbf{q}} ^6 } = 0 \ ,  \notag 
\end{align} 
from which we obtain
\begin{equation*}
E_i =  \frac{ m\dot{q} _i ^2 }{2} + \frac{ \hbar ^2  } {m}  
 \big(  \frac{ \dot{q}_i \dot{\ddot{q}}_i }{ \dot{\textbf{q}} ^4}- \frac{ \ddot{q}_i ^2 }{ 2 \dot{\textbf{q}} ^4} +   \frac{ 2 \ddot{\textbf{q}} ^2 \dot{q}_i^2 }{ \dot{\textbf{q}} ^6} -   \frac{ 4 \dot{\textbf{q}}\ddot{\textbf{q}}  \dot{q}_i \ddot{q}_i  }{ \dot{\textbf{q}} ^6}  
+  \int ^{ \,\, t}  \frac{ 2 ( \dot{\textbf{q}} \ddot{\textbf{q}} \ddot{q}_i^2 - \ddot{\textbf{q}}^2 \dot{q}_i \ddot{q}_i ) }{  \dot{\textbf{q}} ^6 } d\tilde{t}  \,  \big)  = const. 
\end{equation*}
We see $ \sum _{i=1} ^{3}  \int  ^{ \, t}  2 ( \dot{\textbf{q}} \ddot{\textbf{q}} \ddot{q}_i^2 - \ddot{\textbf{q}}^2 \dot{q}_i \ddot{q}_i ) /  \dot{\textbf{q}} ^6 d\tilde{t}  =0$ because $ \sum _{i=1} ^{3} (m \dot{q} _i ^2 / 2 + \hbar ^2 (  \dot{q}_i \dot{\ddot{q}}_i / \dot{\textbf{q}} ^4 - \cdots -  4 \dot{\textbf{q}}\ddot{\textbf{q}}  \dot{q}_i \ddot{q}_i  / \dot{\textbf{q}} ^6  ) /m )= E$. 
We choose consts. of the integrations so as to make $\lim _ { \, t \to + \infty } (t -t_0 )^{-1}  \int _{\, t_0}   ^{ \, t}  2 ( \dot{\textbf{q}} \ddot{\textbf{q}} \ddot{q}_i^2 - \cdots ) /  \dot{\textbf{q}} ^6 d\tilde{t}   =0$ hold true in each direction because it holds true in 1-D space. 
}

\sloppy 

Three dimensional momentum $\textbf{p} = \nabla \mathcal{S}_0 $ along a solution path is obtained as follows. 
 Let $| \textbf{q} |$ be a length from a fixed point $\textbf{q}|_{t _0}$ to $\textbf{q}|_t$ along a solution $\textbf{q}(t)$ of the E-L eq. (\ref{june1308-2apr15}),  where we assume $t > t _0$ to make $| \textbf{q} |$ increase when $t$ increases. 
Then, from $d | \textbf{q} | /dt = | \dot{ \textbf{q} } |$, we have $dt = d | \textbf{q} |  /  | \dot{ \textbf{q} } |$, with which we obtain 
$d  \textbf{q}  =  ( d  \textbf{q} / dt ) dt 
= \dot{ \textbf{q} } d | \textbf{q} |  / | \dot{ \textbf{q} } |$ and 
$d \dot{ \textbf{q} } =  ( d \dot{ \textbf{q} } / dt ) dt 
= \ddot{ \textbf{q} } d | \textbf{q} |  / | \dot{ \textbf{q} } |$. 
Accordingly, $d \mathcal{S}_0 = \textbf{p}_{(1)} d  \textbf{q}  + \textbf{p} _{(2)} d \dot{ \textbf{q} }$ obtained from $d \mathcal{S} = d \mathcal{S}_0 -Et = Ldt$ along trajectory and $E= \textbf{p}_{(1)} \dot{ \textbf{q}}  + \textbf{p} _{(2)}  \ddot{ \textbf{q} } -L$ (cf. (\ref{051506fa}))  is written as 
$d \mathcal{S} _0 = \textbf{p}_{(1)}  \dot{ \textbf{q} } d | \textbf{q} |  / | \dot{ \textbf{q} } | +   \textbf{p} _{(2)}  \ddot{ \textbf{q} } d | \textbf{q} |  / | \dot{ \textbf{q} } | $, from which we have 
$\partial  \mathcal{S} _0 / \partial  | \textbf{q} |  =  ( \textbf{p}_{(1)}  \dot{ \textbf{q} } +  \textbf{p} _{(2)}  \ddot{ \textbf{q} } ) /  | \dot{ \textbf{q} } | $. 
The $x _i$-component $\partial \mathcal{S} _0 / \partial x _i $ of $\partial \mathcal{S} _0 / \partial | \textbf{q} |$ is given as 
\begin{equation}
\frac{ \partial \mathcal{S}_{0}}{\partial x_i} 
= \frac{ \partial \mathcal{S}_{0}}{\partial  | \textbf{q} | } \frac{\dot{q}_i}{| \dot{\textbf{q}} |}  \ |_{ \,  \textbf{q} =  \textbf{x} } 
= \frac{( \textbf{p}_{(1)}\dot{\textbf{q}} +  \textbf{p}_{(2)} \ddot{\textbf{q}}) \dot{q}_i}{  \dot{\textbf{q}} ^2 }    \ |_{ \,  \textbf{q} =  \textbf{x} } 
= m   \dot{q}_i  
+ \frac{  \hbar ^2}{ m} \Big( \frac{ \dot{ \textbf{q}}   \dot{ \ddot{ \textbf{q}}}}{ \dot{ \textbf{q}}^6} +        \frac{  \ddot{ \textbf{q}} ^2  }{ \dot{ \textbf{q}}^6} -  \frac{ 4 ( \dot{ \textbf{q}} \ddot{ \textbf{q}})^2   }{ \dot{ \textbf{q}}^8} \Big)  \dot{q}_i \ |_{ \,  \textbf{q} =  \textbf{x} }  \ .   \label{may0809-q2}
\end{equation}

\fussy

To determine trajectories, we need not know the $Q(\textbf{q})$ because we use 
not (\ref{june1308-2apr15}) but 
(\ref{may0809-q2})  as trajectory determining eq. after inserting a solution of the QSHJE into LHS (\S \ref{jan2107-6}).

A \label{may2516-1} solution $\dot{\textbf{q}}(t)$ of a 3-D free E-L eq., (\ref{june1308-2apr15}) with $V=Q=0$, is written as a Fourier series $\dot{q}_i = v_i + \sum _{n = - \infty} ^{+ \infty} c_{i, \, n} e^{\sqrt{-1} \ n \omega (t -\delta _i) }$, where $i=1,2,3$, $c_{i , \, -n} = \bar{c}_{i, \, n}$, and $\delta _{ \, i} = const. \in \mathbb{R}$, because at times at which $\ddot{q}_i=0$, $\dot{q}_i $ and $\dot{\ddot{q}}_i > 0$ (or $\dot{\ddot{q}}_i < 0$) take the same values according to six eqs.: $ p_{(1) i } ( \dot{q}_1 , \dot{\ddot{q}} _1 , \cdots , \dot{q}_3 , \dot{\ddot{q}} _3 ) = const. $ and $ E_i ( \dot{q}_1 , \dot{\ddot{q}} _1 , \cdots , \dot{q}_3 , \dot{\ddot{q}} _3 ) = const. $ (cf. p.\pageref{aug2114-1}).

$\bullet$ \textit{Stability of a charged particle} \ \   
Even if a particle is charged, 
as long as the amplitude of oscillation is small, 
electromagnetic radiation does not occur: 
The trajectory of the particle is approximated as 
$ \textbf{q}(t) = \textbf{v} t + \textbf{A}_1  \cos (\omega t +  \bm{\theta}) 
$, where $ |  \textbf{v} | = const. >0$ and $ | \textbf{A}_1 \omega | / |  \textbf{v} | \ll 1$.  
Since the energy $E$ of the particle is given as  $E \simeq   m  \textbf{v}^2 /2 +    m \omega ^2 \textbf{A}_1  ^2  /2 $ from (\ref{may0209-a2apr15}) of which $V=Q=0$, for the radiation to occur,  it has to be  $m \omega ^2 \textbf{A}_1^2 /2 \geq \hbar \omega $ or  $ \omega ^2 \textbf{A}_1 ^2 / 2 \textbf{v} ^2 \geq 1 $ because $\hbar \omega \simeq m \textbf{v} ^2$ (see p.\pageref{jun1619-1}).  It  contradicts $ | \textbf{A}_1 \omega | / |  \textbf{v} | \ll 1$.

\newpage

\section{Appearance of Quantum Phenomena} \label{may1821-1}

We elaborate on reasons why energy and angular momentum of 3-D systems are quantized, and why interference appears. 
For concise reasoning of them, of 1-D energy quantization, of uncertainty, and of tunneling, see \S \ref{may0409-8}.

\subsection
{\label{nov1615-1}Energy and angular momentum quantization of a 3-D system}

We give the reason why energy and angular momentum of a 3-D system having a confining central force potential $V(r)$ are quantized.

\sloppy 

For convenience below, we first solve the 3-D QSHJE (\ref{june1308-1}) of which $V=V(r)$: 
\begin{equation}
 \hbar ^2 \nabla ^2 R /( 2mR)  -  ( \partial _i \mathcal{S}_ 0 )^2 / 2m - V(r) +E =0   \ , \ \ \ \ 
(1/m) \partial( R^2 \partial _i \mathcal{S}_0 ) / \partial x^i   =   0  \ .       \label{nov2515-1}     
\end{equation}
For simplicity, we assume that the $V(r)$ is given as 
$V(r) = a_{-1} / r + a_0 + a_1 r + a_2 r^2 + \cdots + a_n r^n$ satisfying $V(r) -E < 0$ for $0 < r < r_0$ and $0 < V(r) -E $ for $ r_0 < r$, where $ n < + \infty$ and $a_{-1} , \cdots , a_n = const.$

We separate (\ref{nov2515-1}) and a 3-D SE $\hbar ^2 \nabla ^2 \psi = 2m (V(r)-E) \psi$ corresponding to (\ref{nov2515-1}) used to solve (\ref{nov2515-1}) on the spherical polar coordinate system ($x = r \sin \theta \cos \varphi $, $y = r \sin \theta \sin \varphi $, $ z = r \cos \theta $). 
The (\ref{nov2515-1}) is separated, if $\mathcal{S}_{0 \varphi}( \varphi ), \mathcal{S}_{0 \theta }( \theta ) , \mathcal{S}_{0 r }(r) \neq const.$, with $\mathcal{S}_0 ( \varphi , \theta ,r ) = \mathcal{S}_{0 \varphi}( \varphi ) + \mathcal{S}_{0 \theta }( \theta ) +\mathcal{S}_{0 r }(r)$ and $R( \varphi , \theta , r ) = R_ \varphi ( \varphi ) R_ \theta ( \theta ) R_r (r)$ as 
\begin{subequations}
\begin{align}
&\frac{1}{2m} ( \frac{ \partial \mathcal{S} _{ 0 \varphi }}{\partial \varphi } )^2 + \frac{ \hbar ^2 }{4m} \{ \mathcal{S} _{ 0 \varphi } , \varphi \} - \frac{ \hbar ^2 \textsf{m}^2 }{2m} = 0      \label{feb1115-1}         \\ 
&\frac{1}{2m} ( \frac{ \partial \mathcal{S} _{ 0 \theta }}{\partial \theta  } )^2 + \frac{ \hbar ^2 }{4m} \{ \mathcal{S} _{ 0 \theta  } , \theta  \} 
+ \frac{ \hbar ^2 }{2m}( 
 -\frac{ \cos ^2 \theta }{4 \sin ^2 \theta } + \frac{  \textsf{m}^2}{  \sin ^2 \theta } - \frac{1}{2} - l(l+1) ) =0   \label{feb1115-2}   \\ 
&\frac{1}{2m} ( \frac{ \partial \mathcal{S} _{ 0 r}}{\partial r } )^2 + \frac{ \hbar ^2 }{4m} \{ \mathcal{S} _{ 0 r  } , r  \} +V(r) - E + \frac{ \hbar ^2 l(l+1) }{2m r^2}  =0 \ ,   \label{feb1115-3}  
\end{align} \label{feb0415-1}
\end{subequations}
which follow from (\ref{nov2515-1}) written on the coordinate system as 
\begin{subequations}
\begin{align}
&0= \frac{\hbar ^2}{2m} \frac{\nabla ^2 R}{R}- \frac{1}{ 2m} ( \partial _i \mathcal{S}_ 0 )^2  - V(r) +E 
= \frac{1}{2m} ( \frac{ \partial \mathcal{S}_{0r}}{ \partial r})^2 
- \frac{\hbar ^2 }{2m} \frac{1}{r^2R_r} \frac{ \partial }{\partial r} ( r^2  \frac{ \partial R_r}{ \partial r})      + V(r) -E  \notag \\  
&+ \frac{1}{2mr^2} \Big( (  \frac{ \partial \mathcal{S}_{0\theta}}{ \partial \theta })^2 - \frac{ \hbar ^2}{ \sin \theta }\frac{1}{ R_\theta} \frac{ \partial}{ \partial \theta } ( \sin \theta \frac{ \partial R_\theta }{\partial \theta }) + \frac{1}{ \sin ^2 \theta } \big(( \frac{ \partial \mathcal{S}_{0 \varphi }}{\partial \varphi })^2 - \frac{\hbar ^2}{ R_\varphi} \frac{ \partial ^2 R_\varphi}{ \partial \varphi ^2} \big) \Big)  \ ,   \label{mar2315-1}  \\ 
&0= \frac{ \partial }{ \partial x^i} ( \frac{R^2}{m}  \partial _i \mathcal{S}_0  ) 
= \frac{2}{R_\varphi}\frac{\partial R_\varphi}{\partial \varphi } \frac{ \partial \mathcal{S}_{0\varphi }}{\partial \varphi } + \frac{\partial ^2\mathcal{S}_{0\varphi }}{ \partial \varphi ^2} + \sin ^2\theta           \notag \\ 
& \times \Big( \frac{2}{R_ \theta }  \frac{ \partial R_\theta }{\partial \theta } \frac{\partial \mathcal{S}_{0\theta }}{\partial \theta } + \frac{1}{\sin \theta } \frac{ \partial }{\partial \theta }( \sin \theta \frac{\partial \mathcal{S}_{0\theta }}{\partial \theta })  +    
 \frac{2r^2}{ R_r} \frac{ \partial R_r}{\partial r} \frac{\partial \mathcal{S}_{0r}}{\partial r} + \frac{\partial }{ \partial r} ( r^2 \frac{\partial \mathcal{S}_{0r}}{\partial r} )  \Big)  \ .     \label{mar2315-2}
\end{align} \label{nov2415-2}
\end{subequations} 
The $R_r$, $R_\theta$, and $R_\varphi$ of (\ref{mar2315-1}) are written with $\mathcal{S}_{0r }$, $\mathcal{S}_{0\theta }$, and $\mathcal{S}_{0\varphi }$ as 
$\partial _r ^2  ( rR_r) /  (rR_r) = -\{ \mathcal{S} _{ 0 r} , r \} /2 $, 
$\partial _\theta  ^2  ( \sqrt{ \sin \theta  } R_\theta ) / ( \sqrt{ \sin \theta  } R_\theta ) = -  \{ \mathcal{S} _{ 0 \theta  } , \theta  \} /2$, and 
$ \partial _\varphi  ^2   R_\varphi / R_\varphi = - \{ \mathcal{S} _{ 0 \varphi  } , \varphi  \} /2$ 
by a theorem (cf. footnote \ref{jan1009-ss}) `$ \, \partial _x ( R^2 \partial _x  \mathcal{S}_0 ) =0 \Leftrightarrow       
\partial _x^2R/R = -\{ \mathcal{S}_0 , x \} /2 $ if $\partial _x \mathcal{S}_0  \neq 0 \, $' through eqs. 
$\partial _r (r^2 R_r ^2 \partial _r \mathcal{S}_{0r}  )  =0$, $\partial _ \theta ( \sin \theta R_\theta  ^2 \partial _ \theta \mathcal{S}_{0\theta}   )   =0$, and 
$\partial _ \varphi (R_\varphi ^2 \partial _ \varphi \mathcal{S}_{0 \varphi }   ) =0$ separated from (\ref{mar2315-2}).  
Separation consts. $\hbar ^2 l(l+1)  \in \mathbb{R}$ and $ \hbar ^2 \textsf{m}^2 \in \mathbb{R}$ in (\ref{feb0415-1}) are equal to  
$(   (   \partial \mathcal{S}_{0\theta}/ \partial \theta )^2 \cdots  ( \cdots ) /  \sin ^2 \theta  )$ and $ ( \partial \mathcal{S}_{0 \varphi }/ \partial \varphi )^2 - ( \hbar ^2 / R_\varphi ) ( \partial ^2 R_\varphi / \partial \varphi ^2)$ in (\ref{mar2315-1}).  
 The consts. $\hbar ^2 l(l+1)$ and $ \hbar \textsf{m}$ are identified as square and $z$-component of angular momentum of the system because they are so in the classical limit. 
\footnote{
In the H-J formalism of CM, angular momentum $\textbf{I}$ is defined as $\textbf{I} = \textbf{r} \times \nabla \mathcal{S}_0$. 
 It is written as \\ 
$\textbf{I} =  \textbf{r} \times 
( \hat{\bm{\varphi}} (1 /  r \sin \theta ) \partial \mathcal{S}_0 / \partial \varphi  + \hat{\bm{\theta}}(1/ r)  \partial \mathcal{S}_0 / \partial \theta  +  \hat{\bm{r}}  \partial \mathcal{S}_0 / \partial r ) 
 = -  \hat{\bm{\varphi}}  \partial \mathcal{S}_0 / \partial \theta   +  \hat{\bm{\theta}} ( 1 / \sin \theta ) (  \partial \mathcal{S}_0 / \partial \varphi ) $  on the spherical polar coordinate system. 
Writing 
$\textbf{I} = I_ \varphi \hat{\bm{\varphi}} + I_ \theta  \hat{\bm{\theta }} + I_r  \hat{\bm{r}} = I_x \hat{\bm{x}} + I_y \hat{\bm{y}} + I_z \hat{\bm{z}} $, 
we have $I_\theta = I_z / \sin \theta $. 
Accordingly, we see $\textbf{I} \cdot \textbf{I} = ( \partial \mathcal{S}_{0 \theta } / \partial \theta )^2 + ( \partial \mathcal{S}_0 / \partial \varphi )^2 / \sin ^2 \theta $ and $I_z = \partial \mathcal{S}_0 / \partial \varphi$. 
} 
 The corresponding 3-D SE is separated, with $\psi  = \Phi (\varphi ) \Theta ( \theta ) \textsf{R}(r)$ as 
\begin{subequations}
\begin{align}
&\frac{ d^2 \Phi }{ d \varphi ^2 } + \textsf{m}^2 \Phi  =0  \label{feb1115-4} \\  
&\frac{d ^2}{d \theta ^2} ( \sqrt{ \sin \theta } \  \Theta  ) + ( \frac{1}{2} + l(l+1) + \frac{\cos ^2 \theta }{4  \sin ^2 \theta } - \frac{ \textsf{m}^2}{ \sin ^2 \theta } )  \sqrt{ \sin \theta } \ \Theta  =0    \label{feb1115-5}    \\  
&\frac{d^2 (r \textsf{R})} {dr^2} + ( \frac{2m}{ \hbar ^2} ( E - V(r) ) - \frac{ l(l+1) }{r^2} ) r \textsf{R} =0 \ ,  \label{feb1115-6} 
\end{align} \label{jun1310-2} 
\end{subequations}
where (\ref{feb1115-5}) is equivalent to 
$\frac{1}{\sin \theta }\frac{d}{d \theta } ( \sin \theta \frac{ d \Theta }{ d \theta } ) + ( l(l+1) - \frac{ \textsf{m}^2}{ \sin ^2 \theta } ) \Theta  =0 $ found in textbooks \cite{Rae(2008)}$^,$ \cite{Griffiths(2005)} of QM. 
Solutions  $ \mathcal{S} _{ 0 \varphi } , \cdots , \mathcal{S} _{ 0 r } \neq const.$ of (\ref{feb0415-1}) are constructed from those $\Phi , \cdots , r \textsf{R}$ of (\ref{jun1310-2}) by (\ref{may0409-s1}). 
From solutions of (\ref{feb0415-1}), a solution $\nabla \mathcal{S}_0$ of (\ref{nov2515-1}) is constructed as 
 $\nabla \mathcal{S}_0 = \hat{\bm{\varphi}} (1 /  r \sin \theta ) \partial \mathcal{S}_{0 \varphi } / \partial \varphi  + \hat{\bm{\theta}}(1/ r)  \partial \mathcal{S}_{0 \theta } / \partial \theta  +  \hat{\bm{r}}  \partial \mathcal{S}_{0 r } / \partial r $, where  $\hat{\bm{\varphi}}$ ($\hat{\bm{\theta}} $, $\hat{\bm{r}}$) is a unit vector in $\varphi$-  ($\theta$-, $r$-) direction.

\fussy


For a system having an atomic scale confining central force potential $V(r)$, the Lag. formalism is ineffective because trajectories do not satisfy the semiclassicality condition: $|\textbf{A}_1 \omega |  / \textbf{v} | \ll 1 $ (see \S \ref{jun1609-1}). 
We however consider that there exists a formalism, of which approximation is the Lag. formalism, effective for the system (see p.\pageref{dec1017-1}). 
We write quantities in the formalism as $\bullet ^ {NW}$. 
In the formalism, the eq. of motion is 
a fourth-order ordinary differential eq.,  
and $ \nabla \mathcal{S}_0 ^{NW} ( \textbf{x}) =  \nabla \mathcal{S}_0 ^{NW} ( \dot{\textbf{q}} , \ddot{\textbf{q}} ,  \dot{\ddot{\textbf{q}}})|_{ \textbf{q} = \textbf{x}}$ constructed from a solution of the eq. of motion is equal to a solution $ \nabla \mathcal{S}_0 ^{HJ}( \textbf{x}) $ of the QSHJE (see \S \ref{may0409-8}).

\begin{wrapfigure}{r}{34 ex}
\includegraphics[width = 34 ex, clip]{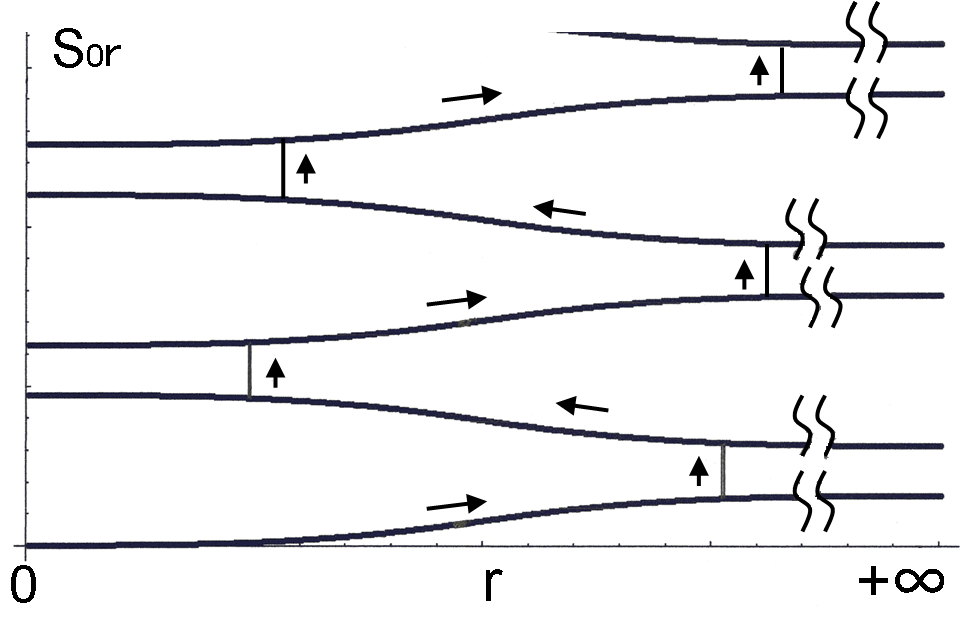}
\caption{A section of $\mathcal{S}_0 ^{ \ HJ\pm}$}
\label{nov2115-2}

\vspace*{- 0.5 \intextsep}
\end{wrapfigure}



We illustrate the equality $ \nabla \mathcal{S}_0 ^{NW}( \textbf{x})  = \nabla \mathcal{S}_0 ^{HJ} ( \textbf{x})$ in $V(r)$. 
Solutions $ \mathcal{S} _0 ^{HJ}( \varphi , \theta , r ) $ of (\ref{nov2515-1}) forms surfaces in $V(r)$. 
In $r$-direction, a component $ \mathcal{S} _{ 0 r  }^{HJ} (r)$ of the solution either monotonically increases or decreases, which is seen from the appearance of the Schwarzian derivative $\{ \mathcal{S}_{0 r } , r \}$ in (\ref{feb1115-3}). 
A section of $ \mathcal{S}_0 ^{HJ} ( \varphi , \theta , r )$ at $\varphi , \theta = const.$ therefore looks like Fig. \nolinebreak \ref{nov2115-2}. 
Spacing between adjacent  $ \mathcal{S} _{ 0 r  }^{HJ+ }$ ($ \mathcal{S} _{ 0 r  }^{HJ- }$) is equal to $| \mathcal{S} _{ 0 \varphi } ( 2 \pi ) - \mathcal{S} _{ 0 \varphi } ( 0) |$, where $ \mathcal{S} _{ 0 r  }^{HJ+ }$ ($ \mathcal{S} _{ 0 r  }^{HJ- }$) is the monotonically increasing (decreasing) one. 
The particle moves around on the surfaces keeping $ \nabla \mathcal{S}_0 ^{NW}( \textbf{x})  = \nabla \mathcal{S}_0 ^{HJ} ( \textbf{x})$. 
Two notes: i) At a point at which the particle turns from $\dot{r} > 0$ ($\dot{r} < 0$) to $\dot{r} < 0$ ($\dot{r} > 0$), the particle aquires an amount of action $ \mathcal{S}_{0 r} $ (see Fig. \ref{nov2115-2}). 
  The amount however is irrelevant to $\partial \mathcal{S}_{0r} ^{NW} / \partial r$. and 
ii) The equality $ \nabla \mathcal{S}_0 ^{NW}( \textbf{x})  = \nabla \mathcal{S}_0 ^{HJ} ( \textbf{x})$ holds true only along trajectory, which resolves a seeming contradiction: 
At the point at which $\dot{r} $ changes sign, $ \partial \mathcal{S}_{0 r } ^{NW} / \partial r =0 $ because $\dot{r} =0$ (see (\ref{may0809-q2})), whereas $\partial \mathcal{S}_{0r} ^{HJ} / \partial r \neq 0$ there. 
It is not a contradiction because the trajectory is perpendicular to $r$-direction at the point. 
The $\theta$-direction equality is similarly illustrated. 
The $\varphi$-direction equality $ \partial  \mathcal{S}_{0 \varphi } ^{NW} / \partial  \varphi  =  \partial  \mathcal{S}_{0 \varphi } ^{HL} / \partial  \varphi$ is obvious since $\dot{ \varphi }$ does not change sign. 
Thus the equality $ \nabla \mathcal{S}_0 ^{NW}( \textbf{x})  = \nabla \mathcal{S}_0 ^{HJ} ( \textbf{x})$ holds true even in $V(r)$.

\sloppy

The value of $ \nabla \mathcal{S}_0 ^{NW}$ averaged along a solution $\textbf{q}(t)$ of the eq. of motion over a macroscopic length $(b-a)$: 
\begin{equation*}
 \frac{1}{ b-a} \int _a^b  \nabla \mathcal{S}_0 ^{NW} d \textbf{q}  = \frac{1}{ b-a}\big( \int  _a^b \frac{\partial  \mathcal{S}_{0 \varphi } ^{NW} }{ \partial  \varphi }  d \varphi +  \int  _a^b \frac{ \partial  \mathcal{S}_{0 \theta } ^{NW} }{ \partial  \theta } d \theta  + \int  _a^b \frac{ \partial  \mathcal{S}_{0 \, r } ^{NW}}{ \partial r } d r \big)  \ , 
\end{equation*}
where  $d \textbf{q} = \hat{\bm{\varphi}} \, r \sin \theta  d\varphi + \hat{\bm{\theta}} \, r d \theta + \hat{\bm{r}} dr$, has to be independent from initial conditions ($\ddot{\textbf{q}}(t_0) , \dot{\ddot{{\textbf{q}}}}(t_0)$) introduced by the extended diff. action because the averaged value is so in CM effective in macroscopic length. 
The independence from the ($\ddot{\textbf{q}}(t_0) , \dot{\ddot{{\textbf{q}}}}(t_0)$) is equivalent to that from parameters $( A , \cdots , D )$ appearing in  (\ref{may0109-1}) at each of the integrals in RHS of 
\begin{equation*}
 \frac{1}{ b-a} \int _a^b  \nabla \mathcal{S}_0 ^{NW} d \textbf{q}  =   \frac{1}{ b-a}  \big( k _ \varphi \int  _0 ^ { 2 \pi } \frac{\partial  \mathcal{S}_{0 \varphi } ^{HJ} }{ \partial  \varphi }  d \varphi +  k_ \theta \int  _0^{ \pi} \frac{ \partial  \mathcal{S}_{0 \theta } ^{HJ} }{ \partial  \theta } d \theta  + k_{ \, r} \int  _0^ { \infty } \frac{ \partial  \mathcal{S}_{0 \, r } ^{HJ}}{ \partial r } d r \big)   
\end{equation*} 
because $ \nabla \mathcal{S}_0 ^{NW}( \textbf{x}) = \nabla \mathcal{S}_0 ^{HJ} ( \textbf{x})$, where $k _ \varphi , k_ \theta , k_{ \, r} = const. \in \mathbb{R}$. 

\fussy

We determine conditions to realize the independence from the $( A , \cdots , D )$.

In $\varphi$-direction, since the domain is homeomorphic to a circle,  a solution $\partial \mathcal{S}_{0 \varphi }^{HJ} /\partial \varphi$ of (\ref{feb1115-1}) is required to be single-valued. 
Therefore we first determine a condition for which a solution $\partial \mathcal{S}_{0 \varphi }^{HJ}/\partial \varphi$ of (\ref{feb1115-1}) be single-valued. 
 Assume $0 \neq \textsf{m}  \in \mathbb{R}$. 
Then, $\partial \mathcal{S}_{0 \varphi }^{HJ}/\partial \varphi $ constructed with (\ref{may0409-s1}) from two linearly independent solutions  $\Phi ^C = \cos \textsf{m} \varphi $ and $\Phi ^D = \sin \textsf{m} \varphi $ of (\ref{feb1115-4}) as 
\begin{equation*}
\frac{\partial \mathcal{S}_{0 \varphi }^{HJ}}{\partial \varphi }
= \frac{ - \hbar \textsf{m} l_1}{ ( \cdots )^2 + l_1 ^2 \cos  \textsf{m} \varphi  }   
= \frac{ - 2 \hbar \textsf{m} l_1}{ (l_1^2+l_2^2 +1)+ (l_1^2+l_2^2 -1) \cos 2 \textsf{m} \varphi + 2 l_2  \sin 2 \textsf{m} \varphi }  
\end{equation*}
is single-valued if $ \textsf{m} = \pm 1 , \pm 2, \cdots $. 
Assume $\textsf{m}  =0$. 
Then, $\partial \mathcal{S}_{0 \varphi }^{HJ} /\partial \varphi $ constructed with (\ref{may0409-s1}) from $\Phi ^C = \textbf{1} $ and $\Phi ^D = \varphi $ as  $ \partial \mathcal{S}_{0 \varphi }^{HJ}/\partial \varphi  =  -  \hbar l_1 \textbf{1}  / ( (\varphi + l_2 \textbf{1} )^2 + (l_1 \textbf{1}  ) ^2 )$, where  $\textbf{1}$ is a constant function, is not single valued. 
  For $ \textsf{m} = 0$, however, $\partial \mathcal{S}_{0 \varphi }^{HJ} / \partial \varphi  \equiv 0$, which is single-valued, is allowed because $\nabla \mathcal{S}_0 \neq 0$ is satisfied---a solution of the 3-D QSHJE has to satisfy $\nabla \mathcal{S}_0 \neq 0$ like a solution of the 1-D QSHJE has to satisfy $\partial \mathcal{S}_0 / \partial x \neq 0$---according to $\partial \mathcal{S}_{0 \theta }^{HJ} / \partial \theta \neq 0$ and $\partial \mathcal{S}_{0 \, r }^{HJ} / \partial r \neq 0$. 
For $ \textsf{m} = \pm 1 , \pm 2, \cdots $, $ \int  _0 ^ {2 \pi } ( \partial  \mathcal{S}_{0 \varphi } ^{HJ} / \partial  \varphi ) d \varphi $ is independent from  the parameters because $ \int  _ {\varphi _0}^{\varphi _0 + \pi / \textsf{m}  } ( \partial  \mathcal{S}_{0 \varphi } ^{HJ} / \partial  \varphi )  d \varphi $ is; see p.\pageref{may0315-2}.  
For $ \textsf{m} = 0$, $ \int  _0 ^ {2 \pi } ( \partial  \mathcal{S}_{0 \varphi } ^{HJ} / \partial  \varphi ) d \varphi $ is trivially independent from  the parameters. 
Thus, allowed values of $\textsf{m}$ are $\textsf{m} = 0, \pm 1 ,\pm 2, \cdots$.

In $\theta$-direction, $\int _0^{ \pi} \partial \mathcal{S}_{0 \theta } ^{HJ} / \partial \theta d \theta = \mathcal{S}_{0 \theta }^{HJ}( \pi ) - \mathcal{S}_{0 \theta }^{HJ}( 0) + const.$ of a solution of (\ref{feb1115-2}) is independent from the parameters for discrete values of $l$ in (\ref{feb1115-2}), which we see as follows. 
A solution $\mathcal{S}_{0 \theta}^{HJ} ( \theta )$ of (\ref{feb1115-2}) is constructed from two linearly independent  solutions  $\sqrt{ \sin \theta } \  \Theta ^C  $ and $\sqrt{ \sin \theta } \  \Theta ^D$ of (\ref{feb1115-5}) as 
$e^{ \frac{ 2 i}{ \hbar }  \mathcal{S}_{0 \theta}^{HJ} ( \theta ) }
=  (A  \Theta ^D (\theta ) /\Theta ^C (\theta )   + B ) /( C \Theta ^D (\theta )/\Theta ^C (\theta ) + D ) $---see (\ref{apr0915-3}).  
From the last eq., we have, through 
\begin{equation*}
e^{ \frac{ 2 i}{ \hbar } ( \mathcal{S}_{0 \theta}^{HJ} ( \pi ) - \mathcal{S}_{0 \theta} ^{HJ} ( 0 ) )}
= \frac{ e^{ \frac{ 2 i}{ \hbar }  \mathcal{S}_{0 \theta}^{HJ} ( \pi ) }}{ e^{ \frac{ 2 i}{ \hbar }  \mathcal{S}_{0 \theta} ^{HJ} ( 0 ) }} =  \frac{(A  \Theta ^D (\pi ) /\Theta ^C (\pi )   + B )}{( C  \Theta ^D (\pi )/\Theta ^C (\pi ) + D ) } \frac{ (C \Theta ^D (0) /\Theta ^C (0) + D )}{(A  \Theta ^D (0) /\Theta ^C (0)   + B )} \ , 
\end{equation*}
\begin{equation*}
\mathcal{S}_{0 \theta}^{HJ} ( \pi ) - \mathcal{S}_{0 \theta} ^{HJ} ( 0 ) + const.  
= \frac{\hbar}{2i} \ln \frac{(A  \Theta ^D (\pi ) /\Theta ^C (\pi )   + B )}{( C \Theta ^D (\pi )/\Theta ^C (\pi ) + D ) } \frac{ (C  \Theta ^D (0) /\Theta ^C (0) + D )}{(A  \Theta ^D (0) /\Theta ^C (0)   + B )} + const.  \ , 
\end{equation*} 
of which RHS is independent of the parameters iff, allowing $\lim _{\theta \to 0 , \, \pi} \Theta ^D (\theta )/\Theta ^C (\theta )  = \pm \infty$, 
\footnote{
If $\lim _{\theta \to 0} \Theta ^D (\theta )/\Theta ^C (\theta ) = \pm \infty$,   (\ref{may1515-1}) is satisfied for $\lim _{\theta \to  \pi} \Theta ^D (\theta )/\Theta ^C (\theta ) = \mp \infty$, where double sign corresponds.  
If  $\lim _{\theta \to 0 } \Theta ^D (\theta )/\Theta ^C (\theta ) = \lim _{\theta \to  \pi} \Theta ^D (\theta )/\Theta ^C (\theta )  \in \mathbb{R}$,  $\Theta ^D /\Theta ^C  $ blows up to $\pm \infty$ somewhere in $\mathbb{R}$. 
These phenomena follow from monotonicity of $\Theta ^D /\Theta ^C $ seen from (\ref{apr0915-3}) and (\ref{may0409-s1}).
}
\begin{equation} \textstyle 
\lim _{\theta \to \pi}    \Theta ^D (\theta )/\Theta ^C (\theta ) 
= \lim _{\theta \to 0}    \Theta ^D (\theta )/\Theta ^C (\theta ) \ . \label{may1515-1}
\end{equation}
The eq. (\ref{may1515-1}) is satisfied only for $l \in \mathbb{Z}$ satisfying $|l| \geq | \textsf{m} |  $; see \ref{apr1615-1}.

In $r$-direction, likewise, $\int  _0^ { \infty }  \partial  \mathcal{S}_{0 \, r } ^{HJ} / \partial r d r = \mathcal{S}_{0 r}^{HJ}( + \infty ) - \mathcal{S}_{0 r}^{HJ}( 0) + const.$ of a solution of (\ref{feb1115-3}) is independent from the parameters iff, allowing   
$\lim _{ r \to 0, \, +\infty} \textsf{R} ^D (r)/ \textsf{R} ^C (r) = \pm \infty$, 
\begin{equation} \textstyle 
  \lim _{ r \to +\infty} \textsf{R} ^D (r)/  \textsf{R} ^C (r) = \lim _ { r \to 0 }       \textsf{R} ^D(r) /  \textsf{R}^C(r) \ . \label{may1515-5}  
\end{equation}  
It is satisfied only for discrete values of $E$; see \ref{apr1615-1}.

Thus, physically allowed values of $\textsf{m}$, $l$, and $E$ are discrete. 
The values for $\textsf{m}$ is, as we saw above, the same as those of QM. 
The values for $l$ and $E$ are also the same as those of QM because the conditions (\ref{may1515-1}) and (\ref{may1515-5}) are satisfied iff the corresponding SE has a $L^2$-solution (see \ref{apr1615-1}).


\subsection{Interference} \label{jan2107-6}

We illustrate interference in GCM with interference of electrons by an electron biprism.


\begin{figure}[b]
\vspace*{- \intextsep}
\centering
\includegraphics[width = 80ex ,height = 26 ex, clip]{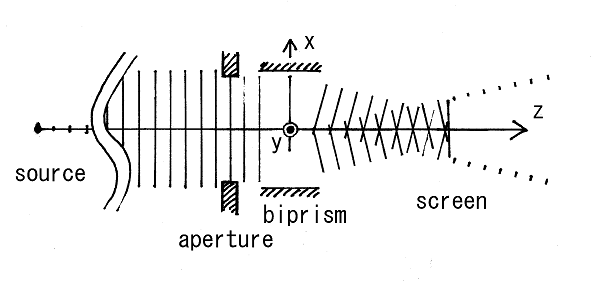}
\vspace*{- \intextsep}
\caption{Experimental setting: Electrons emitted from a source at the left side end run toward the screen. Those passed through the upper (lower) slot  change moving direction downward (upward) because of the voltage applied to the central filament.}
\label{fig-interference}
\end{figure}

$\bullet$ \textit{Experimental setting} \label{mar0612-1} \ \ We study with the apparatus of Fig.\ref{fig-interference}, where filament diameter is $4.15 \times 10^{-4}$ mm and distance from the filament to the observation screen is 33.77mm. 
With these values, we simulate the apparatus of Ref. \refcite{Komrska(1967)}.



\fussy 
$\bullet$ \textit{System on the left side} \ \ \label{may3009-23} 
An ensemble of electrons is involved in the experiment. 
The behavior of each electron is described by the H-J eq. (\ref{june1308-1}) and the eq. of motion (\ref{may0809-q2})---we call (\ref{may0809-q2}) so. 
The solutions of (\ref{june1308-1}) and (\ref{may0809-q2}) in the left side region of the biprism are given as follows.

A solution $\nabla \mathcal{S}_0 (\textbf{x})$ of (\ref{june1308-1}) in the left side region is constructed from a solution of the corresponding SE with (\ref{may0609-1}). 
In the present section, we call a solution of the SE a beam. 
An incoming beam, a solution of the SE in the left side region, is a solution of a free SE. 
Plane wave, \label{nov2917-1} Gaussian beam, and Bessel beam \cite{Durnin(1987)} are candidates. 
\footnote{
The incoming beam is not a wave packet, \cite{Messiah(1961)} that is amplitude in $z$-direction of the beam is uniform; 
the wave packet is unnecessary because in GCM particle distribution is not given by $| \Psi |^2$, and contradictory because the wave packet does not give a definite value to the energy of a system, whereas GCM gives a definite value to that. 
} 
We exclude the last one because it is not generated by usual electron sources such as a thermionic cathode. \cite{Grillo(2014)} 
We exclude also the plane wave for a reason below. 
Behind the biprism, we expect $x$-direction momentum $p_x^{EL} \simeq m \dot{q}_x$ is nonzero because of a voltage applied to the filament. 
Then $p_x^{HJ}$, too, is nonzero since $p_x^{EL} \simeq  p_x^{HJ} $. 
For  $p_x^{HJ}$ to be nonzero, $\psi$ in (\ref{may0609-1}) has to have a bell shaped transverse profile. 
Accordingly the incoming beam is Gaussian beam $\Psi _G$ \cite{Svelto(1998)} of which transverse profile is given by a Gaussian function  $e^{-(r -r_0)^2/w^2}$, where $ r -r_0 =\surd ( ( x -x_0 )^2 + (y - y_0)^2 )$ and $w$ is a slowly varying function of $z$. 
\footnote{\label{feb1921-7} 
The Gaussian beam $\Psi _G$ is a solution of the paraxial form of the Helmholz eq. $( \nabla ^2 + k^2 ) \Psi =0$. 
It is written as
\begin{equation}
 \Psi_G = \frac{w_0}{w(z)} \exp ( -\frac{(x- x_0)^2 +(y - y_0)^2}{w^2(z)} ) \exp ik ( z   + \frac{(x- x_0)^2 +(y - y_0)^2}{2 \mathcal{R}(z) } + \frac{ \hat{\phi} (z) }{k} )  \ , 
 \label{may0809-e5}
\end{equation}
where $w_0$ is the beam width at the beam waist position, $w$ is the beam width at $z$, 
$(x_0 , y_0) $ is the beam center coordinates, 
$k= \surd (k_x^2 + k_y ^2 +k_z ^2 )$, 
$\mathcal{R}$ is the radius of curvature of the wavefront, and $\hat{\phi}$ is the Gouy phase. The $z$-dependence of $w$, $\mathcal{R}$, and $\hat{\phi}$ is written using the Rayleigh range $z_R := k w_0^2 /2$ as 
$w^2 (z) = w_0 ^2(1+( z/z_R )^2 )$, $\mathcal{R} (z) = z (1+( z_R/z )^2 )$, and $\hat{\phi} (z) = \tan ^{-1} ( - z/z_R)$. 
Note that a free SE $( \hbar ^2 \nabla ^2 / 2m + E ) \Psi =0$ is written as $( \nabla ^2 + k^2 ) \Psi =0$ since $\hbar k = \surd ( 2mE )$. 
}

\begin{figure}[b]
\begin{tabular}{c}
\begin{minipage}{0.5\hsize}
\includegraphics[width=40 ex ,  height = 24 ex ]{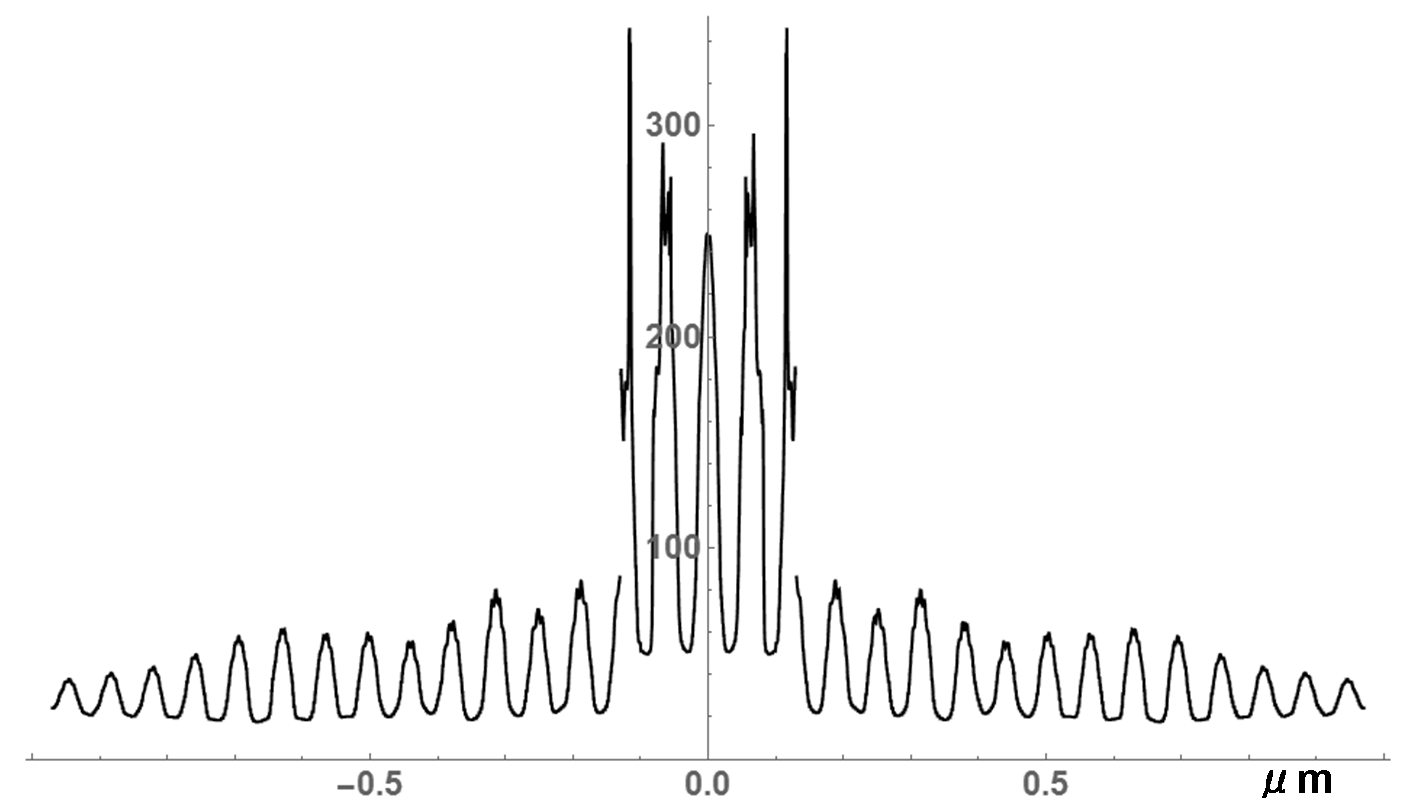} 
\caption{Density distribution on the screen of electrons accompanied by Gaussian beams of which $w = 2.2 \mu$m. Spikes are artifacts.} 
\label{jan1418-1}
\end{minipage}
 \begin{minipage}{0.5\hsize}
\includegraphics[width=40 ex ,  height = 24 ex ]{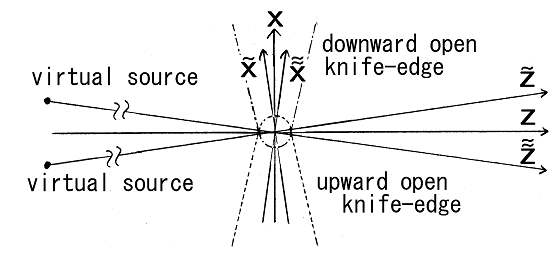} 
\caption{The coordinate systems} 
\label{fig-coordinates} 
\end{minipage}
\end{tabular}
\end{figure}

The Gaussian beam given above however does not reproduce observed interference patterns on the screen. 
Indeed when beam width represented by the $w$ is large enough to form a reasonable interference contrast on the screen, electrons located near the center of the $\Psi _G$ and passed near the filament accumulate around the center ($x=0$) of the screen because almost equal values of amplitude of lower and upper slot passed part, $\tilde{U}$ and $\tilde{\tilde{U}}$ respectively, of the $\Psi _G $ leads to $p_x^{HJ} ( \simeq p_x^{EL} \simeq m \dot{x})  \approx 0 $ (see p.\pageref{jan1518-1}) at electron position; see Fig.\ref{jan1418-1}. 
\footnote{
Calculation procedure to derive the dens. distr. of Fig.\ref{jan1418-1} is, except for the transverse profile of the beam, essentially the same as that given on p.\pageref{jan1518-2} used to derive the dens. distr. of Fig.\ref{fig-ifcmpr}. 
} 
It contradicts experiments. \cite{Komrska(1967)} \cite{Yamamoto(2000)} 
We therefore consider that the transverse profile of the incoming Gaussian beam is close to $e^{-|r -r_0|/w}$. 
\footnote{
The transverse profile differs from that considered in electron optics, in which wave front of an incoming beam is considered to be spherical, almost plane at the biprism, with equal amplitude on the wave front. \cite{Laty(2017)} 
Accordingly, in electron optics, fringe visibility \cite{Born(1999)} of field intensity on the screen  is perfect (100\%) for  a point source, while in GCM, even for a point source, the visibility is not perfect. 
} 
Such a profile is constructed from a sum of Gaussian functions having various widths. 
We assume, for calculational convenience, that the incoming beam is a two-component Gaussian beam $\Psi _s$ described near $z=0$, at which $\mathcal{R}$ (footnote \ref{feb1921-7}) is adjusted to $\mathcal{R} = \infty $ with lenses on the source side, as
\begin{equation}
\Psi _s = \big( 0.8 \exp ( - \frac{(x-x_0)^2 + ( y - y_0)^2 }{ 0.34 ^{ \ 2}} )  + 0.2 \exp ( - \frac{(x-x_0)^2 + ( y - y_0)^2 }{ 2.2 ^{ \ 2}} )  \big)  
  \times  \exp ( ik_z  z ) \ , \label{jul0808-1}
\end{equation}
where the unit of length in $(x,y)$-plane is  $\mu m$. 
We assume that wave number $\textbf{k} $ of incoming beams is $\textbf{k} = ( k_x, k_y , k_z) = ( 0,0,  1.45 \times 10^9 mm ^{-1} )$, which changes to $( \pm 4.99 \times 10^4 mm ^{-1} , 0 , 1.45 \times 10^9 mm ^{-1})$ behind the biprism because of a positive voltage applied to the filament.

A solution $\dot{\textbf{q}}(t)$ of the eq. of motion on the left side region is $\dot{\textbf{q}}(t) =  ( 0,0, \hbar k_z /m )$. 
We assume in the present section that the electron is located at the center of the beam in the region because we consider it natural for the electron to be located at the center of the beam when transverse profile of the beam is close to the $e^{-|r -r_0|/w}$. 
However, for comparison, we examine in \ref{sep0811-1} the case that density distribution of electron in the $\Psi _s $ is given by $ | \Psi _s |^2 $.

Electrons characterized by the above solutions uniformly illuminate the aperture.

$\bullet$ \textit{Momentum field behind the biprism} \label{nov0717-1}  \ \ 
We henceforth refer to $\nabla \mathcal{S}_0 $, $\Psi _G $ and so on as a field in the sense  that it has a value at every point of domain. 
The momentum field $\textbf{p}^{HJ} ( \textbf{x} ) = \nabla \mathcal{S}_0^{HJ} ( \textbf{x} )$  behind the biprism caused by (\ref{jul0808-1}) is constructed  from a solution $\Psi _{\textrm{diff}}$ of the corresponding SE using (\ref{may0609-1}). 
\footnote{
Although we find the formula which gives the   Gaussian beam field diffracted by the knife-edge in  Ref. \refcite{Pearson(1969)}, it is too elaborate for the present purpose. We therefore determine the diffracted field in a simpler setting.
} 
The $\Psi _{\textrm{diff}}$ is a superposition of two beams: $\tilde{U}$ and $\tilde{ \tilde{U}}$, one which  originates from a virtual source located at the lower left and was diffracted by the knife-edge  open  downward  at $z=0$   (no voltage applied), the other is obvious from symmetry; see Fig.\ref{fig-coordinates} and Ref. \refcite{Komrska(1967)}. 
We determine the  $\tilde{U}$ and $\tilde{ \tilde{U}}$ applying Fresnel approximation to the Fresnel-Kirchhoff diffraction formula \cite{Born(1999)} (see \ref{nov1608-4}). 
  Inserting $\Psi _{\textrm{diff}} = \tilde{U}  + \tilde{ \tilde{U}}$ into $\psi$ of (\ref{may0609-1}), we have the momentum field $ \nabla \mathcal{S}_0 $ behind the biprism. 
The $ \nabla \mathcal{S}_0 $ is given explicitly in (\ref{aug1411-1}).

$\bullet$ \textit{Trajectory determining eq.} \ \ 
We regard (\ref{may0809-q2}) as a differential eq. for $\textbf{q}(t) $ after inserting  $ \nabla \mathcal{S}_0 $ of (\ref{aug1411-1}) into the LHS, though we later modify it to (\ref{may1309-31}). 
Note we cannot use the E-L eq. (\ref{june1308-2apr15}) because we do not know the exact $Q$ in it.

We \label{jun1619-2} simplify (\ref{may0809-q2}). 
In $y$-direction, $p_y^{HJ} = 0 = m \dot{y} $ since we consider $\Psi _{\textrm{diff}}$ on $y = y_0 =0$. 
In $z$-direction, $p_z ^{HJ} \simeq \hbar k_z \simeq m \dot{z} $ because $k_x \ll k_z$. 
In $x$-direction, $p_x ^{HJ} \simeq m \dot{x}$, which we see as follows. 
We assume that $p_x ^{HJ} = p_x ^{EL} = m \dot{q}_x  = const.$ before entering the interference (i.f.) region. 
In the i.f. region, the $ \dot{q}_x $ is forced to oscillate with angular frequency  $\omega _x := 2 k_x v_x $ because \label{jan1518-1} 
 $p_x^{EL} = p_x^{HJ}  \simeq  \hbar k_x (| \tilde{U} |^2 -| \tilde{\tilde{U}} |^2) / (| \tilde{U} |^2 +|  \tilde{\tilde{U}}  |^2 - 2 | \tilde{U} || \tilde{\tilde{U}}  | \cos 2 k_x x)$; see (\ref{jul0508-1}) and (\ref{aug1411-1}). 
Writing the $ \dot{q} _x$ as  $ \dot{q} _x \simeq  v_x -  a_0 \omega _x \sin ( \omega _x t + \theta )$, where $v_x$ and $a_0$ are slowly changing functions, we see $\hbar ^2 ( \cdots )/m $-terms of (\ref{may0809-q2}) is ignorable because $| \hbar ^2 \dot{ \ddot{q}} _x / m \dot{ \textbf{q}}^4 | \simeq m a _0 \omega _x \cdot  \hbar ^2  \omega _x ^2  /  m ^2 \dot{ \textbf{q}}^4 \simeq  m a _0 \omega _x \cdot E_x ^2 / E_z^2  \ll m a _0 \omega _x $ and so on. 
Thus,  (\ref{may0809-q2}) is simplified to
\begin{equation}
\dot{x}  = \frac{1}{m} \frac{  \partial \mathcal{S}_0 }{ \partial x} =  \frac{1 }{m} \  \frac{ \hbar }{ 2 i } 
\frac{  \bar{ \Psi } _{\textrm{diff}} \partial _x   \Psi _{\textrm{diff}} - \Psi _{\textrm{diff}}  \partial _x  \bar{ \Psi } _{\textrm{diff}} }{\Psi _{\textrm{diff}} \bar{\Psi} _{\textrm{diff}} } \ ,  \ \ \ \ \dot{y} =0 \ , \ \ \ \ \dot{z} = \frac{ \hbar k _z }{m}  \ . \label{may1009-f3}
\end{equation}

If we use (\ref{may1009-f3}), the electron may  reverse the moving direction from upward (downward) to downward (upward) macroscopically at some point in the i.f. region. 
Indeed, if the beam center passes the upper (lower) slot while an electron at the skirt of the beam passes the lower (upper) slot, then,  
 in the approximate form of (\ref{may1009-f3}):  $ \dot{x} = m^{-1} \hbar k_x (| \tilde{U} |^2 -| \tilde{\tilde{U}} |^2) / (| \tilde{U} |^2 +|  \tilde{\tilde{U}}  |^2 - 2 | \tilde{U} || \tilde{\tilde{U}}  | \cos 2 k_x x)$, 
the $| \tilde{U} |^2 -| \tilde{\tilde{U}} |^2$ may change sign at the particle position in the i.f. region. 
\footnote{
The moving direction reversal occurs in Bohm's interpretation in which  (\ref{may1009-f3}) is used; see Ref. \refcite{Holland(1993)} \S 5.1.      
} 
It means that a particle comes in the i.f. region with $p_x^{EL} >0$ ($p_x^{EL} <0$) and goes out with $p_x^{EL} <0$ ($p_x^{EL} >0$), which is prohibited by the law of conservation of momentum since no potential $V $ to which $p_x^{EL}$ is transfered exists in the i.f. region---the $Q$ in the Lagrangian does not store momentum because it is not real potential (see p.\pageref{apr1912-1}).

The GCM has a mechanism to prevent the moving direction reversal. 
In 1-D space,  $p = \partial \mathcal{S}_0 / \partial x $  is determined, giving two linearly independent solutions 
 $\Psi  , \Psi ^D     \in \mathbb{R}$ of the corresponding SE \textit{and}  $c_1, c_2$ in $  c_1 \Psi + c_2 \Psi ^D  = R e ^{ i \mathcal{S}_0 / \hbar }$; see sentences below (\ref{may0609-1}). If we replace $c_1$ and $c_2$ in the eq. with their complex conjugates, the $p$ is reversed. 
Likewise, if we replace $c_1$ and $c_2$  in $\Psi _{\textrm{diff}x} = c_1 \Psi + c_2 \Psi ^D$ of $\Psi _{\textrm{diff}} = \Psi _{\textrm{diff}x} (x) \Psi _{\textrm{diff}y}(y) \Psi _{\textrm{diff}z} (z)$ as above, the $p$ is reversed. 
We thus avoid the moving direction reversal. 
We note:  
A solution $\partial \mathcal{S}_0 / \partial x $ of the 1-D QSHJE is monotonic over a domain of the eq. 
Accordingly, if $\partial \mathcal{S}_0 / \partial x =0 $ is required at some point of 1-D space, a single solution cannot cover both intervals devided by the point, which means that values of $c_1$ and $c_2$ may differ in each interval.

We thus modify (\ref{may1009-f3}) to 
\begin{equation}
\dot{x}  = \frac{1}{m} \frac{  \partial \mathcal{S}_0 }{ \partial x} = \pm \frac{1 }{m} \ |  \frac{ \hbar }{ 2 i } 
\frac{  \bar{ \Psi } _{\textrm{diff}} \partial _x   \Psi _{\textrm{diff}} - \Psi _{\textrm{diff}}  \partial _x  \bar{ \Psi } _{\textrm{diff}} }{\Psi _{\textrm{diff}} \bar{\Psi} _{\textrm{diff}} }| \ ,  \ \ \ \ \dot{y} =0 \ , \ \ \ \ \dot{z} = \frac{ \hbar k _z }{m}  \ , \label{may1309-31}
\end{equation}
where the `$+ (-)$' sign is for the particle which passed the lower (upper) slot.

If the electron runs at the center of the incoming beam, (\ref{may1009-f3}) and (\ref{may1309-31}) gives the same trajectory since moving direction reversal does not occur. 

\label{oct1311-1}
If the   electron dens. distr. in an ensemble of incoming beams is assumed as $|\Psi _s |^2$ and the velocity is given with (\ref{may1009-f3}), the density of electrons on the screen is the same as that of QM  because, if $ m \dot{\textbf{q}} =  \nabla \mathcal{S}_0 $, according to the second eq. of (\ref{june1308-1}), $R^2 = | \Psi_{\textrm{diff}}  |^2$ represents electron density.

$\bullet$ \textit{Calculation of trajectory, electron distribution on the screen, and field intensity on the screen} \ \ 
We calculate trajectories  for two cases of electron distr. in the incoming beam $ \Psi _s$      (\ref{jul0808-1}): i) an electron is located at the center of the $ \Psi _s$, and ii) dens. distr. of electrons in an ensemble of $ \Psi _s$'s is given as $| \Psi _s|^2$ like QM.
 In this subsection, we examine the case i); we examine the case ii) in \ref{sep0811-1}. As was mentioned, we consider it  natural for the incoming beam of which transverse profile is close to $e^{-|r-r_0 |}$ to accompany an electron at its center.

\begin{wrapfigure}{r}{38ex}
\vspace*{- 0.5 \intextsep}
\centering
\includegraphics[width = 36ex , height = 20ex , clip]{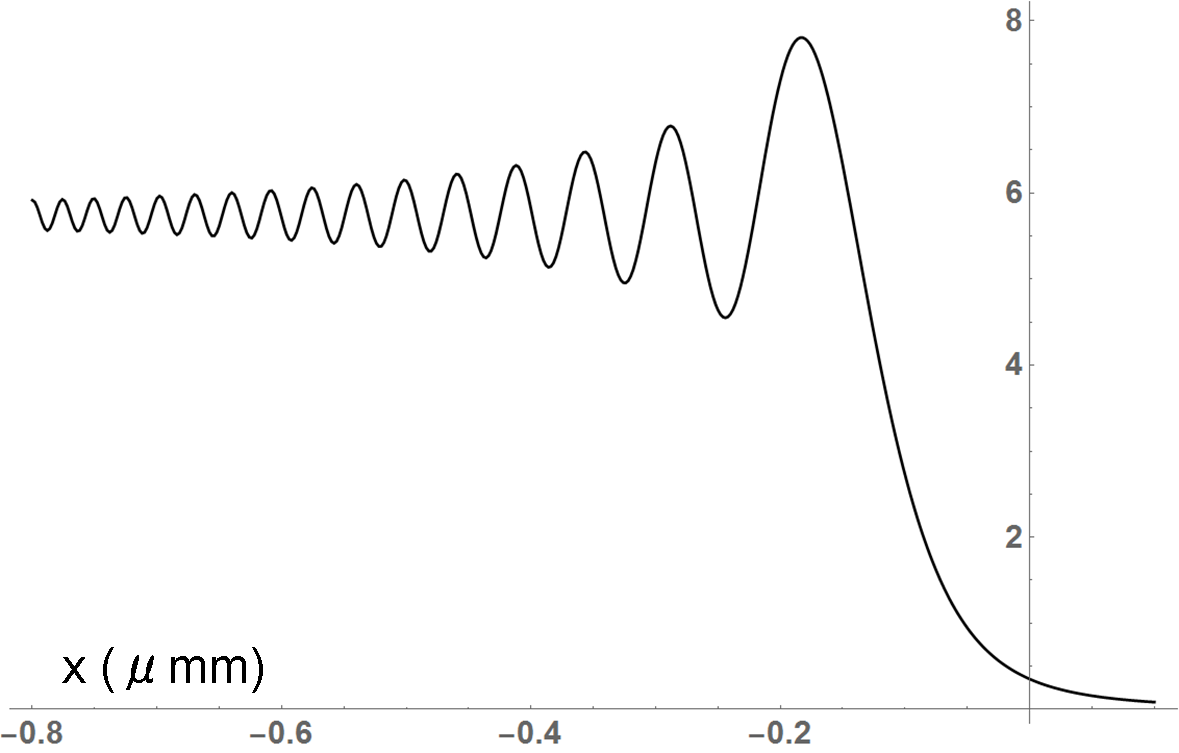}
\caption{Assumed electron density at $z=4mm$} 
\label{fig-z4mmdns1}
\vspace*{- 0.5 \intextsep}
\end{wrapfigure}

For \label{jan1518-2} the trajectory calculation, we have to assume some initial electron dens. distr. in the Fresnel region behind the biprism because the Fresnel approximation fails near the biprism. 
We assume the dens. distr. of the lower-slot-passed electron at $z = 4mm$  is 
equal to the field intensity $I_{ \textrm{ini}} (x)$  caused by the uniform flux of the $\Psi _s$  chopped by the downward open knife-edge, which is given as $I_{ \textrm{ini}} ( x ) = \int _{- \infty}^{+\infty} | \tilde{U}|^2 (x; x_0 ) dx_0 $; 
\footnote{All numerical calculations in the present paper were made with Mathematica on a personal computer.} 
\footnote{
To obtain $I_{ \textrm{ini}} ( x ) $, we summed up $(  | 0.8 \textsf{M}_1 + 0.2 \textsf{M}_{11}|^2 + | 0.8 \textsf{N}_1 + 0.2 \textsf{N}_{11}|^2 ) (x; x_0)$ 
over $x_0 = -6.4 $, $-6.2 , \cdots ,  4.2$, $4.4\mu m$, where $\textsf{M}_1$ is (\ref{nov2517-1}) of which $w_0 = 0.34 \mu m$ and   $\textsf{M}_{11}$ is that of which $w_0 = 2.2 \mu m$; likewise for $\textsf{N}_1$ and $ \textsf{N}_{11}$. 
} 
see  Fig.\ref{fig-z4mmdns1}.

We assume the electron at  $(x_{ini} (< 0) , y_{ini} , z_{ini} (= 4mm) )$ interacts with the field which originates the $\Psi _s$ of which beam center is $(x_0 =  ( x_{ini}- 4000 \, k_x / k_z )  \mu m , \ y_0 = y_{ini})$.

\sloppy 

We calculate trajectories of the lower-slot-passed electrons of which $x_{ini} = -2.000 , -1.998 , \cdots , -0.108 , -0.106 \mu m$, $y_{ini} = 0 \mu m$, and $z_{ini} = 4mm$ with (\ref{may1309-31}).

\fussy 



The electron density $D_{\textrm{lower}} ( x)$
 on the screen caused by the lower-slot-passed electrons 
is constructed as follows. Let initial $x$-coordinates of $n$ sufficiently many electrons be $x_{1 \cdot ini}  < \cdots < x_{i \cdot ini} < \cdots < x_{n \cdot ini}$. Corresponding hitting $x$-coordinates be $x_{1 \cdot hit}  < \cdots < x_{i \cdot hit} < \cdots < x_{n \cdot hit}$. Then  $D_{\textrm{lower}}$ at $x_{i \cdot hit}$ is given as 
\begin{equation}
D_{\textrm{lower}}( x_{i \cdot hit} ) \propto I_{ \textrm{ini}} ( x_{i \cdot ini}) \frac{( x_{(i+1) \cdot ini} -  x_{(i-1) \cdot ini}) }{ ( x_{(i+1) \cdot hit} -  x_{(i-1) \cdot hit})}  \ \ \ \ \ ( 2 \leq i \leq n-1 ) \ .  \label{jul-0108-1}
\end{equation}
The  $D_{\textrm{upper}} (x) $ by  upper-slot-passed electrons  is given as  $D_{\textrm{upper}} (x) = D_{\textrm{lower}} (- x)  $. 
The electron density $D (x)$ on the screen is given as $D (x) = D_{\textrm{lower}} ( x) + D_{\textrm{upper}} (x) $.

Field intensity $I(x) $ on the screen is obtained as $I(x) = \int_{-\infty}^\infty  I(x;x_0) dx_0$, where $I(x;x_0)$ is the field intensity $| \Psi _{\textrm{diff}} |^2$ caused by the two-component Gaussian beam $\Psi _s$ of which center  $(x,y)$ is $ (x_0 ,0) $. 
\footnote{\label{sep1411-1}
We obtain $I(x;x_0)$ from (\ref{may2608-e4}) replacing $\textsf{M}_1 , \cdots ,  \textsf{N}_2$ with $  0.8 \textsf{M}_1 + 0.2\textsf{M}_{11} , \cdots ,  0.8 \textsf{N}_2 + 0.2 \textsf{N}_{21} $, where what $\textsf{M}_{11}, \cdots , \textsf{N}_{21}$ represent are obvious.  
In actual calculation, we summed up $I(x;x_0)$ over $x_0 = -6.4 $, $-6.3 , \cdots , 6.3 $, $6.4 \mu m$. 
}

\begin{figure}
\centering
\includegraphics[width = 74ex ,clip]{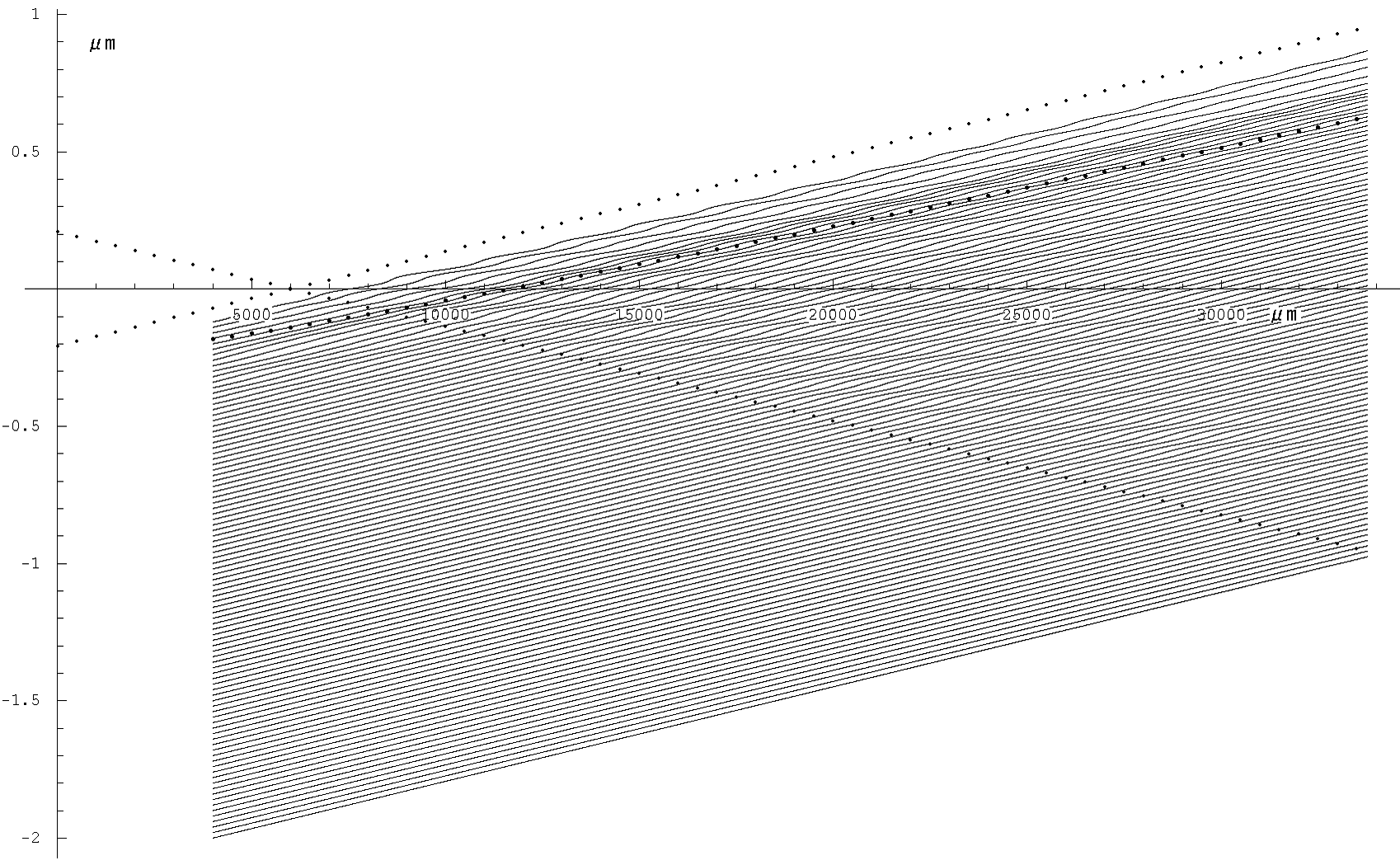}
\caption{Trajectories of  electrons which passed the lower slot.  
The electron is assumed to be located at the center of the incoming beam. 
 Crossed dotted lines border on the geometrical shadows of the filament. The dotted curve is the path along which the top of the largest peak of Fig.\ref{fig-z4mmdns1} runs.}
\label{fig-trajdiff}
\end{figure}

$\bullet$ \textit{Results} \ \ 
Calculated trajectories  are shown in Fig.\ref{fig-trajdiff}. 
Calculated field intensity and dens. distr. on the screen  are shown in Fig.\ref{nov2317-1} and \ref{fig-ifcmpr} respectively.

In QM, the field intensity and electron density agree each other.  In GCM, they  disagree as is seen from Fig.\ref{nov2317-1} and \ref{fig-ifcmpr}. 
In particular, the largest Fresnel peaks of the  dens. distr.  of Fig.\ref{fig-ifcmpr} shift  outwards from those of the field intensity  of Fig.\ref{nov2317-1} at arrowed positions. 
The electron at the center of the incoming beam runs almost linearly parallel to the geometrical border, while the largest Fresnel peaks run along $\tilde{x} = const.  \sqrt{ \tilde{z}}$ and $\tilde{\tilde{x}} = const.  \sqrt{ \tilde{\tilde{z}}}$ (Ref. \refcite{Born(1999)} \S 8.7.3), which leads to the disagreement (Fig.\ref{fig-trajdiff}). 
Even if electron density in the incoming beam is assumed to be $| \Psi _s |^2$,   they disagree because the eq. of motion is given not by (\ref{may1009-f3}) but by (\ref{may1309-31}); 
see p.\pageref{oct1311-1} and \ref{sep0811-1}.

Envelope shape of dens. distr. on the screen GCM predicts and that QM predicts disagree: 
In QM, the dens. distr. is equal to the field intensity. 
The  field intensity QM gives results from an ensemble of incoming beams of which transverse profile is not clearly known. 
Yet, unlike fringe visibility, the envelope shape of the field intensity is almost identical for all practical beams 
\footnote{
The beam is either plane wave or Gaussian beam (see p.\pageref{nov2917-1}). 
The momentum $\hbar \textbf{k}$ and  beam width (in the case of the Gaussian beam) may slightly vary from beam to beam. 
} 
 regardless of the details of the transverse profile.  
Indeed, the ones resulting from plane wave and $\Psi _s$ are almost identical as is seen with vertical scaling in Fig.\ref{nov2317-1}. 
We therefore regard the envelope shape of field intensity in Fig.\ref{nov2317-1} as that of dens. distr. QM predicts. 
As we saw, it disagrees with the envelope shape of dens. distr. GCM predicts in Fig.\ref{fig-ifcmpr}.

On graphs in articles \cite{Komrska(1967)} \cite{Yamamoto(2000)} comparing  electron density observed in experiments and field intensity calculated with QM, we notice Fresnel peak shifts like Fig.\ref{fig-ifcmpr}, which makes us feel that GCM better reproduces observed electron density.

\section{Concluding Remarks}

Using extended diffeomorphism group action, 
we generalized classical analytical mechanics (CM) of a particle in a time-independent potential field to that effective in both classical and quantum mechanical scales. 
Particle distribution in an ensemble the generalized CM (GCM) gives disagrees with that quantum mechanics gives. 
The disagreement is detectable with present-day technology. 
The GCM thus is testable with present-day technology. 
If GCM be verified by experiments, we would have an option to unify quantum and classical physics within conceptual framework of classical one.

\section*{Acknowledgments}

The author  would like to express  special gratitude to

A. E. Faraggi and M. Matone: If I had not happened to find their article, the present work did not even  started. I appreciate their seminal work.

Toshio Ishigaki, Emeritus Professor of Hokkaido University: For guiding me through the world of interpretations of Quantum Mechanics.

Kenzo Ishikawa, Emeritus Professor of Hokkaido University: For providing me a research environment after T. Ishigaki's retirement. For close readings of the manuscripts and discussions at various stages.

\clearpage

\begin{figure}
\centering
\includegraphics[width = 76ex , height = 25ex ,clip]{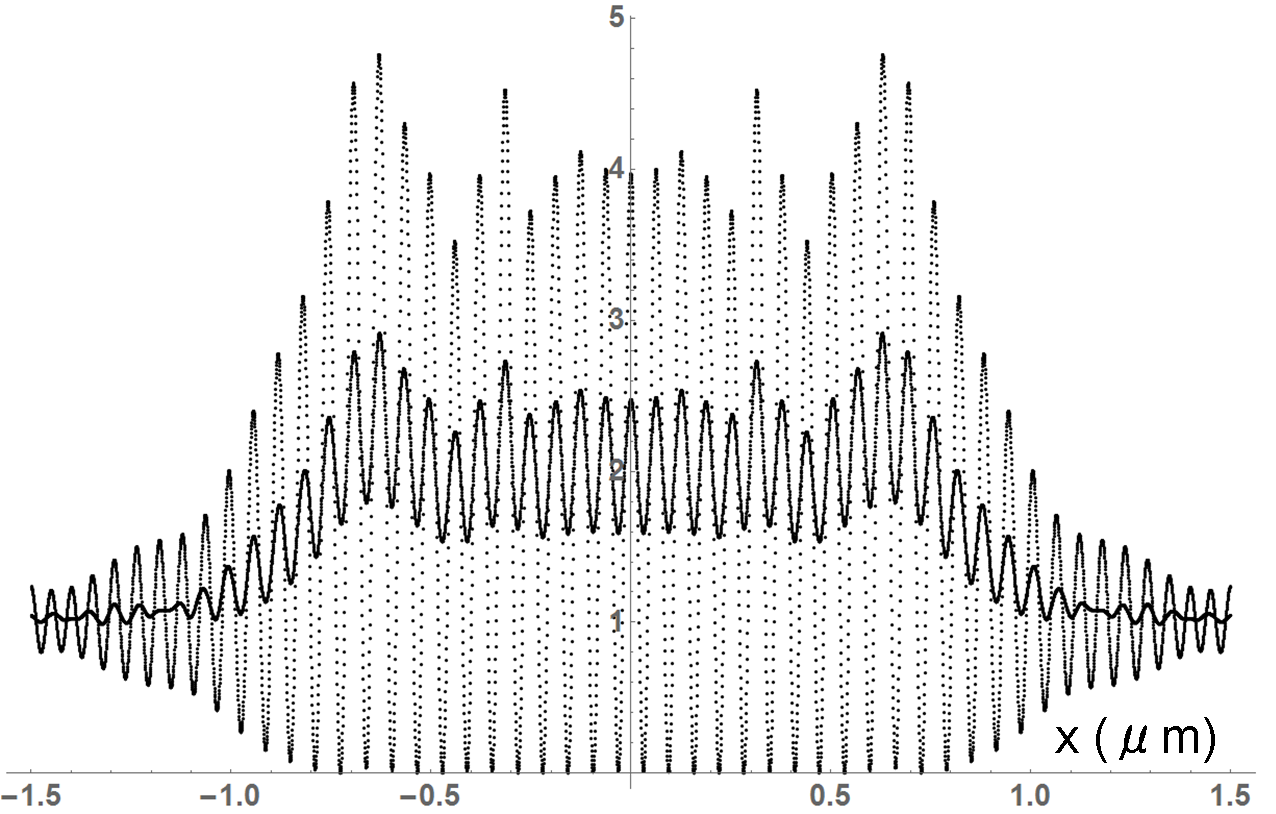}
\caption{
Field intensity on the screen in the cases that the incoming beam is the Gaussian beam $\Psi _s$ (line) and a plane wave (dotted line).
}
\label{nov2317-1}
\end{figure}

\begin{figure}
\centering
\includegraphics[width = 76ex , height = 21ex ,clip]{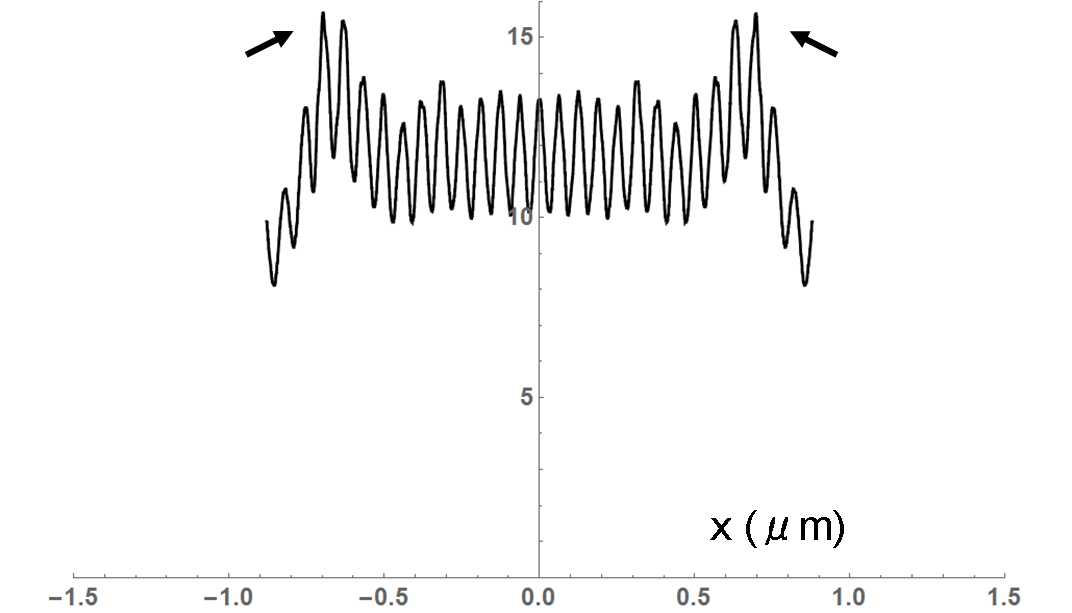}
\caption{Electron density on the screen in the case that the  electron is located at the center of the incoming beam. 
Peaks shift outwards at arrowed positions from those of Fig.\ref{nov2317-1}. Marks on vertical axis are arbitrary.}
\label{fig-ifcmpr}
\end{figure}

\clearpage

\appendix

\fussy

\section{Derivation of the Relativistic H-J Equation of GCM}

We derive the relativistic H-J eq. (\ref{aug0608-3}) of GCM. 
Materials of \ref{jun1418-1} except that of final paragraph are found in Refs. \refcite{Rotman(1995)}, \refcite{Dummit(2004)}, \refcite{Brown(1982)}, \refcite{Ovsienko(2005)}, and \refcite{Neeb(2004)}.

\subsection{Extension of groups} \label{jun1418-1}

\sloppy

We first determine a group operation rule in an extension $E$ of a group $G$ by an abelian group $A$. 
The extension is a short exact sequence 
$0 \to A \stackrel{i}{\to} E \stackrel{\pi}{\to} G \to 1$, where $i$ and $\pi$ are group homomorphisms. 
We use symbols $a , a _1 , a_2 , b \in A$, $e , e_1 , e_2 \in E$ and $g , g_1 , g_2 , g_3 \in G$ below. 
With them, we write the homomorphisms as $i(a _1 + a _2 ) = i( a _1 ) + i( a _2 )$ and $\pi ( e_1 e_2 ) = \pi ( e_1 ) \pi ( e_2 )$. 
Since the sequence is exact, $\operatorname{Im} i = \operatorname{Ker} \pi$, where $\operatorname{Im}$ and $\operatorname{Ker}$ signifies image and kernel respectively. 
Since  $i(A)$ is a normal subgroup ($e i( a ) e^{-1} \in i(A)$), $E$ acts on $i(A)$ by conjugation $a \mapsto e i( a ) e^{-1} \in i(A)$. 
Since conjugation action of $i(A)$ on itself is trivial ($i(b)i(a)i^{-1}(b) = i(a)$), the conjugation action induces an action of $E/A \simeq G$ on $i(A) \simeq A$. 
Let $s(g) \in E$ be a section, a function $G \to E$ satisfying $\pi (s(g)) =g$. 
Then, an action $T_{g}$ of $G$ on $A$ is defined as $i(T_{g} a) = s (g) i (a) s  ^{-1} (g)$, which is written also as $i(T_{g} a) s(g) = s (g) i (a)$. 
Since $s(g_2g_1 )$ and $s(g_2)s(g_1)$ belong to the same coset of $E$, we have $s(g_2)s(g_1) = i( \frak{f} ( g_2,g_1 )) s(g_2g_1)$, where $\frak{f} ( g_2,g_1 )$ is a function from $G \times G \to A$. 
Group operation in $E$ thus is given as 
\begin{equation}
i(a_2) s(g_2) \cdot i(a_1) s(g_1) = i(a_2) i(T_{g_2} a_1) s(g_2) s(g_1)  
= i ( a_2 + T_{g_2} a_1 + \frak{f} ( g_2,g_1 ) )s(g_2 g_1) \ . \label{may1621-1} 
\end{equation}  
Since group operation is associative, an equation 
$\bm{(} i(a_3) s(g_3) i(a_2)s (g_2) \bm{)} i(a_1)s(g_1) = i(a_3) s(g_3) \bm{(} i(a_2)s (g_2) i(a_1)s(g_1) \bm{)}$ holds true, from which we have 
$\frak{f} (g_3,g_2) + \frak{f}(g_3g_2 , g_1) = T_{g_3} \frak{f}(g_2 ,g_1 ) + \frak{f} ( g_3,g_2 g_1 ) \ \ \cdots (*1)$.

An element $e = i(a) s(g) \in E$ corresponds one-to-one to $ (\tilde{a} , \tilde{g} ) \in A \times G$. 
The correspondence is given as $i(a) \mapsto (a , 1)$, $s(g) \mapsto (dg , g )$, and  $i(a) s(g) \mapsto (a + dg , g)$. 
The operation rule on $A \times G$ is given from (\ref{may1621-1}) as 
\begin{equation*}
( a_2 + dg_2 , g_2 ) \cdot ( a_1 + dg_1 , g_1 ) = ( a_2 + dg_2  + T_{g_2} ( a_1 + dg_1 ) + \frak{f} ( g_2,g_1 )  , g_2 g_1 ) \ . 
\end{equation*} 
If $s(g_2)s(g_1) = s(g_2g_1)$, that is if $\frak{f} (g _2, g_1) = 0$, we have $dg_2 + T_{g_2} dg_1 = d(g_2g_1) \ \ \cdots (*2)$ because $s(g_2)s(g_1) = (dg_2 ,g_2 )( dg_1 , g_1 ) = ( dg_2 + T_{g_2} dg_1 , g_2 g_1 )$ and $s(g_2g_1) = ( d( g_2g_1 ) , g_2g_1 )$.

In group cohomology theory, we rediscover the above eqs., $(*1)$ and $(*2)$, as follows. 
Let $q$-cochain $C^q (G , A )$ be a set of functions $c( g_q , \cdots , g_1 )$'s from  $G \times \cdots \times G$ to $A$. 
We define $C^0 (G,A)$ as $C^0 (G,A) =A$. 
Differentials $d_q : C^q \to C^{q+1}$ ($q = 0,1,2$) are given as 
$d_0 c(g_1)= T_{g_1} a _0 - a _0 $, 
$d_1 c (g _2, g_1 ) = T_{g_2} c(g_1 ) - c(g_2g_1) + c(g_2)$, and 
$d_2 c (g_3 , g_2 , g _1 ) = T_{g_3} c(g_2 ,g_1 ) -c(g_3g_2,g_1 ) + c(g_3,g_2g_1)  -c(g_3,g_2)$, where $a_0$ is an arbitrary fixed element of $A$. 
Let $Z^1 = \operatorname{Ker} d_1$  ($Z^2 = \operatorname{Ker} d_2$). 
We call elements of $Z^1$ ($Z^2$) 1-cocycles (2-cocycles). 
Let $B^1 = \operatorname{Im}  d_0$  ($B^2 = \operatorname{Im}  d_1$). 
We call elements of  $B^1$ ($B^2$) 1-coboundaries (2-coboundaries). 
We call $H^1 = Z^1 / B^1$ ($H^2 = Z^2 / B^2$) the first (second) cohomology group. 
The $Z^2 $ and $Z^1 $ correspond to $(*1)$ and $(*2)$ respectively.


We give 1-cocycles and 2-cocycles for $G$ being the diffeomorphism group $\textrm{Diff}(\mathbb{RP}^1)$ of 1-D real projective space $\mathbb{RP}^1$ and $A$ being the additive group of covariant tensor fields of order two $\mathcal{F}_2 (\mathbb{RP}^1)$ on $\mathbb{RP}^1$. 
Let $ f^{-1} , g^{-1} \in \textrm{Diff}(\mathbb{RP}^1)$, where $y = f(x)$, $z=g(y)$. 
Let $ \phi _x = \phi (x) dx \otimes dx , \phi _y = \phi (y) dy \otimes dy , \cdots \in  \mathcal{F}_2 (\mathbb{RP}^1)$, where $\phi (x), \phi (y),  \cdots $ are scalars ($\phi  (y) |_p = \phi (z) |_p$). 
The action $T_{g_{-1}}$ of $g_{-1} \in \textrm{Diff}(\mathbb{RP}^1)$ on $\phi _z \in \mathcal{F}_2 (\mathbb{RP}^1)$ is given as $T_{ g ^{-1}} \phi _z = T_{g^{-1}} \phi (z) dz \otimes dz |_p = (\partial g/ \partial y )^2 \phi (z) dy \otimes dy |_p$, where $p$ is a point of $\mathbb{RP}^1$. 
The 2-cocycle satisfying $(*2)$ is an element of $H^2$ or that of $B^2$. 
The element of the former is given as $ \frak{f} ( f^{-1} , g ^{-1} )  = \beta ( f^ * \operatorname{log} (\partial g / \partial y) ) \{ f(x) , x \} dx \otimes dx $, where $\beta = const. \in \mathbb{R}$ and $f^*$ is pullback by $f$, \cite{Neeb(2004)} \cite{Roger(1998)} 
\footnote{
The $ \frak{f} ( f^{-1} , g ^{-1} )  = ( f^ * \operatorname{log} (\partial g / \partial y) ) \{ f(x) , x \} dx \otimes dx$ satisfies 
$\frak{f} ( f^{-1} , g ^{-1} ) + \frak{f} ( f^{-1} g ^{-1} , h^{-1} ) = T_{f^{-1}} \frak{f} ( g^{-1} , h ^{-1} )+ \frak{f} ( f^{-1} , g ^{-1}h^{-1} )$, where $z = h^{-1} (w) $, $y = g^{-1}(z)$, and $x = f^{-1}(y)$ ($x \stackrel{f}{\to} y \stackrel{g}{\to} z \stackrel{h}{\to} w $). 
Using $\{ g(f(x)) , x \} = ( \partial f / \partial x )^2 \{ g(y) , y \} + \{ f(x) , x \}$ (see footnote \ref{dec0316-1}), we have 
\begin{align}
&\frak{f} ( f^{-1} , g ^{-1} ) = ( f^ * \operatorname{log} (\partial g / \partial y) ) \{ f(x) , x \} dx \otimes dx  \ , \ \ \ \ 
\frak{f} ( f^{-1}  g ^{-1} , h ^{-1} )  = ( (gf) ^ * \operatorname{log} (\partial h / \partial z) ) \{ g(f(x)) , x \} dx \otimes dx \ ,  \notag \\ 
&T_{f^{-1}} \frak{f} ( g^{-1} , h ^{-1} )  = T_{f^{-1}} (( g^ * \operatorname{log} (\partial h / \partial z) ) \{ g(y) , y \} dy \otimes dy )   \notag \\ 
&  \ \ \ \ \ \ \ \ \ \ \ \  \ \ \ \ \ \ \ \ \ \ \ \ \ \  =  ( (gf)^ * \operatorname{log} (\partial h / \partial z) )( \{ g(f(x)) , x \} - \{ f(x) , x \}) dx \otimes dx ,  \notag \\ 
&\frak{f} ( f^{-1} , g ^{-1}h ^{-1} )  =  ( f^ * \operatorname{log} (\partial h(g(y)) / \partial y) ) \{ f(x) , x \} dx \otimes dx  
= ( f^ * \operatorname{log} (\partial h / \partial z)(\partial g / \partial y) ) \{ f(x) , x \} dx \otimes dx  \notag \\ 
& \ \ \ \ \ \ \ \ \ \ \ \  \ \ \ \ \ \ \ \ \ \ \ \ \ = ( (gf) ^ * \operatorname{log} (\partial h / \partial z) + f^* \operatorname{log} (\partial g / \partial y) ))  \{ f(x) , x \} dx \otimes dx \ , \notag 
\end{align} 
from which the eq. follows.  
} 
while that of the latter is the $d_1 c ( f^{-1} , g ^{-1} )$. 
The 1-cocycle $df^{-1} $ satisfying $(*2)$ is an element of $H^1$ or that of $B^1$. 
The element of the former is $df^{-1} = \alpha \{ f(x) , x \} dx \otimes dx$, where $\alpha = const. \in \mathbb{R}$ and $\{ f(x) , x \}$ is the Schwarzian derivative defined on p.\pageref{jul1320-1}, while that of the latter is $df^{-1} = (\partial f / \partial x )^2 \phi _0 - \phi _0 $, where $\phi _{ \, 0 }$ is an arbitrary fixed scalar.

We incorporate the extension of $\operatorname{Diff} (\mathbb{RP}^1)$ by $ \mathcal{F}_2 (\mathbb{RP}^1)$ into CM as follows. 
We consider that $g^{-1} \in \operatorname{Diff}(\mathbb{R}^1)$, $\phi _z \in \mathcal{F}_2 (\mathbb{R}^1)$, and diff. action $T_{g^{-1}} \phi _z \in \mathcal{F}_2 (\mathbb{R}^1)$ of CM correspond to $s(g) \in E$, $i( \phi _z) \in E$, and $s(g ^{-1})i( \phi _z ) \in E$ of extension incorporated CM respectively.  
The associative law  $T_{f^{-1}} ( T_{ g^{-1}} \phi _z ) = ( T_{f^{-1}g^{-1}} ) \phi _z$ of CM accordingly corresponds to $s(f^{-1})s(g^{-1})i( \phi _z ) = s(f^{-1}g^{-1} ) i( \phi _z )$,  that is, to $s(f^{-1})s(g^{-1}) = s(f^{-1}g^{-1} ) $, from which $ \frak{f} (f^{-1} , g^{-1} ) = 0$ follows. 
In an operation $s(g^{-1} ) i( \phi _z ) = ( dg^{-1} , g^{-1} ) ( \phi _z , 1 ) = ( dg^{-1} + T_{g^{-1}} \phi _z + T_{g^{-1}} \phi _{ \, 0} - \phi _{ \, 0} , g^{-1})$ corresponding to the diff. action $T_{g^{-1}} \phi _z $ of CM, we write $dg^{-1} + T_{g^{-1}} \phi _z + T_{g^{-1}} \phi _{ \, 0} - \phi _{ \, 0}$ 
as $T _{g^{-1}} ^{\textrm{ext}} \phi _z$ and call it an extended diff. action. 
Explicitly, the $T _{g^{-1}} ^{\textrm{ext}} \phi _z$ is 
\begin{equation}
T _{g^{-1}} ^{\textrm{ext}} \phi (z) dz \otimes dz |_p
= \Big( (\frac{ \partial g}{ \partial y })^2 \phi (z) + \alpha \{ g(y) , y \} + (\frac{ \partial g}{ \partial y })^2 \phi _ { \, 0} - \phi _ { \, 0} \Big) dy \otimes dy |_p \ .   \label{jul0920-1}
\end{equation}

\fussy

\subsection{Derivation of the Relativistic H-J Equation of GCM} \label{may0309-a3}
\sloppy 

We show that, under $T ^{\textrm{ext}}$, the relativistic 4-D H-J eq. (\ref{oct1609-a1}) of CM is modified to the relativistic 4-D H-J eq. (\ref{aug0608-3}) of GCM. 
As was mentioned in \S \ref{jun3009-1}, i) our Minkowski metric is $ \textrm{diag}(1,-1,-1,-1)$, 
ii) we use the Einstein summation convention, and 
iii) we work with the Gaussian system of electromagnetic units. \cite{Landau(1975)} 

\fussy 

First, we determine a modified-by-$T ^{\textrm{ext}}$ form of $x^1$-direction component of 4-D free H-J eq. of CM. 
We set $A^{\mu} =0$ in the 4-D H-J eq. (\ref{oct1609-a1}) to obtain the 4-D free H-J eq. 
In the 4-D free eq., variables separate with $ \mathcal{S} =  \mathcal{S} _{x^0}( x^0 ) + \cdots  + \mathcal{S}_ {x^3} (x^3)$. 
We accordingly have four eqs. in four directions. 
In $x^1$-direction, we have $( \partial \mathcal{S} _{x^1} / \partial x^1 ) ^2 = 2m E_{x^1}$. 
We regard the eq. as an abbreviation of 
\begin{equation*}
 \big( \frac{\partial}{\partial x^1} \otimes  \frac{\partial}{\partial x^1} \big)   \big( \frac{\partial \mathcal{S}}{\partial x^1}  \big) ^2  dx^1 \otimes dx^1 
  = 2 m E_{x^1}    \ , 
\end{equation*}
where $(\partial /\partial x^1) \otimes (\partial /\partial x^1)$ is a component of the inverse metric tensor. \cite{Szekeres(2004)}  
We sometimes write $\mathcal{S}$ for $\mathcal{S} _ {x^1}$ to save space.  
Let $ q^a \to q^b \to q^c $ be  transformations of coordinate systems.
For a coordinate transformation  from $q^c $ to $q^b$, the extended action $T _{  q^{c \ -1} }^{\textrm{ext}}$ of $q^{c \ -1}$ upon $ ( \partial \mathcal{S} / \partial q^c  ) ^2 dq^c \otimes dq^c$ is given as 
\begin{equation*}
T_{  q^{c \ -1} } ^{\textrm{ext}} ( \frac{\partial \mathcal{S}}{\partial q^c}  ) ^2 dq^c \otimes dq^c
= \Big( (\frac{ \partial q^c}{ \partial q^b })^2   (\frac{\partial \mathcal{S} }{\partial q^c})^2   + \alpha \{ q^c,q^b \}  +   (\frac{ \partial q^c}{ \partial q^b })^2   \phi _ { \, 0} - \phi _ { \, 0} \Big) dq^b \otimes dq^b  \ .   
\end{equation*}
Successive transformation  from $q^b $ to $q^a$ takes the above eq. to
\begin{align}
&T_{ q^{b \ -1}} ^{\textrm{ext}} T_{ q^{c \ -1} } ^{\textrm{ext}} ( \frac{\partial \mathcal{S}}{\partial q^c}  ) ^2  dq^c \otimes dq^c  \notag \\ 
&= \Big ((\frac{ \partial q^b}{ \partial q^a })^2  \big( (\frac{\partial \mathcal{S} }{\partial q^b})^2   + \alpha \{ q^c,q^b \}  +   (\frac{ \partial q^c}{ \partial q^b })^2   \phi _ { \, 0} -\phi _ { \, 0}  \big)     
+ \alpha \{ q^b,q^a \}  +   (\frac{ \partial q^b}{ \partial q^a })^2   \phi _ { \, 0} - \phi _ { \, 0}  \Big)    \notag \\ 
 &\times  dq^a \otimes dq^a = \Big( (\frac{\partial \mathcal{S}}{\partial q^a})^2   + \alpha \{ q^c,q^a \} + (\frac{ \partial q^c}{ \partial q^a })^2 \phi _ { \, 0}  - \phi _ { \, 0}  \Big)  dq^a \otimes dq^a         \ , \label{mar2507-1} 
\end{align}
where we used the chain rule of the Schwarzian derivative 
$  \{ q^c,q^a \} = ( \partial q^b / \partial q^a)^2  \{ q^c,q^b  \} +     \{ q^b,q^a \}  $; see footnote \ref{dec0316-1}. 
We thus see that transformed quantities take the same form 
\begin{equation} 
\Big( (\frac{\partial \mathcal{S}}{\partial q^i})^2   + \alpha \{ q^ \star , q^i \}  +  (\frac{ \partial q^ \star }{ \partial q^i })^2 \phi _ { \, 0} - \phi _{ \, 0}  \Big)  dq^i \otimes dq^i  \label{may1321-1}
\end{equation} 
on any coordinate $q^i$, where we renamed $q^c$ to $q^ \star $ to highlight that $\alpha \{ q^\star , q \} + ( \partial q^ \star / \partial q )^2 \phi _ {\, 0} - \phi _ {\, 0} = 0 $ on $q ^c = q^ \star $. 
Contracting (\ref{may1321-1}) on $q^i $ with $(\partial /\partial x^1) \otimes (\partial /\partial x^1) = ( \partial q^i / \partial x^1 )^2  (\partial /\partial q^i ) \otimes (\partial /\partial q^i )$, then equating with  $2 m E_{x^1} $, which is irrelevant to $T^ {\textrm{ext}}$, we have 
\begin{equation*}
 ( \frac{\partial q^i  }{ \partial x ^1 })^2   \Big( (\frac{\partial \mathcal{S}}{\partial q^i})^2   + \alpha \{ q^ \star , q^i \}  +  (\frac{ \partial q^ \star }{ \partial q^i })^2 \phi _ { \, 0} - \phi _{ \, 0}  \Big)  = 2 m E_{x^1}       \ . 
\end{equation*} 
It looks like 
\begin{multline}
 \big( ( \frac{\partial \mathcal{S} ( g ^ \star (q^i (x^1 )) ) }{ \partial x^1 })^2  + \alpha \{ q^ \star (q^i(x^1)) , x^1 \} - \alpha \{ q^i (x^1) , x^1 \} +  (\frac{ \partial q^ \star }{ \partial x^1  })^2 \phi _ { \, 0} -     ( \frac{\partial q^i  }{ \partial x ^1 })^2    \phi _{ \, 0} \big) \\  
= 2 m E_{x^1}  \ .    \label{jan0718-1}
\end{multline} 
on $x^1$, where we used the chain rule of the SD. 
Unless $\phi  _{ \, 0} = const.$ and $q^i (x^1) = Ax^1$ ($A = const. $), the (\ref{jan0718-1}) explicitly depends on coordinate systems of the 1-D space. 
The $\{ q^i (x^1) , x^1 \}$ satisfies $\{ q^i (x^1) , x^1 \} = const.$ with a diffeomorphism $q^i (x^1)$ of $\mathbb{RP}^1$ iff $q^i ( x^1 ) =( A x^1 + B )/(C x^1 + D)$ with $\{ q^i (x^1) , x^1 \} = 0$ (see footnote \ref{may2121-1}). 
The (\ref{jan0718-1}) with $\phi  _{ \, 0} = const.$ and $q^i ( x^1 ) =Ax$ ($A \neq 0$) 
determines $\partial \mathcal{S} / \partial x^1 $ by itself if $q^ \star$ is a function $G$ of $ \mathcal{S} , \cdots , \mathcal{S} ^{(n-3)}$, $q^ \star = G (  \mathcal{S} , \cdots , \mathcal{S} ^{(n-3)})$. 
The (\ref{jan0718-1}) with $\phi  _{ \, 0} = const.$, $q^i ( x^1 ) = Ax^1$, and  $q^ \star = G( \cdots )$ is independent of coordinate systems of the 1-D space. 
In addition, iff $q^ \star =G( \cdots ) = \mathcal{S}$, solutions of (\ref{jan0718-1}) are stable (see p.\pageref{may0121-1}). 
The (\ref{jan0718-1}) depending on the coordinate systems or not having stable solutions is unqualified as a H-J eq. 
Thus the modified-by-$T ^{\textrm{ext}}$ form of $x^1$-direction component of 4-D free H-J eq. of CM is given, noting  $\mathcal{S} ( \mathcal{S} (q^i (x^1 )) ) = \mathcal{S} (q^i (x^1 ))$ and renaming $\mathcal{S} (q^i (x^1 ))$ to $\mathcal{S} (x^1 )$ (see p.\pageref{may0121-1}), as 
\begin{equation}
 \big( ( \frac{\partial \mathcal{S} (x^1 ) }{ \partial x^1 })^2 + \alpha \{ \mathcal{S} (x^1) , x^1 \} + (\frac{ \partial \mathcal{S}  }{ \partial x^1  })^2 \phi _ { \, 0} - A^2 \phi _{ \, 0} \big) = 2 m E_{x^1} \ ,  \label{may1721-1} 
\end{equation} 
where $\phi _{ \, 0} = const.$ 
We set $\phi _{ \, 0} = 0$ because phenomena of QM---energy quantization, tunneling, and so on---are described with $\phi _{ \, 0} = 0$ in GCM (see \S \ref{may0409-8} and \S \ref{may1821-1}). 
We thus have the $x^1$-component of 4-D free H-J eq. of CM modified by $T ^{\textrm{ext}} $ 
\begin{equation}
( \partial \mathcal{S}/ \partial x^1 )^2 + \alpha \{ \mathcal{S} , x^1 \} = 2m E_{x^1}  \ .            \label{apr0519-1} 
\end{equation}

\sloppy

Next, we generalize (\ref{apr0519-1}) to the 4-D H-J eq. (\ref{aug0608-3}). 
We first determine a 4-D eq. effective for electromagnetic potential $A$'s taking a form: $A= (A^0 (x^0),  \, A^1(x^1), \, A^2(x^2), \, A^3(x^3))$ on a Lorentz frame. 
  For such $A$'s, the 4-D eq. to be determined separates to four 1-D eqs. on the frame. 
Each 1-D eq. has a form similar to (\ref{apr0519-1}). Collecting the eqs.,
 we have
\begin{equation}
  \Big( ( \frac{\partial \mathcal{S}}{\partial x^0 } + \frac{e}{c} A^0 ) ^2      + \alpha _0 Q^0  \Big)       -   \Big(  \sum _ {i = 1}^3 (\frac{\partial \mathcal{S}}{\partial x^i } - \frac{e}{c} A^i  ) ^2 + \alpha _1 Q^1 + \alpha _2 Q^2+\alpha _3 Q^3 \Big) = m ^2c^2   \ ,  \label{dec3108-1243}
\end{equation}
where $\alpha _0 , \cdots , \alpha _3 \in \mathbb{R}$ and 
\begin{subequations}
\begin{align}
Q^0 &=  \frac{ \partial _0^2 (\partial _0 \mathcal{S} +  \frac{e}{c} A^0)  }{\partial _0 \mathcal{S} +  \frac{e}{c} A^0 } - \frac{3}{2} \big( \frac{ \partial _0 (\partial _0 \mathcal{S} +  \frac{e}{c} A^0) }{\partial _0 \mathcal{S} +  \frac{e}{c} A^0 } \big)^2               \\ 
 Q^i &=  \frac{ \partial _i^2 (\partial _i \mathcal{S} -  \frac{e}{c} A^i) }{\partial _i \mathcal{S} -  \frac{e}{c} A^i} - \frac{3}{2} \big( \frac{ \partial _i (\partial _i \mathcal{S} -  \frac{e}{c} A^i)}{\partial _i \mathcal{S} -  \frac{e}{c} A^i } \big)^2  \ ,   \label{feb2714-1added}
\end{align}     \label{feb2714-1}
\end{subequations}
where $i= \{1,2,3 \}$.  
We take no sum of $i $ in (\ref{feb2714-1added}). 
The $Q^1$  is written as $Q^1 =    - 2 \partial _1  \partial _ 1 R _1/R_1   =    2 \partial _1  \partial ^1 R _1/R_1 $ 
with a  solution $R_1$ of $\partial _1 (R_1^2  (\partial _1 \mathcal{S} -  e A^1 /c  ) ) =0$. 
\footnote{\label{jan1009-ss}
Let $ \mathcal{S}' >0$. 
We see $(R^2 \mathcal{S}')' =0  \Leftrightarrow R^2  \mathcal{S}' = const. \Leftrightarrow R = const. / \surd \mathcal{S}'$. 
From $R = 1 / \surd \mathcal{S}'$, 
through $R' = - (1/2) \mathcal{S}'' / ( \mathcal{S}' )^{3/2}$, 
we have $ R'' = - (1 / 2) ( \mathcal{S}''' / ( \mathcal{S}' )^{3/2} - ( 3 / 2 ) ( \mathcal{S}'' )^2 / ( \mathcal{S}' )^{5/2} ) $, 
from which $R'' / R = - \{ \mathcal{S} (x) , x \} /2$ follows. 
} 
Likewise for  $Q^2 $ and $Q^3$. 
The $Q^0$ is written as $Q^0 = - 2 \partial _0 \partial _ 0 R _0/R_0 = -2 \partial _0  \partial ^0 R _0/R_0 $.  
We therefore have 
$\alpha _0 Q^0 - \alpha _i Q^i = - 2 \alpha _0  \partial _0  \partial ^0 R/R     - \cdots -  2 \alpha _3 \partial _3  \partial ^3 R/R $, 
where $R= R_0(x^0)R_1(x^1)R_2(x^2)R_3(x^3)$. 
Thus, if (\ref{dec3108-1243}) is true, an eq. system:
\begin{subequations}
\begin{align}
( \partial _{\mu}  \mathcal{S}  + \frac{e}{c} A_{\mu}) ( \partial ^{\mu}  \mathcal{S}  + \frac{e}{c} A^{\mu}) &= m^2  c^2 +    \frac{ 2 \alpha _0 \partial _0 \partial ^0 R}{R }  +    \frac{ 2 \alpha _1  \partial _1  \partial ^1 R}{R }  +  \cdots +    \frac{ 2 \alpha _3  \partial _3  \partial ^3 R}{R }        \label{feb0407aug2905-3d}                   \\
\partial _{\mu} j^{\mu} &= 0 \ , \ \ \ \ \ \ \ \    j^{\mu} = R^2 ( \partial ^{\mu}  \mathcal{S} + \frac{e}{c} A^{\mu} )    \label{feb0407aug2905-4d}     
\end{align}      \label{apr2907-7d} 
\end{subequations}
is true. 
If (\ref{apr2907-7d}) is Lorentz invariant, $\alpha _0 = \alpha _i (= \alpha )$. 
A 4-D eq. effective for $A^{\mu} = A^{\mu} (x^0, x^1,x^2,x^3)$ has the same form as (\ref{apr2907-7d}) because i) the forms of (\ref{apr2907-7d}) are meaningful for inseparable $A ^{\mu} (x^0, \cdots ,x^3)$ and $R (x^0, \cdots ,x^3)$, and ii) functional forms of dynamical eqs. are irrelevant to the separability of the $A$. 
Thus, inserting $\alpha = \hbar ^2 /2 $ determined by atomic scale experiments into (\ref{apr2907-7d}), we have (\ref{aug0608-3}).

\fussy

\newpage

\section{Stability of Solutions of the Prototype 1-D Free H-J Eq.} \label{may1618-1}

Let $q^ \star (x) $ be $q^ \star (x) = \sum _{ a_0 , \cdots , a_{n-3}} c_{ a_0 , \cdots , a_{n-3}} ( \tilde{\mathcal{S}}_0 (x) ) ^{a_0}( \tilde{\mathcal{S}}_0 ' (x) ) ^{a_1}  \cdots ( \tilde{\mathcal{S}}_0^{(n-3)}(x) ) ^{a_{n-3}}$, where $ 3 \leq  n   \in \mathbb{Z} $, $a_0 , \cdots , a_{n-4} \in \mathbb{Z}$, $a_{n-3} \in \{ 0 , 1 \} $, and $c_{ a_0 , \cdots , a_{n-3}} = const. \in \mathbb{R}$. 
We show that solutions of an $n$-th order differential eq. for  $ \tilde{\mathcal{S}}_0 ' = \partial \tilde{\mathcal{S}} _0 / \partial x$  (see p.\pageref{may0121-1}) 
\begin{equation}
(  \tilde{\mathcal{S}}_0 ' )^2 + \alpha \{ q^ \star (x) , x \} =     (  \tilde{\mathcal{S}}_0 ' )^2 + \alpha \big({ q^ \star } ''' / { q^ \star } ' - 3 ({q^ \star } ''/ {q^ \star } ')^2 /2 \big) = 2mE  \ ,    \label{apr0219-1} 
\end{equation} 
where $\alpha = const. >0$, is unstable ($| \tilde{\mathcal{S}}_0 ' (x)| \to + \infty$ for $ x \to - \infty$ or $x \to + \infty$) for some values of $E >0$ if $n \geq 4$,

We write the $q^ \star (x)$ as a function $G$ of $( \tilde{\mathcal{S}}_0 , \tilde{\mathcal{S}}_0 ' , \cdots , \tilde{\mathcal{S}}_0^{(n-3)} )$ as 
$q^\star (x) = G( \tilde{\mathcal{S}}_0 , \cdots , \tilde{\mathcal{S}}_0^{(n-3)})$. 
The $q^\star (x) = G ( \cdots )$ contains $\tilde{\mathcal{S}}_0$ as an additive term, $ q^\star (x)  = k \tilde{\mathcal{S}}_0 + \bar{\bar{G}}( \tilde{\mathcal{S}}_0 ' , \cdots , \tilde{\mathcal{S}}_0^{(n-3)} )$, where $k=const.$, because (\ref{apr0219-1}) is free from $\tilde{\mathcal{S}}_0$. 
The $k$ is nonzero for $\{ q^ \star (x) , x \}$ to be definable ($ { q^\star } ' \neq 0 $) for $\tilde{\mathcal{S}}_0 ' = const. \neq 0$ for which $\bar{\bar{G}} ' = 0$, and set to one without losing generality since $\{ q^\star (x) , x \} = \{q^\star (x) / k , x \} $. 
We thus have  $q^\star (x) = \tilde{\mathcal{S}}_0 (x) + \bar{\bar{G}}( \tilde{\mathcal{S}}_0 ' , \cdots , \tilde{\mathcal{S}}_0^{(n-3)}  ) $.

Let $n \geq 4$. 
Then, $q^\star (x)$ is written as $q^\star (x) = \tilde{\mathcal{S}}_0 + \bar{\bar{G}} (  \cdots ) $    
$ =  \tilde{\mathcal{S}}_0 + \bar{\bar{G}}_0 ( \tilde{\mathcal{S}}_0 ' , \cdots , \tilde{\mathcal{S}}_0^{(n-4)} ) +  \tilde{\mathcal{S}}_0 ^{(n-3)}  \bar{\bar{G}} _{ \neq 0} ( \tilde{\mathcal{S}}_0 ' , \cdots , \tilde{\mathcal{S}}_0^{(n-4)} ) $, 
where $\bar{\bar{G}}_0 ( \cdots ) $ is a function of $( \tilde{\mathcal{S}}_0 ' , \cdots , \tilde{\mathcal{S}}_0^{(n-4)} )$ and $\bar{\bar{G}} _{ \neq 0} $ is a nonzero function ($\bar{\bar{G}} _{ \neq 0} \neq 0$) of $( \tilde{\mathcal{S}}_0 ' , \cdots , \tilde{\mathcal{S}}_0^{(n-4)} )$. 
Indeed, if $\bar{\bar{G}} _{ \neq 0} =0$ for some boundary values $( \tilde{\mathcal{S}}_0 ' , \cdots , \tilde{\mathcal{S}}_0^{(n-1)} )$ at, say, $x = x_0$ , the $\tilde{\mathcal{S}}_0^{(n)}$ is undetermined.  
If $n=4$, we have $\bar{\bar{G}}_0 =0$ and $\bar{\bar{G}} _{ \neq 0} = const.$ 
From the expression of $q^\star (x)$, we have  \\ 
${q^ \star } ' = \tilde{\mathcal{S}}_0 ' + \bar{\bar{G}} ' ( \tilde{\mathcal{S}}_0 ' , \cdots , \tilde{\mathcal{S}}_0^{(n-2)} ) =  \tilde{\mathcal{S}}_0 ' + \bar{\bar{G}}_1 ' ( \tilde{\mathcal{S}}_0 ' , \cdots , \tilde{\mathcal{S}}_0^{(n-3)} ) + \tilde{\mathcal{S}}_0^{(n-2)} \cdot \bar{\bar{G}}_{ \neq 0}( \tilde{\mathcal{S}}_0 ' , \cdots , \tilde{\mathcal{S}}_0^{(n-4)} )$, 
${q^ \star} '' = \cdots $, and \\ 
${q ^ \star } ''' =  \tilde{\mathcal{S}}_0 ''' + \bar{\bar{G}} ''' ( \tilde{\mathcal{S}}_0 ' , \cdots , \tilde{\mathcal{S}}_0^{(n)} ) = \tilde{\mathcal{S}}_0 ''' + \bar{\bar{G}}_3 ''' ( \tilde{\mathcal{S}}_0 ' , \cdots , \tilde{\mathcal{S}}_0^{(n-1)} ) + \tilde{\mathcal{S}}_0^{(n)} \cdot \bar{\bar{G}} _{ \neq 0} ( \tilde{\mathcal{S}}_0 ' , \cdots , \tilde{\mathcal{S}}_0^{(n-4)} )$. 
With them, we rewrite (\ref{apr0219-1}) as  
\begin{equation}
\tilde{\mathcal{S}}_0 ^{(n)} = ( \bar{\bar{G}} _{ \neq 0}) ^{-1} \big( ( \frac{( 2mE - ( \tilde{\mathcal{S}}_0' )^2)( \tilde{\mathcal{S}}_0' + \bar{\bar{G}} ' )}{\alpha } + \frac{3}{2} \frac{( \tilde{\mathcal{S}}_0 '' +\bar{\bar{G}} '' )^2}{( \tilde{\mathcal{S}}_0' + \bar{\bar{G}} ' )}) - \tilde{\mathcal{S}}_0 ''' - \bar{\bar{G}}_3 ''' \big) \ . \label{mar2020-1} 
\end{equation} 
Introducing unknowns $ y_1 , \cdots , y_{n}$ defined as $y_1 = \tilde{\mathcal{S}}_0 ',  \cdots \ , y _{n} = \tilde{\mathcal{S}}_0 ^{(n)} =  y_{n-1} '$, we convert (\ref{mar2020-1}) to a system of first-order differential eqs. 
\begin{align} 
y_2 = y_1 ' \ , \ \  \cdots \ ,  \ \ 
 y_n = &y_{n-1} '     = (  \bar{\bar{G}}_{ \neq 0} ( y_1 , \cdots , y_{n-4} ) ) ^{-1} \big(  \frac{2mE - y_1^2}{ \alpha} ( y_1 + \bar{\bar{G}}'(y_1 , \cdots , y_{n-2} )  )          \notag \\ 
&+ \frac{3}{2} \frac{ ( y_2 + \bar{\bar{G}}''(y_1 , \cdots , y_{n-1} ))^2 }{y_1 + \bar{\bar{G}}'(y_1 , \cdots , y_{n-2} ) } - y_3 -  \bar{\bar{G}}_3 '''(y_1 , \cdots , y_{n-1} ) \big) \ .  \label{feb1920-5} 
\end{align} 
For boundary values $( \tilde{\mathcal{S}}_0 ' , \cdots , \tilde{\mathcal{S}}_0^{(n-1)} )|_{x = x_0} = ( \surd (2mE) , 0 , \cdots , 0 )$  giving $\tilde{\mathcal{S}}_0^{(n)} |_{x = x_0} = 0$, $ \tilde{\mathcal{S}}_0 ' (x)  = \surd (2mE) $ is a solution of (\ref{apr0219-1}).  
At the solution $\textbf{y}_0 = ( y_1 , y_2 , \cdots , y_{n-1} ) ^T = ( \surd (2mE) , 0 , \cdots , 0)^T$, we linearize \cite{Perko(2001)} (\ref{feb1920-5}) as 
\begin{equation}
\begin{pmatrix}
\delta _1 ' \\ \vdots \\ \delta _{n-2} '  \\ \delta _ {n-1} ' 
\end{pmatrix}
= 
\begin{pmatrix} 
0 & 1 & \cdots & 0      \\ 
\vdots & \vdots & \ddots & \vdots  \\ 
0 & 0 & \cdots & 1  \\ 
\eta _1  &   \eta _2  & \cdots & \eta _{n-1}   
\end{pmatrix} 
\begin{pmatrix}
\delta _1 \\ \vdots \\ \delta _{n-2} \\ \delta _ {n-1}  
\end{pmatrix} 
  \ ,   \label{apr0419-1}
\end{equation}
where $ \delta _1 = y_1 - \surd (2mE), \delta _2 =  y_2 , \cdots , \delta _{n-1} = y_ {n-1} $ and 
$\eta _1 =\partial y_{n-1}' / \partial y_1 |_{\textbf{y}_0} , \cdots , \eta _{n-1} =\partial y_{n-1}' / \partial y_{n-1} |_{\textbf{y}_0}$. 
We have $\bar{\bar{G}}' |_ {\textbf{y}_0} = \bar{\bar{G}}'' |_ {\textbf{y}_0} = \bar{\bar{G}} _3 ''' |_ {\textbf{y}_0} =0$ because $\bar{\bar{G}}' = y_2 \partial \bar{\bar{G}} / \partial y_1 + \cdots + y_{n-2} \partial \bar{\bar{G}} / \partial y_{n-3}$ and so on. 
We have $\partial \bar{\bar{G}} '/ \partial y_1 \ |_ {\textbf{y}_0} =\partial \bar{\bar{G}} ''/ \partial y_1 \ |_ {\textbf{y}_0} =\partial \bar{\bar{G}} _3 '''/ \partial y_1 \ |_ {\textbf{y}_0} =0$ for similar reason. 
We have 
$\eta _1 = - 4mE (  \alpha \bar{\bar{G}} _{ \neq 0} ) ^{-1} |_ {\textbf{y}_0} \neq 0$ according to (\ref{feb1920-5}).  
If the solution $y_1 = \surd ( 2mE )$ is stable, any eigenvalue $\lambda _i \in \mathbb{C}$ given as a solution of $| M- \lambda I_{n-1}| = 0$ or $ \lambda ^{n-1} - \eta _{n-1} \lambda ^{n-2} - \cdots - \eta _1 =0$, where $I_{n-1}$ is the identity matrix and $M$ is the matrix of (\ref{apr0419-1}), satisfies $\operatorname{Re}(\lambda _i ) =0$. \cite{Perko(2001)} 
The eq. $\lambda ^{n-1} - \cdots - \eta _1 =0$ of which $\eta _1 = - 4mE (  \alpha \bar{\bar{G}} _{ \neq 0} )^{-1} |_ {\textbf{y}_0} $ is rewritten with a variable $\xi =\lambda / ( (4mE) / ( \alpha \bar{\bar{G}} _{ \neq 0} |_ {\textbf{y}_0}) )  ^{1/(n-1)}$ as 
\begin{equation}
\xi  ^{n-1} - (\frac{4mE}{\alpha})^{ \frac{n-2}{n-1} -1} \eta _{n-1} \xi ^{n-2} - \cdots - (\frac{4mE}{\alpha})^{ \frac{1}{n-1} -1} \eta _2 \xi +1 =0 \ , \label{mar2120-2} 
\end{equation} 
in which coefficients of $\xi , \xi ^2 , \cdots , \xi ^ {n-2}$ are much smaller than unity for some values of $E$. 
For such coefficients, solutions of (\ref{mar2120-2}) are close to those of an  eq. $\xi ^{n-1} + 1 =0 $. 
If $n \geq 4$ therefore $\operatorname{Re}(\xi _i) \neq 0$ for some $\xi _i$. It means $\operatorname{Re}(\lambda _i) \neq 0$ for some $\lambda _i$ since $\xi =\lambda / ( (4mE) / ( \alpha \bar{\bar{G}} _{ \neq 0} |_ {\textbf{y}_0}) )  ^{1/(n-1)}$.

Let $n=3$. 
Then, we have $y_2 = y_1'$ and $y_3 = y_2 ' =  (2mE -y_1^2)y_1 /\alpha  + ( 3/ 2) y_2^2 / y_1 $. 
We linearize them at $\textbf{y}_0 = ( y_1 , y_2 ) ^T = ( \surd (2mE) , 0 )^T$ as 
\begin{equation}
\begin{pmatrix}
\delta _1 '   \\ \delta _ 2 ' 
\end{pmatrix}
= 
\begin{pmatrix} 
0 & 1      \\ 
\zeta _1  &   \zeta _2   
\end{pmatrix} 
\begin{pmatrix}
\delta _1 \\ \delta _ 2 
\end{pmatrix} 
   \ ,    \label{oct2319-1}
\end{equation}
where $\zeta _1 =\partial y_2' / \partial y_1 |_{\textbf{y}_0} = -4mE / \alpha $ and $ \zeta _2 =\partial y_2' / \partial y_2 |_{\textbf{y}_0} =0$. 
We have $| M- \lambda I_2| = \lambda ^2 + 4mE / \alpha =0$. 
Accordingly, $\operatorname{Re}(\lambda _i ) =0$ since $\alpha >0$.

Thus, if $n \geq 4$, solutions of (\ref{apr0219-1}) is unstable for some values of $E$.

\fussy

\section{Higher-Order Analytical Mechanics}  \label{0428ab}

We summarize third order analytical mechanics, in which  Lagrangian $L$ is given as $L = L (  \textbf{q} , \dot{ \textbf{q} } , \ddot{ \textbf{q} }   , \dot{\ddot{ \textbf{q} } })  $, as a representative of higher-order one. 
 Materials in this section are found in Ref. \refcite{Leon(1985)} and \refcite{Miron(1997)}. Reference \refcite{Leon(1989)} is useful as a support reading.

$\bullet$ \textit{Lagrangian formalism}  \ \ 
As in CM, action $\mathcal{S}$ is defined as 
$ \mathcal{S} :=   \int _a ^b L(  \textbf{q} ,\dot{ \textbf{q} },\ddot{ \textbf{q} } ,  \dot{\ddot{ \textbf{q} } }     )dt  
$. 
Let $ \delta \textbf{q} $ be  the variation of the curve $ \textbf{q} (t) $. Then, we have
\begin{align}
\delta \mathcal{S} 
&= \int  L(  \textbf{q}  +  \delta   \textbf{q}  , \dot{ \textbf{q} }+  \delta  \dot{ \textbf{q} },\ddot{ \textbf{q} }+  \delta  \ddot{ \textbf{ \textbf{q} } } , \dot{\ddot{ \textbf{q} }}+  \delta  \dot{\ddot{ \textbf{ \textbf{q} } }} )dt - \int L(  \textbf{q} ,\dot{ \textbf{q} },\ddot{ \textbf{q} }  , \dot{ \ddot{ \textbf{q} }})dt       \notag \\ 
= &\int ( \frac{ \partial L}{\partial  \textbf{q} }  \delta   \textbf{q}  + \frac{ \partial L}{\partial \dot{ \textbf{q} }}  \delta  \dot{ \textbf{q} } + \frac{ \partial L}{\partial \ddot{ \textbf{q} }}  \delta  \ddot{ \textbf{q} } + \frac{ \partial L}{\partial \dot{\ddot{ \textbf{q}} }}  \delta  \dot{\ddot{ \textbf{q} }}  ) dt    
= \int ( \frac{ \partial L}{\partial  \textbf{q} }  - \frac{d}{dt} \frac{ \partial L}{\partial \dot{ \textbf{q} }}  + \frac{d^2}{dt^2} \frac{ \partial L}{\partial \ddot{ \textbf{q} }} - \frac{d^3}{dt^3} \frac{ \partial L}{\partial \dot{\ddot{ \textbf{q} }}}  )  \delta   \textbf{q}  dt   \notag \\  
& \ \ \ \  
+ (\frac{ \partial L}{\partial \dot{ \textbf{q} }}  - \frac{d}{dt} \frac{ \partial L}{\partial \ddot{ \textbf{q} }}  + \frac{d^2}{dt^2} \frac{ \partial L}{\partial \dot{\ddot{ \textbf{q} }}}    )  \delta   \textbf{q}  |_a ^b + (\frac{ \partial L}{\partial \ddot{ \textbf{q} }}  - \frac{d}{dt} \frac{ \partial L}{\partial \dot{\ddot{ \textbf{q} }}})  \delta  \dot{ \textbf{q} } | _a ^b + (\frac{ \partial L}{\partial \dot{\ddot{ \textbf{q} }}}   )  \delta  \ddot{ \textbf{q} } | _a ^b  \ .
  \label{051606-k}    
\end{align} 
From the first term in the RHS of (\ref{051606-k}), we obtain the E-L equation:
\begin{equation}
\frac{ \partial L}{\partial  \textbf{q} }  - \frac{d}{dt} \frac{ \partial L}{\partial \dot{ \textbf{q} }}  + \frac{d^2}{dt^2} \frac{ \partial L}{\partial \ddot{ \textbf{q} }} - \frac{d^3}{dt^3} \frac{ \partial L}{\partial \dot{\ddot{ \textbf{q} }}} =0 \ . \label{42606va}
\end{equation}
We define $\textbf{p}_{(1)}$, $\textbf{p}_{(2)}$, and $\textbf{p}_{(3)}$ as ( See (\ref{051606-k}) )
\begin{equation}
\textbf{p}_{(1)} :=  \frac{ \partial L}{\partial \dot{ \textbf{q} }}  - \frac{d}{dt} \frac{ \partial L}{\partial \ddot{ \textbf{q} }}  + \frac{d^2}{dt^2} \frac{ \partial L}{\partial \dot{\ddot{ \textbf{q} }}}     \ , \ \ \ \ \    
\textbf{p}_{(2)} := \frac{ \partial L}{\partial \ddot{ \textbf{q} }}  - \frac{d}{dt} \frac{ \partial L}{\partial \dot{\ddot{ \textbf{q} }}} \ ,   \ \ \ \ \ \textbf{p}_{(3)} := \frac{ \partial L}{\partial \dot{\ddot{ \textbf{q} }}} \ .    \label{jun0207-a}
\end{equation} 
The $\textbf{p}_{(1)}$, $\textbf{p} _ { (2)}$, and $\textbf{p} _ { (3)}$ are called  Jacobi-Ostrogradsky (J-O) momenta. We define $ \stackrel{3}{\mathcal{E}} _c (L) $ as 
\begin{multline}
 \stackrel{3}{\mathcal{E}} _c (L)    
= \textbf{p}_{(1)} \dot{ \textbf{q} } + \textbf{p} _{(2)} \ddot{ \textbf{q} }     + \textbf{p} _{(3)} \dot{\ddot{ \textbf{q} }} -L  \\ 
= (    \frac{ \partial L}{ \partial \dot{ \textbf{q} }} -  \frac{d}{dt} \frac{ \partial L}{ \partial \ddot{ \textbf{q} }}  + \frac{d^2}{dt^2} \frac{ \partial L}{\partial \dot{\ddot{ \textbf{q} }}}    ) \dot{ \textbf{q} } + (    \frac{ \partial L}{ \partial \ddot{ \textbf{q} }}   - \frac{d}{dt} \frac{ \partial L}{\partial \dot{\ddot{ \textbf{q} }}}   ) \ddot{ \textbf{q} }  + \frac{ \partial L}{\partial \dot{\ddot{ \textbf{q} }}}  \dot{\ddot{ \textbf{q} }} -L \ .
    \label{051506fa}
\end{multline} 
The $\stackrel{3}{\mathcal{E}} _c (L)$, which is reduced to $(  \partial L / \partial \dot{ \textbf{q} } )  \dot{ \textbf{q} } -L$ if $L = L( \textbf{q} ,\dot{ \textbf{q} })$, 
 is const. along a solution curve of the E-L eq. (\ref{42606va}) as is seen as follows:
\begin{align}
&\frac{d}{dt} \stackrel{3}{\mathcal{E}} _c (L)
= \frac{d}{dt} ( \textbf{p}_{(1)} \dot{ \textbf{q} } + \textbf{p} _{(2)} \ddot{ \textbf{q} }   + \textbf{p} _{(3)} \dot{\ddot{ \textbf{q} }}  -L ) \notag \\  
&= (  \frac{d}{dt}   \frac{ \partial L}{ \partial \dot{ \textbf{q} }} -  \frac{d^2}{dt^2} \frac{ \partial L}{ \partial \ddot{ \textbf{q} }}  + \frac{d^3}{dt^3} \frac{ \partial L}{\partial \dot{\ddot{ \textbf{q} }}}    ) \dot{ \textbf{q} } + (    \frac{ \partial L}{ \partial \dot{ \textbf{q} }} -  \frac{d}{dt} \frac{ \partial L}{ \partial \ddot{ \textbf{q} }}  + \frac{d^2}{dt^2} \frac{ \partial L}{\partial \dot{\ddot{ \textbf{q} }}}    ) \ddot{ \textbf{q} }  \notag \\ 
&+ (   \frac{d}{dt}   \frac{ \partial L}{ \partial \ddot{ \textbf{q} }}   - \frac{d^2}{dt^2} \frac{ \partial L}{\partial \dot{\ddot{ \textbf{q} }}}   ) \ddot{ \textbf{q} }  
+ (    \frac{ \partial L}{ \partial \ddot{ \textbf{q} }}   - \frac{d}{dt} \frac{ \partial L}{\partial \dot{\ddot{ \textbf{q} }}}   ) \dot{\ddot{ \textbf{q} }}  + (   \frac{d}{dt}     \frac{ \partial L}{\partial \dot{\ddot{ \textbf{q} }}} ) \dot{\ddot{ \textbf{q} }}   + (\frac{ \partial L}{\partial \dot{\ddot{ \textbf{q} }}} ) \ddot{\ddot{ \textbf{q} }}   \notag \\  
 &- \dot{ \textbf{q} } \frac{ \partial L}{ \partial  \textbf{q} } - \ddot{ \textbf{q} } \frac{ \partial L}{ \partial \dot{ \textbf{q} }} -  \dot{\ddot{ \textbf{q} }} \frac{ \partial L}{ \partial \ddot{ \textbf{q} }}  -  \ddot{\ddot{ \textbf{q} }} \frac{ \partial L}{ \partial \dot{\ddot{ \textbf{q} }}}   
= - \dot{ \textbf{q} } \left(  \frac{ \partial L}{ \partial  \textbf{q} } -  \frac{d}{dt} \frac{ \partial L}{ \partial \dot{ \textbf{q} }} +  \frac{d^2}{dt^2} \frac{ \partial L}{ \partial \ddot{ \textbf{q} }}   - \frac{d^3}{dt^3} \frac{ \partial L}{\partial \dot{\ddot{ \textbf{q} }}}    \right) =0 \ . \label{may1311-1}  
\end{align}
We sometimes use symbols $E$ and $H$ for $  \stackrel{3}{\mathcal{E}} _c (L)$. 
Since the E-L eq. (\ref{42606va}) is written as 
$   \partial L /  \partial  \textbf{q}  - d \textbf{p}_{(1)} / dt =  0 $, 
if the Lagrangian does not contain $  \textbf{q} $ explicitly, we have $d\textbf{p}_{(1)} / dt = 0$ along a solution curve of (\ref{42606va}) .

$\bullet$ \textit{Hamiltonian formalism}  \ \ 
Coordinate system  $(  \textbf{q} ,   \dot{ \textbf{q} } ,   \ddot{ \textbf{q} }  , \textbf{p}_{(1)} , \textbf{p} _{(2)} , \textbf{p} _{(3)}  ) $    of Hamiltonian formalism is  obtained from that  $( \textbf{q} , \stackrel{(1)}{ \textbf{q} } , \cdots , \stackrel{(5)}{ \textbf{q} } )$  of the Lagrangian formalism with a Legendre transformation, \cite{Leon(1985)} which we assume to be a diffeomorphism. 
From (\ref{051506fa}) and $H= H (  \textbf{q} ,   \dot{ \textbf{q} } ,   \ddot{ \textbf{q} }  ,     \textbf{p}_{(1)} , \textbf{p} _{(2)} , \textbf{p} _{(3)}  ) $, we have
$  dH = - (\partial L / \partial  \textbf{q} ) d  \textbf{q} +  \stackrel{(n)}{  \textbf{q}} d  \textbf{p}_{(n)} +    ( \textbf{p}_{(n)} -  \partial L / \partial \stackrel{(n)}{  \textbf{q}}  ) d  \stackrel{(n)}{ \textbf{q}}  $ and 
$ dH = ( \partial H /  \partial  \stackrel{(n-1)}{\textbf{q}}  ) d   \stackrel{(n-1)}{\textbf{q}} + ( \partial H / \partial  \textbf{p}_{(n)} )  d  \textbf{p}_{(n)}   $, where   $n=1 , 2, 3$, and  $  \stackrel{(0)}{\textbf{q}} = \textbf{q}$. 
Comparison of $dH$'s in cooperation with   (\ref{42606va}) and (\ref{jun0207-a}) leads to canonical eqs.:  \\ 
\hspace*{ 5 em} 
$  d \stackrel{(n-1)}{ \textbf{q}} / dt = \partial H / \partial  \textbf{p}_{(n)}   \  , \ \ \ \ \ \ d  \textbf{p}_{(n)} / dt = - \partial H / \partial \stackrel{(n-1)}{ \textbf{q}} $ \ .


\newpage

\section{
Determination of $\bm{c_{2b}}$ in the 1-D  Lagrangian
}
 \label{mar2009-01}

We determine $c_{2b}$  in the 1-D  Lagrangian (\ref{jun1211-1}) of which $c_1 = c_{2a} =0$.


We expand $p^{HJ}$ of (\ref{may2009-pp5}) as 
\begin{align}
 p^{HJ}
&=   \sqrt{2 m E} \big(1- \epsilon_1 \cos 2kx - \epsilon_2 \sin  2 kx      + \mathcal{O}( \epsilon _1^2) + \mathcal{O}( \epsilon _2^2) + \mathcal{O}(   \epsilon _1   \epsilon _2)     \big)    \notag \\ 
&=   \sqrt{2 m E} \big(1- \epsilon_1 \cos 2kx - \epsilon_2 \sin  2 kx      + \mathcal{O}( \epsilon ^2)   \big)    \label{oct2210-3}    \ . 
\end{align}
Because we are in the semiclassical world, $\epsilon_1 , \epsilon_2 \ll  1$.  
We write the $\mathcal{O}( \epsilon _1^2) + \mathcal{O}( \epsilon _2^2) + \mathcal{O}(   \epsilon _1   \epsilon _2) $ as $\mathcal{O}( \epsilon ^2) $ with $\epsilon = \surd (\epsilon _1 ^2 + \epsilon _2 ^2  )$, which is allowed since $\epsilon _1^2 , \, \epsilon _2 ^2 , \, |\epsilon _1   \epsilon _2   | \leq ( \epsilon _1 ^2 + \epsilon _2 ^2 )$.

We expand $p^{EL}$ of (\ref{jun1211-3}), inserting $q = v t + f = vt + A_1 \cos ( \omega t + \theta _1 ) + \cdots $ (see (\ref{may0409-cv3})), as
\begin{equation}
p^{EL} 
=  p_{(1)}   +  \frac{ 2 c_{2b} \hbar ^2} {m} \frac{  \ddot{f}   ^{ \, 2}}{ (v +  \dot{f}  )^5} + \mathcal{O} ( \hbar ^3 )  \ .        \label{oct2210-1} 
\end{equation}
We note that $\mathcal{O}( \hbar ^3 )$ of (\ref{oct2210-1}) is equal to $ \mathcal{O}( ( \dot{f} / v )^3 )$; see  (\ref{jun1211-3}). 
We know, among  oscillatory terms in (\ref{may0409-cv3}),  $A_1 \cos ( \omega t + \theta _1 ) $ has the lowest inf. order with respect to (w.r.t.)  $\dot{f}/v$ (see p.\pageref{feb1521-1}).  
The angular freq. of the oscillation of $p^{EL}$ having the lowest inf. order w.r.t. $\dot{f}/v$ is therefore $2 \omega $ generated from $A_1 ^2 \cos ^2( \omega t + \theta _1 ) $.

From i) $p^{HJ} \simeq p^{EL}$, ii) velocity of the particle in the moving direction is $v$, iii) wave number of undulation of $p^{HJ}$ is $2k$ at the lowest inf. order w.r.t. $\epsilon $, and iv) angular frequency of oscillation of $p^{EL}$ is $2 \omega$ at the lowest inf. order w.r.t. $\dot{f}/v$, 
we have $\omega  =kv = v \sqrt{2mE^{HJ}} / \hbar \ \ \cdots (*1)$. 
 In passing, we see that $\mathcal{O}( \epsilon ) =  \mathcal{O}( ( \dot{f} / v )^2 )$.

We restrict $\omega$ with the E-L eq. 
We insert $q(t)$ of 
(\ref{may0409-cv3}) into the E-L eq. (\ref{jun1311-1}) of which $c _1 = c_{2a} =0$ to have $ 0= B_1 \cos ( \omega t + \theta _1 ' ) + B_2 \cos ( 2 \omega t + \theta _2 ' ) + \cdots $, where $B_1 \cos ( \omega t + \theta _1 ' )$ is given  as
\begin{align}
B_1 & \cos ( \omega t + \theta _1 ' )   
  =  -m \omega ^2 A_1    \big( \  \big( 1+ \frac{ 2 c_{2b} \hbar ^2 \omega ^2}{m^2v^4} + \frac{ 10 c_{2b} \hbar ^2 \omega ^2}{m^2v^4}  ( \frac{ \omega A_1 }{v} )^2 + \frac{3 c_4 \hbar ^4 \omega ^4 }{m^4v^8} ( \frac{ \omega A_1 }{v} )^2 \big)  \notag   \\  
& \ \ \ \  \times  \cos ( \omega t + \theta _1 )      - \frac{ 12 \hbar ^2 \omega ^2}{m^2 v^4}  \frac{ \omega A_2 }{v}      
   \big( 2 c_{2b} \sin ( \omega t - \theta _1 + \theta _2 ) + \frac{ c_3 \hbar \omega }{ mv^2}  \cos  ( \omega t - \theta _1 + \theta _2 ) \big)   \notag \\  
&  \ \ \ \ \ \ \ \ \ \ \ \ \ \ \  \ \ \ \ \ \ \ \ \ \ \ \ \ \ \  \ \ \ \ \ \ \ \ \ \ \ \ \ \ \  \ \ \ \ \ \ \ \ \ \ \ \ \ \ \  \ \ \ \ \ \ \ \ \ \ \ \ \ \ \   +    \mathcal{O} (  ( \dot{f} / v ) ^4 ) \cos ( \omega t + \theta _1 {'}{ '} ) \  \big)        \ .      \label{oct0712-1}
\end{align}
If $q(t)$ is a solution of the E-L eq., $B_1 =0$, from which we have 
$1+  2 c_{2b} \hbar ^2 \omega ^2 / m^2v^4 + \mathcal{O} ( ( A_1 \omega  /v) ^2 )= 0$ or $\omega  = mv^2 / \hbar \sqrt{-2 c_{2b} } +  \mathcal{O} ( ( A_1 \omega  /v) ^2 ) \ \ \cdots (*2)$.

Lastly, replacing $E^{HJ}$ in $(*1)$ with 
\begin{equation}
 E^{EL} = \frac{mv^2}{2} \Big( 1 + \frac{1}{2} ( \frac{ A_1 \omega }{v} )^2 - \frac{ c_{2b} \hbar ^2 \omega ^2 }{ m^2 v^4} ( \frac{ A_1 \omega }{v} )^2 + \mathcal{O} ( ( \frac{ \dot{f} }{v} )^4 ) \Big)  \label{mar2116-1}
\end{equation}
 obtained from (\ref{jun2711-2}), we have \label{aug0414-8} $\omega = mv^2 / \hbar + \mathcal{O} ( ( A_1 \omega / v )^2 ) \ \ \cdots (*3)$. Comparing $(*2)$ and $(*3)$, we have $c_{2b} = -1/2$.

\newpage

\section{The 
$v$, $a_1$, $b_1$ and $\omega$ of $q(t)$ as functions of $\epsilon _1$, $\epsilon _2$, and $k$ of $p^{HJ}$  \label{oct0512-5} }

We  determine $v$, $a_1$, $b_1$ and $\omega$ of  $q(t)$ (\ref{may0409-cv3}) as functions of $\epsilon _1$, $\epsilon _2$, and $k$ of $p^{HJ}$ (\ref{may2009-pp5}). We assume $v>0$.

$\bullet$ \textit{Comparison of const. parts of $p^{HJ}$ and $p^{EL}$} \ \  
The const. part of $p^{HJ}$ is $\sqrt{2mE}+  \mathcal{O}( \epsilon ^2 )$ from (\ref{oct2210-3}), while the const. part of $p^{EL}$ is obtained from  the const. part of 
\begin{align}
p^{EL} 
&=  p_{(1)}   +  \frac{ 2 c_{2b} \hbar ^2} {m} \frac{  \ddot{q}   ^{ \, 2}}{  \dot{q}  ^5} + \mathcal{O} ( \hbar ^3 ) 
= m ( v + \dot{f} ) + \frac{2 c_{2b} \hbar ^2}{m} \big( - \frac{  \dot{\ddot{f}}}{( v + \dot{f} )^4 } + \frac{ 3  \ddot{f}^2 }{( v + \dot{f} ) ^5} \big)  + \mathcal{O} ( \delta  ^3  )   \notag \\ 
&=  m ( v + \dot{f} ) + \frac{2 c_{2b} \hbar ^2}{m} \big( - \frac{  \dot{\ddot{f}}}{v^4 } + \frac{4 \dot{f} \dot{\ddot{f}}}{v^5 } + \frac{ 3  \ddot{f}^2 }{ v  ^5} \big)  + \mathcal{O} ( \delta ^3  )  \ ,  \label{sep2612-2}  
\end{align}
 where $\delta := \dot{f}/v$, 
as $ mv - c_{2b} \hbar ^2 A_1 ^2 \omega ^4 / m v^5 + \mathcal{O}( \delta ^3 )$.  Because   $\delta ^3  $-terms in the 
$\mathcal{O}( \delta ^3 )$  contain no const. term of which inf. order is $ \delta ^3 $, we replace $\mathcal{O}( \delta ^3 )$ with $\mathcal{O}( \delta ^4 )$.  
We therefore  have 
\begin{equation}
\sqrt{2mE} +  \mathcal{O}( \epsilon ^2 )   = mv -  \frac{c_{2b} \hbar ^2 A_1 ^2 \omega ^4 }{ m v^5} + \mathcal{O}( \delta ^4 )  \ .  \label{sep2612-9}
\end{equation}

$\bullet$ \textit{Comparison of oscillatory parts of $p^{HJ}$ and $p^{EL}$} \ \   We compare oscillatory parts of (\ref{oct2210-1}) written in the form:
\begin{align}
 p^{EL} &= p_{(1)} +  \frac{2 c_{2b} \hbar ^2 \ddot{f} ^2 } {m v^5}   + \mathcal{ O}(  \delta ^3 )  \notag \\ 
&= p_{(1)}   +  \frac{2 c_{2b} \hbar ^2   \omega ^4 } {mv^5}   ( \frac{ a_1 ^2 + b_1 ^2}{2} + a_1 b_1 \sin 2 \omega t + \frac{ a_1 ^2 - b_1 ^2}{2} \cos 2 \omega t )  + \mathcal{ O}( \delta ^3 )   \label{oct2310-kh1}
 \end{align}
with that of (\ref{oct2210-3}) written in the form:
\begin{align}
 &p^{HJ} = \sqrt{2 m E} \big(1- \epsilon_1 \cos 2k(vt +f) - \epsilon_2 \sin  2 k(vt +f) + \mathcal{O}( \epsilon ^2) \big) = \sqrt{2 m E} \, \times \notag \\ 
 &\big(1- \epsilon_1 \cos 2 \omega t - \epsilon_2  \sin 2 \omega t   
 -  2kf (\epsilon_2 \cos 2 \omega t -  \epsilon_1  \sin 2 \omega t ) + \mathcal{O}((kf)^2 \epsilon ) + \mathcal{O}( \epsilon ^2)   \big)  \label{jul0211-2}
\end{align}
to have
\begin{subequations}
\begin{align} 
 \frac{2 c_{2b} \hbar ^2   \omega ^4 } {mv^5}  a_1b_1  + \mathcal{ O}( \delta ^4 ) &= - \sqrt{2 m E}  \ ( \epsilon _2 + \mathcal{O}( \epsilon ^2)) \label{oct0510-1}          \\ 
 \frac{2 c_{2b} \hbar ^2   \omega ^4 } {mv^5} \frac{ a_1 ^2 - b_1 ^2}{2}  + \mathcal{ O}( \delta ^4 ) 
&= - \sqrt{2 m E}  \  ( \epsilon _1  +  \mathcal{O}( \epsilon ^2 ))  \ .  
\end{align}    \label{feb0807-u}
\end{subequations}  
Note $ kf \simeq \dot{f} /v ( \ll 1)$ according to $\omega = kv$.  
Because  no term has angular freq. $2 \omega$ at the inf. order $\delta ^3$ in $p^{HJ}$ of (\ref{jul0211-2})  or in $p^{EL}$ of (\ref{oct2310-kh1}), we  converted $\mathcal{ O}( \delta ^3 )$ in (\ref{oct2310-kh1}) to $\mathcal{ O}( \delta ^4 ) $ in (\ref{feb0807-u}).
 From (\ref{feb0807-u}) we have, using $  c_{2b} =   -1 /  2 $ and $\omega = v \sqrt{2mE} / \hbar $, 
\begin{equation}
a_1^2 = \frac{ \hbar ^2 m v  } { (2 m E ) ^ \frac{3}{2}  } (  \epsilon _1 + \epsilon + \mathcal{O}(\epsilon ^2) )  , \ \  \    b_1^2 = \frac{ \hbar ^2 mv } {( 2 m E) ^ \frac{3}{2}   }  ( - \epsilon _1 + \epsilon  + \mathcal{O}(\epsilon ^2) )  \ . \label{jun0307-k}
\end{equation}

$\bullet$ \textit{Expressing  $v$, $\omega$, $a_1$, and $b_1$ as  functions of $\epsilon _1, \epsilon _2 $, and $E$} \ \ 
Inserting $  c_{2b} = -1/2 $ and 
(\ref{jun0307-k}) into (\ref{sep2612-9}), and solving for $v$, 
 we have $v$ and $\omega $ as 
\begin{equation}
v=  \frac{ \sqrt{2mE}}{m} ( 1 - \epsilon   + \mathcal{O}(\epsilon ^2)) \  , 
\ \ \ \ \ \ \ \ 
\omega = kv =  \frac{2E}{\hbar } ( 1- \epsilon + \mathcal{O}(\epsilon ^2)) \ .        \label{may0909-rr3}   
 \end{equation}
Inserting $v$ of (\ref{may0909-rr3}) into $a_1^2 , b_1^2$ of (\ref{jun0307-k}), we have 
\begin{subequations}
\begin{align}
a_1^2 &= - \frac{ \hbar ^2  } {2mE  } ( 1-  \epsilon ) ( -\epsilon _1 - \epsilon )  
=  \frac{ \hbar ^2  } {2mE  }  (  \epsilon _1 + \epsilon   + \mathcal{O}(\epsilon ^2)) 
   \\  
    b_1^2 &= - \frac{ \hbar ^2  } {2mE  }  ( 1- \epsilon ) ( \epsilon _1 - \epsilon ) 
=  \frac{ \hbar ^2  } {2mE  }     ( - \epsilon _1 + \epsilon   + \mathcal{O}(\epsilon ^2)) \ ,      
\end{align} \label{sep0912-1}
\end{subequations}
from which we have, according to  (\ref{oct0510-1}):   
$
 - \epsilon_2 \sqrt{ 2 mE} \simeq - \hbar ^2   \omega ^4  a_1b_1 / mv^5 $, if $\epsilon_2 \geq 0$ (the double sign corresponds)
\begin{equation}
a_1 = \pm       \frac{ \hbar    } { \sqrt{2mE}   }   \sqrt{   \epsilon _1 + \epsilon }    + \mathcal{O}(\epsilon ^\frac{3}{2})  , \     b_1 = \pm       \frac{ \hbar    } { \sqrt{2mE}   }   \sqrt{  - \epsilon _1 + \epsilon }  + \mathcal{O}(\epsilon ^\frac{3}{2})   \ ;  \label{oct1410-7}
\end{equation} 
if $\epsilon_2 < 0$  (the double sign corresponds)
\begin{equation}
a_1 = \pm       \frac{ \hbar    } { \sqrt{2mE}   }   \sqrt{   \epsilon _1 + \epsilon }    + \mathcal{O}(\epsilon ^\frac{3}{2})  , \     b_1 = \mp       \frac{ \hbar    } { \sqrt{2mE}   }   \sqrt{  - \epsilon _1 + \epsilon }  + \mathcal{O}(\epsilon ^\frac{3}{2})  \ . 
\label{062906ab}
\end{equation}


\section{Inconsistency between $\bm{p^{HJ}}$  and $\bm{p^{EL}}$ at $\bm{\mathcal{O}((\dot{f} /v)^3})$ \label{jun2511-1}
}

No value of $c_3 \in \mathbb{R}$ in the Lagrangian (\ref{jun1211-1}) with $c_1 = c_{2a} =0$ and $c_{2b} = -1/2$ makes  $p^{EL}=p^{HJ}$ true at $\mathcal{O}(\epsilon ^{3/2}) = \mathcal{O}((\dot{f} /v)^3)$, which we show comparing $p^{EL}$ and $p^{HJ}$ at $\mathcal{O}(\epsilon ^{3/2})$.


We expand $p^{HJ}$ of (\ref{may2009-pp5}), replacing $x$ in which with $ q(t) = vt + A_1 \cos ( \omega t + \theta _1 ) + \cdots$ of (\ref{may0409-cv3}) of which $q_0 =0$, as 
\begin{align} 
&p^{HJ} = \hbar k \Big( 1 - \big( \ \ \epsilon _1 \cos   2 \omega t  + \epsilon _2  \sin 2  \omega t    
 + kA_1 ( \ \epsilon _2 \cos ( \omega t - \theta _1  ) - \epsilon _1 \sin ( \omega t - \theta _1  ) \ )     \notag \\ 
&+  kA_1 ( \ \epsilon _2 \cos (3 \omega t + \theta _1  ) - \epsilon _1 \sin ( 3 \omega t +  \theta _1  ) \ ) \ \ \big) + \mathcal{O}(\epsilon ^2 ) \Big) = \hbar k -\hbar k  \epsilon  \cos  ( 2 \omega t  + \zeta  )       \notag \\  
&- \hbar k ^2 A_1  \epsilon  \cos ( \omega t - \theta _1  + \eta  ) - \hbar k^2 A_1  \epsilon  \cos (3 \omega t + \theta _1  + \eta ) + \mathcal{O}(\epsilon ^2 )  \ , 
 \label{nov1214-1} 
\end{align}
where $\epsilon = \surd ( \epsilon _1^2 + \epsilon _2^2 )$, $\cos \zeta  = \epsilon _1 / \epsilon $, $\sin \zeta  =- \epsilon _2 / \epsilon $,  $\cos \eta = \epsilon _2 / \epsilon $, and $\sin \eta = \epsilon _1 / \epsilon $.

\sloppy 

We determine the $3\omega$-term at $\mathcal{O}((\dot{f} /v)^3)$ of $p^{EL}$ (\ref{jun1211-3}) to be compared with that of (\ref{nov1214-1}). 
Inserting the $q(t)$ into  (\ref{jun1211-3}) with $c_{2b} = -1/2$, we have $3 \omega$-terms of $p^{EL}$ at $\mathcal{O}((\dot{f} /v)^3)$ as
\begin{equation}
  - \frac{ 4 \hbar ^2     \omega ^4 A_1 A_2 }{mv^5} \cos ( 3 \omega t + \theta _1 + \theta _2 )
 -  \frac{5 \hbar ^2       \omega ^5 A_1 ^3 }{4 mv^6} \sin ( 3 \omega t + 3 \theta _1  )      
    - \frac{ 3 c_3 \hbar ^3 \omega ^6 A_1 ^3 }{4m^2v^8} \cos ( 3 \omega t + 3 \theta _1  )  . 
   \label{jun2411-2}
\end{equation}
To express $( \omega A_2 /v , \theta _2)$ in   (\ref{jun2411-2})   in terms of $(\omega A_1 /v , \theta _1)$, we insert the $q(t)$ into the E-L eq. (\ref{jun1311-1}), then collect $2 \omega $-terms at $\mathcal{O}((\dot{f} /v)^2)$ to have
\begin{align}
-4 A_2 m \omega ^2 ( 1 - \frac{4 \hbar ^2 \omega ^2 }{ m^2 v^4} )  
 \cos ( 2 \omega t +  \theta _2 )    
+ \frac{ 6 c_3 A_1 ^2 \hbar ^3 \omega ^6 }{ m^2v^7 }  \cos ( 2 \omega t + 2 \theta _1 )&  \notag \\  
+ \frac{6 A_1^2 \hbar ^2 \omega ^5 }{mv^5} \sin ( 2 \omega t + 2 \theta _1 )       &=0  \ . \label{oct0512-3}  
\end{align}
Using $\hbar \omega / mv^2 = 1+  \mathcal{O} (( \dot{f} /v )^2)$ (see (\ref{may0909-rr3})), we rewrite (\ref{oct0512-3}) as 
\begin{align}
\frac{\omega A_2}{v} \cos ( 2\omega t + \theta _2 ) 
&= - \frac{1}{2} (\frac{ \omega A_1 }{v})^2 \big( c_3 \cos ( 2 \omega t + 2 \theta _1 ) + \sin ( 2 \omega t + 2 \theta _1 ) \big)      \notag \\  
&= - \frac{ \surd ( c_3^2 + 1 )}{2} (\frac{ \omega A_1 }{v})^2 \cos ( 2 \omega t + 2 \theta _1 + \xi  )   
                 \ ,     \label{nov1414-1} 
\end{align}
where $\cos \xi = c_3 / ( c_3^2 + 1 ) ^ {1/2} $, and $\sin \xi = -1 / ( c_3^2 + 1 ) ^ {1/2} $. 
From (\ref{nov1414-1}), we see $\omega A_2 /v =  (\omega A_1/v) ^2 ( c_3^2 + 1 ) ^ {1/2} / 2$, $\theta _ 2 = 2 \theta _1 + \xi \pm \pi$.    Using these expressions,  we rewrite (\ref{jun2411-2}) as 
\begin{align}
\textrm{(\ref{jun2411-2})} =  &mv (\frac{ \omega A_1 }{v})^3 \Big( 2  ( c_3^2 + 1 ) ^ {1/2}  \cos ( 3 \omega t + 3 \theta _1 + \xi )  -\frac{3c_3}{4} \cos ( 3 \omega t + 3 \theta _1  )                   \notag \\ 
&-\frac{5}{4}  \sin ( 3 \omega t + 3 \theta _1  ) \Big)      = \frac{mv}{4} ( \frac{\omega A_1}{v})^3 \sqrt{25c_3^2 + 9 } \cos  ( 3 \omega t + 3 \theta _1 + \kappa  )  \  , \label{jun2411-3}
\end{align}
where $\cos \kappa = 5c_3 / ( 25 c_3^2 + 9 ) ^ {1/2}$ and $\sin \kappa = -3 / ( 25 c_3^2 + 9 ) ^ {1/2}$. 

\fussy

Comparing (\ref{jun2411-3}) with (\ref{nov1214-1}), we have 
$\hbar k^2 A_1 \epsilon =(mv/4) ( \omega A_1 / v)^3 \surd (25c_3^2 + 9 )$, which is simplified to $\surd ( 25 c_3^2 + 9 ) /2  =1$ with (\ref{may0909-rr3})  and (\ref{sep0912-1}). 
The eq. is satisfied by no $c_3 \in \mathbb{R}$. 
That is, no $c_3 \in \mathbb{R}$ makes $p^{HJ}= p^{EL}$ true  at  $\mathcal{O}( \epsilon ^{3/2})$.

\label{sep2212-19} 
\begin{figure}[b]  
\centering
\includegraphics[width = 50ex ,  height = 21ex , clip]{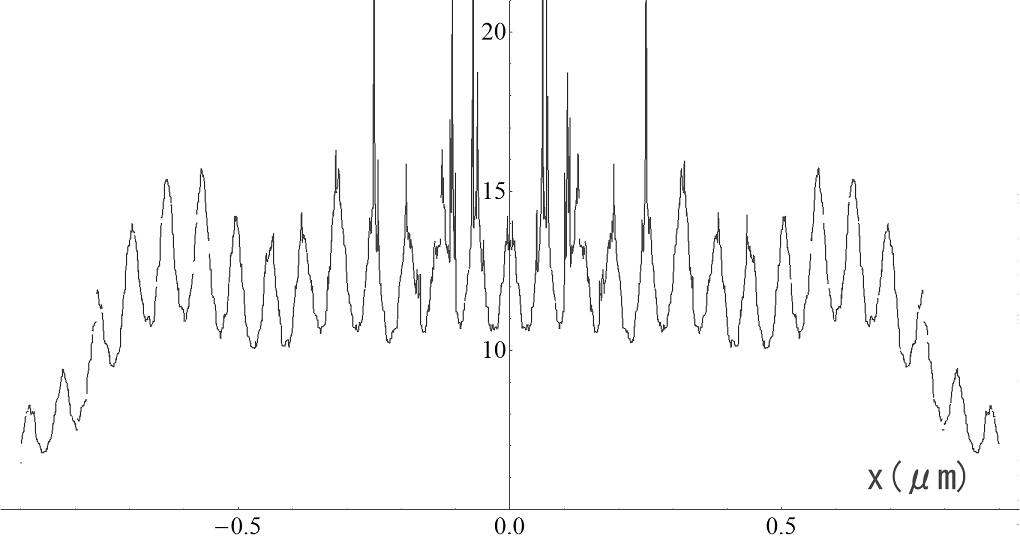}


\includegraphics[width = 50ex ,  height = 21ex , clip]{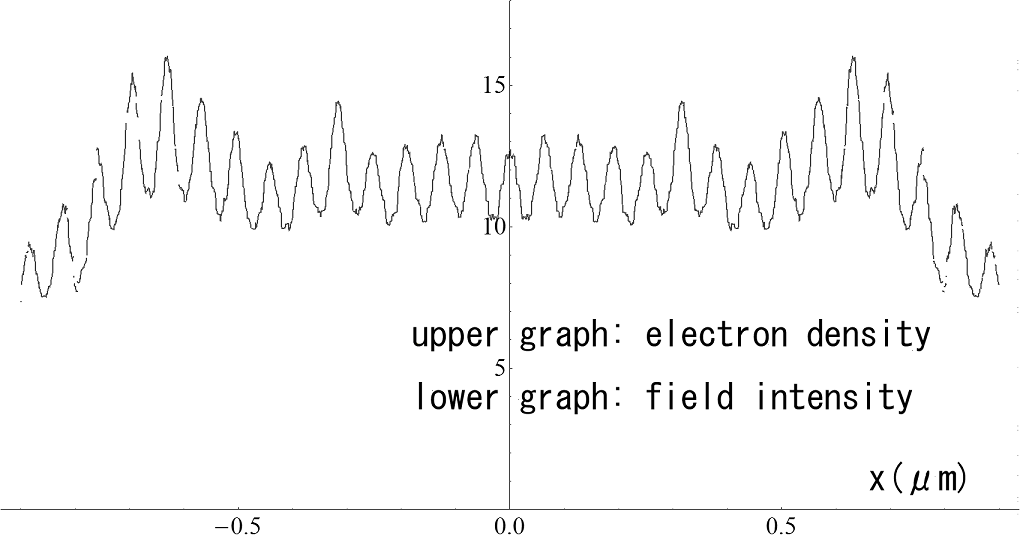}
\caption{---for \ref{sep0811-1}--- \ \ Comparison of electron density and field intensity on the screen in the case that the  electron density in an ensemble of incoming beams is assumed as $| \Psi _s |^2$.   The  largest Fresnel peaks around $x = \pm 0.6 \mu m$ of the upper graph are located more inwardly than those of  the lower graph. Spikes are artifacts.}
\label{fig-psi-square}
\end{figure}

\newpage

\section{A formalism of particle motion constructed upon higher-order d'Alembert's principle \label{sep2212-730}}

We construct a formalism of particle motion of a 1-D free particle upon higher-order d'Alembert's principle. 

\sloppy 

We generalize the classical d'Alembert's principle $(F -m \ddot{q} ) \delta q =0$ \cite{Lanczos(1986)} of the system to $G_0 \delta q + G_1 \delta \dot{q} + \cdots =0$, where $G_i = G_i ( q , \dot{q} , \ddot{q} , \cdots )$ for  $i= 0,1, \cdots $. 
Then, we transform it to an integral form (cf. Ref. \refcite{Lanczos(1986)} chap.5) 
\begin{equation}
\delta \mathcal{S} = \int (   \delta q \cdot G_0  +  \delta \dot{q} \cdot G_1  + \delta \ddot{q} \cdot G_2  + \cdots )dt =0 \ .  \label{oct1312-1}
\end{equation} 
Note that (\ref{oct1312-1}) reduces to the  Hamilton's principle: $ \delta \mathcal{S} =   \int ( \delta q \cdot \partial L / \partial q    + \delta \dot{q} \cdot \partial L / \partial \dot{q}   + \cdots ) dt = 0 $ if $G_0 ,G_1 , \cdots $ are written as $G_0 =  \partial L / \partial q , G_1 =  \partial L / \partial \dot{q} , \cdots $ for a function $L = L ( q , \dot{q} , \ddot{q} , \cdots )$;
 cf. (\ref{051606-k}). 
 We have an eq. of motion
\begin{equation}
G_0 -\dot{G}_1 + \ddot{G_2} - \cdots = 0   \label{sep1812-1}
\end{equation}
through 
\begin{multline}
\delta \mathcal{S} = \int _a ^b  ( G_0 - \dot{G}_1 + \ddot{G}_2 - \cdots ) \delta q dt   
+ ( G_1 - \dot{G}_2 + \ddot{G}_3 - \cdots ) \delta q |_a^b \\ 
 + (  G_2 - \dot{G}_3 +\ddot{G}_4 - \cdots ) \delta \dot{q}|_a^b  + ( G_3 - \dot{G}_4 + \ddot{G}_5 - \cdots )  \delta \ddot{q} |_a^b + \cdots \ .  \notag  
\end{multline}
We   define 
$ p_{(1)} = G_1 - \dot{G}_2 +  \cdots $,   $p_{(2)} = G_2 - \dot{G}_3 + \cdots $, $ p_{(3)} = G_3 - \dot{G}_4 + \cdots$, 
and so forth; cf. (\ref{jun0207-a}). 
 Along a solution $q(t)$ of (\ref{sep1812-1}), we have $dp_{(1)}/dt = G_0$. 
We define $p^{NW}$ corresponding to $p^{EL}$ as
\begin{align} 
p^{NW} =  \delta \mathcal{S} /  \delta q =
 &p_{(1)} +  p_{(2)}   \ddot{q} /  \dot{q} +  p_{(3)}   \dot{\ddot{q}} / \dot{q}   + \cdots     = (   G_1 - \dot{G}_2 + \ddot{G} _3 - \cdots  )   \notag  \\  
&+  (G_2 - \dot{G}_3 + \ddot{G} _4 - \cdots )   \ddot{q} / \dot{q}   + (G_3 - \dot{G}_4 + \ddot{G}_5 - \cdots )   \dot{ \ddot{q} } / \dot{q}   + \cdots \ ;   \label{sep1912-21}
\end{align}
 cf. (\ref{jun1211-3}). We define $ E ^{NW}(x)$ as
\begin{equation}
E^{NW} (x) =   \dot{q}   p^{NW} |_{q=x}- \int ^ {x = q} (  G_0 \dot{q} +  G_1 \ddot{q} + G _2 \dot{\ddot{q}} +  \cdots )dt    \ ;   \label{sep1812-6}
\end{equation}
cf. (\ref{051506fa}). Along a solution curve of (\ref{sep1812-1}), we have $ dE^{NW}(x) /dt = \big( (d/ dt)( \dot{q}p_{(1)} + \ddot{q}p_{(2)} + \cdots ) - (  G_0 \dot{q} +  G_1 \ddot{q} + \cdots ) \big) |_{q=x} =0 \, $; cf. (\ref{may1311-1}).

We see $G_0 =0$ for the present system because $\mathcal{S}$ is invariant under spatial translation of the system: 
Let $q(t)$ and $q(t) + \delta q$, where $\delta q = const. \neq 0$, be solutions of (\ref{sep1812-1}). 
Let $\mathcal{S}_{q(t)} $ ($\mathcal{S}_{q(t)+ \delta q} $)  be the action $\mathcal{S}$  along $q(t)$ ($q(t) + \delta q$) between $t_0$ and $t_1$. 
 Then, we see $  \mathcal{S}_{q(t)+ \delta q} -\mathcal{S}_{q(t)} =  \int       G_0 ( \dot{q} , \ddot{q} , \cdots ) \delta q  dt =0 $ because $ \mathcal{S}_{q(t)+ \delta q} - \mathcal{S}_{q(t)} = 0$ from translational invariance and $ \mathcal{S}_{q(t)+ \delta q} -\mathcal{S}_{q(t)} =  \int   G_0 \delta q  dt$ from $ \delta \dot{q} =  \delta \ddot{q} =  \cdots =0$ in (\ref{oct1312-1}).  
Since $t_0$ and $t_1$  are arbitrary, we see $G_0 =0$.

We restrict $G_1 , G_2, \cdots $. 
The $G_1 , G_2, \cdots $ do not depend on $q$ because they are invariant under spatial translation. 
We have $G_1 = G_1( \dot{q} , \ddot{q} , \dot{\ddot{q}} )$, $G_2 = G_2( \dot{q} , \ddot{q} )$,  $G_3 = G_3( \dot{q} )$, and $G_4 , G_5 , \cdots =0$ for the eq. of motion to be of fourth order (\S \ref{jun1609-1}). 
We expand the $G_i$'s as power series of $\hbar$ as $G_i = G_{i0} + \hbar G_{i1} + \hbar ^2 G_{i2} + \cdots $. 
Then, with dimensional analysis,  we see 
$ \hbar ^k G_{1 \, k} =   c_{1 \, k \, \alpha} \hbar ^k  \ddot{q}^{ k - 2 \alpha } \dot{\ddot{q}}^\alpha / m^{k-1}  \dot{q} ^{ 3k - \alpha -1 }  $
 and  $ \hbar ^k G_{2 \, k} =    c_{2k} \hbar ^k  \ddot{q}^{k-1} / m^{k-1} \dot{q} ^{3k -2}  $, 
where $k = 0,1, \cdots $ and $ c_{1 \, k \, \alpha} , c_{2 \, k} = const. \in \mathbb{R}$. Exponents of  $\ddot{q}$ and $ \dot{\ddot{q}}$, which may take value zero, are nonnegative since $G_1 , G_2 , \cdots = \pm \infty$ is unallowed. 
We assume $\alpha \in \mathbb{Z}$. 
We have, \label{jan2817-1} for example, 
\begin{align}
&G_{10} =  m \dot{q}  , \ 
\hbar G_{11} = \hbar c_{110} \frac{ \ddot{q}}{ \dot{q}^2}  , \ 
 \hbar ^2 G_{12} = \hbar ^2 (   \frac{ c_{1  2  0}  }{m}   \frac{ \ddot{q}^{ 2 } }{ \dot{q} ^{5}} +   \frac{ c_{1  2  1} }{m} \frac{ \dot{\ddot{q}}  }{ \dot{q} ^{ 4 }} ), \   \notag \\ 
  &\hbar ^3 G_{13} =  \hbar ^3  ( \frac{ c_{1  3  0}  }{m^2}   \frac{ \ddot{q}^{ 3 }  }{ \dot{q} ^{ 8 }} +   \frac{ c_{1  3  1}  }{m^2}   \frac{ \ddot{q} \dot{\ddot{q}}  }{ \dot{q} ^{7 }} )  ,  \ 
 \hbar ^4 G_{14} = \hbar ^4 ( \frac{ c_{1  4  0}  }{m^3}   \frac{ \ddot{q}^{ 4 }  }{ \dot{q} ^{ 11 }} +   \frac{ c_{1  4  1}  }{m^3}   \frac{ \ddot{q} ^2 \dot{\ddot{q}}  }{ \dot{q} ^{10  }}  +   \frac{ c_{1  4  2}  }{m^3}   \frac{  \dot{\ddot{q}} ^2  }{ \dot{q} ^{9}}      ) ,   \notag \\ 
&\hbar ^5 G_{15}= \hbar ^5 ( \frac{c_{150}}{m^4}\frac{ \ddot{q}^5}{ \dot{q}^{14}} + \frac{c_{151}}{m^4} \frac{\ddot{q}^3\dot{\ddot{q}}}{ \dot{q}^{13}} + \frac{ c_{152}}{m^4}\frac{\ddot{q}\dot{\ddot{q}}^2}{ \dot{q}^{12}} ) , \ \ \ \  \ \   \ \ \ \ \ \ \ \   G_{20} = 0  , \ 
 \hbar G_{21} = \hbar c_{21} \frac{1}{ \dot{q}}  ,   \notag \\  
&\hbar ^2 G_{22} =  \hbar ^2   \frac{ c_{2  2} }{m}  \frac{ \ddot{q} }{ \dot{q} ^4}   ,  \ 
 \hbar ^3 G_{23} =     \hbar ^3   \frac{ c_{2  3} }{m^2}  \frac{ \ddot{q}^2}{ \dot{q} ^7}   , \ 
  \hbar ^4 G_{24} =     \hbar ^4  \frac{ c_{2  4} }{m^3}  \frac{ \ddot{q}^3}{ \dot{q} ^{10}}  , \ 
\hbar ^5G_{25} = \hbar ^5  \frac{ c_{25}}{m^4} \frac{ \ddot{q}^4}{ \dot{q}^{13}}          .  \notag 
\end{align}
The $G_3$, which consists of a term $ \hbar ^2 G_{32} = c_{3  2 } \hbar ^2 / m \dot{q}^3$, vanish ($G_3 =0$) because, for $p^{NW}$ to be a function of $( \dot{q} , \ddot{q} , \cdots )$ at particle position, $ p_{(3)} = G_3 - \dot{G}_4 + \cdots = G_3 =0$ has to hold true like the case of $p^{EL}$ (see \S \ref{jun1609-1}). 
\footnote{\label{may1316-1}
The 3-D form of, say, $G_{12}$ is given as 
$ G_{12} = c_{1  2  0} \ddot{\textbf{q}}^{ 2 } \dot{q}_i / m \dot{\textbf{q}} ^{6} + c_{1  2  1}  \dot{\ddot{q}}_i  / m \dot{\textbf{q}} ^{ 4 } $; cf. \S \ref{jun1609-1}. 
}

The eq. of motion $0 = G_0 - \dot{G}_1 + \ddot{G_2} - \cdots = - G_{10} +  \hbar ( - \dot{G}_{11} + \ddot{G}_{21} )
+  \hbar ^2 ( - \dot{G}_{12} + \ddot{G}_{22} - \dot{\ddot{G}}_{32} ) 
+\hbar ^3 ( -\dot{G}_{13} + \ddot{G}_{23} ) + \hbar ^4 ( -\dot{G}_{14} + \ddot{G}_{24} ) + \cdots  $, 
$p_{(1)} = G_1 - \dot{G}_2 $, 
$p_{(2)} = G_2 $, 
$p^{NW} = p_{(1)} + p_{(2)}\ddot{q} / \dot{q} + p_{(3)} \dot{\ddot{q}} /\dot{q}$, and 
$E^{NW} = \dot{q} p^{NW} -\int (G_1 \ddot{q} + G_2 \dot{\ddot{q}} + G_3 \ddot{\ddot{q}} ) dt$ are given as power series of $\hbar$ as 
\begin{align}
&0 = - m\ddot{q} +  \hbar \big( ( -c_{110} -c_{21}) \dot{\ddot{q}} / \dot{q}^2  + 2( c_{110} + c_{21})   \ddot{q} ^2 / \dot{q} ^3 \big) +                                         \hbar ^2   \times            \notag \\  
  &\big( \frac{ -c_{121} + c_{22}  }{m} \frac{\ddot{\ddot{q}}}{ \dot{q} ^4}   - \frac{ 2( c_{120} - 2 c_{121} + 6 c_{22} )}{m} \frac{ \ddot{q}\dot{\ddot{q}}}{\dot{q}^5}
+ \frac{ 5 (c_{120} + 4 c_{22} ) }{m} \frac{ \ddot{q}^3}{ \dot{q}^6}    \big)                                \notag \\ 
& + \hbar ^3  \big( \frac{-c_{131} + 2c_{23}}{ m^2} \frac{( \dot{\ddot{q}}^2 + \ddot{q} \ddot{\ddot{q}} ) }{ \dot{q}^7} 
+ \frac{ -3 c_{130} + 7c_{131} -35c_{23}}{m^2} \frac{ \ddot{q}^2 \dot{\ddot{q}}}{ \dot{q}^8} 
+ \frac{8(c_{130} + 7c_{23}) }{m^2}\frac{ \ddot{q}^4}{ \dot{q}^9} \big)   \notag \\  
&+ \hbar ^4 \big( 
 \frac{ -2 c_{142}}{ m^3} \frac{\dot{\ddot{q}} \ddot{\ddot{q}}}{ \dot{q}^9} 
  + \frac{ -c_{141} + 3 c_{24}}{m^3} \frac{ \ddot{q} ^2 \ddot{\ddot{q}}}{ \dot{q}^{10}}
+ \frac{ - 2 c_{141} + 9c_{142} + 6 c_{24} }{ m^3} \frac{ \ddot{q} \dot{ \ddot{q}}^2}{\dot{q} ^{10}}   \notag \\  
& \ \ \ \ \  \ \ \ \ \   + \frac{ -4 c_{140} + 10 c_{141} -70 c_{24}}{m^3} \frac{ \ddot{q}^3 \dot{\ddot{q}}}{ \dot{q}^{11}} 
+ \frac{ 11 (c_{140} + 10 c_{24}) }{m^3} \frac{ \ddot{q}^5}{ \dot{q}^{12}} 
\big)      \notag \\ 
&+ \hbar ^5 \big( 
 - \frac{ c_{152}( \dot{\ddot{q}}^3 + 2 \ddot{q} \dot{\ddot{q}} \ddot{\ddot{q}})}{ m^4 \dot{q}^{12}} 
 + \frac{ (- 3 c_{151} + 12c_{152} + 12 c_{25} ) \ddot{q}^2 \dot{ \ddot{q}}^2 + ( -c_{151}+4c_{25}) \ddot{q}^3 \ddot{\ddot{q}}   }{m^4 \dot{q} ^{13}}   \notag \\  
& \ \ \ \ \  \ \ \ \ \   + \frac{( -5 c_{150} + 13 c_{151} -117 c_{25})}{m^4} \frac{ \ddot{q}^4 \dot{\ddot{q}}}{ \dot{q}^{14}} 
+ \frac{( 14 c_{150} + 182 c_{25}) }{m^4} \frac{ \ddot{q}^6}{ \dot{q}^{15}} 
\big)  + \cdots  \  , \label{jul3014-5}
\end{align} 
\begin{align}
&p_{(1)}
=  m\dot{q} + \hbar ( c_{110} + c_{21}) \frac{\ddot{q}}{ \dot{q}^2} 
+ \hbar ^2     \big( \frac{ c_{121} - c_{22} }{m} \frac{\dot{\ddot{q}}}{ \dot{q} ^4}   + \frac{ c_{120}  +4 c_{22} }{m} \frac{ \ddot{q}^2}{\dot{q}^5}    \big)      + \hbar ^3 \times                      \notag \\ 
 &\big( \frac{c_{131} - 2c_{23}}{ m^2} \frac{  \ddot{q} \dot{\ddot{q}} }{ \dot{q}^7} 
+ \frac{ c_{130} +7c_{23}}{m^2} \frac{ \ddot{q}^3 }{ \dot{q}^8}  \big)  
+ \hbar ^4 \big(  \frac{ c_{142}}{ m^3} \frac{\dot{\ddot{q}} ^2}{ \dot{q}^9} 
  + \frac{ c_{141} - 3 c_{24}}{m^3} \frac{ \ddot{q} ^2 \dot{\ddot{q}}}{ \dot{q}^{10}}    + \frac{  c_{140} + 10 c_{24} }{m^3} \frac{ \ddot{q}^4 }{ \dot{q}^{11}} \big)  \notag \\ 
&+ \hbar ^5  \big(  \frac{ c_{152}}{ m^4} \frac{\ddot{q} \dot{\ddot{q}} ^2}{ \dot{q}^{12}} 
  + \frac{ c_{151} - 4 c_{25}}{m^4} \frac{ \ddot{q} ^3 \dot{\ddot{q}}}{ \dot{q}^{13}}    + \frac{  c_{150} + 13 c_{25} }{m^4} \frac{ \ddot{q}^5 }{ \dot{q}^{14}} \big)  + \cdots \ ,  \notag \\ 
&p_{(2)} 
= \hbar \frac{c_{21}}{\dot{q}} + \hbar ^2 \frac{ c_{22}  \ddot{q}}{ m \dot{q} ^4} + \hbar ^3 \frac{c_{23} \ddot{q} ^2}{ m^2 \dot{q} ^7} 
+ \hbar ^4 \frac{c_{24} \ddot{q} ^3 }{m^3 \dot{q}^{10}}
 + \hbar ^5 \frac{c_{25} \ddot{q}^4}{m^4 \dot{q}^{13}} + \cdots \ .  \notag 
\end{align} 
\begin{align}
&p^{NW} =  m\dot{q} +  \hbar ( c_{110} +2 c_{21}) \frac{\ddot{q}}{ \dot{q}^2}   +    \hbar ^2     \big( \frac{ c_{121} - c_{22} }{m} \frac{\dot{\ddot{q}}}{ \dot{q} ^4}   + \frac{ ( c_{120}  +5 c_{22} )}{m} \frac{ \ddot{q}^2}{\dot{q}^5}    \big)  + \hbar ^3               \notag \\ 
&\times \big( \frac{c_{131} - 2c_{23}}{ m^2} \frac{  \ddot{q} \dot{\ddot{q}} }{ \dot{q}^7} + \frac{  c_{130} +8c_{23}}{m^2} \frac{ \ddot{q}^3 }{ \dot{q}^8}  \big)                  
 + \hbar ^4 \big( \frac{ c_{142}}{ m^3} \frac{\dot{\ddot{q}} ^2}{ \dot{q}^9} 
  + \frac{ c_{141} - 3 c_{24}}{m^3} \frac{ \ddot{q} ^2 \dot{\ddot{q}}}{ \dot{q}^{10}}    + \frac{ c_{140} + 11 c_{24} }{m^3} \frac{ \ddot{q}^4 }{ \dot{q}^{11}} \big)  \notag \\ 
&+ \hbar ^5\big( 
 \frac{  c_{152}}{ m^4} \frac{\ddot{q} \dot{\ddot{q}} ^2}{ \dot{q}^{12}} 
  + \frac{ c_{151} - 4 c_{25}}{m^4} \frac{ \ddot{q} ^3 \dot{\ddot{q}}}{ \dot{q}^{13}}    + \frac{ c_{150} + 14 c_{25} }{m^4} \frac{ \ddot{q}^5 }{ \dot{q}^{14}} \big) + \cdots \ ,       \label{nov0314-1}    
\end{align}
\begin{align}
&E^{NW} = m\dot{q} ^2 +  \hbar ( c_{110} +2 c_{21}) \frac{\ddot{q}}{ \dot{q}}   +    \hbar ^2     \big( \frac{ c_{121} - c_{22} }{m} \frac{\dot{\ddot{q}}}{ \dot{q} ^3}   + \frac{ ( c_{120}  +5 c_{22} )}{m} \frac{ \ddot{q}^2}{\dot{q}^4}    \big)  + \hbar ^3             \notag \\ 
&\times  \big( \frac{c_{131} - 2c_{23}}{ m^2} \frac{  \ddot{q} \dot{\ddot{q}} }{ \dot{q}^6} + \frac{  c_{130} +8c_{23}}{m^2} \frac{ \ddot{q}^3 }{ \dot{q}^7}  \big)                  
 + \hbar ^4 \big( \frac{ c_{142}}{ m^3} \frac{\dot{\ddot{q}} ^2}{ \dot{q}^8} 
  + \frac{ c_{141} - 3 c_{24}}{m^3} \frac{ \ddot{q} ^2 \dot{\ddot{q}}}{ \dot{q}^9}    + \frac{ c_{140} + 11 c_{24} }{m^3} \frac{ \ddot{q}^4 }{ \dot{q}^{10}} \big)  \notag \\ 
&+ \hbar ^5\big( 
 \frac{  c_{152}}{ m^4} \frac{\ddot{q} \dot{\ddot{q}} ^2}{ \dot{q}^{11}} 
  + \frac{ c_{151} - 4 c_{25}}{m^4} \frac{ \ddot{q} ^3 \dot{\ddot{q}}}{ \dot{q}^{12}}    + \frac{ c_{150} + 14 c_{25} }{m^4} \frac{ \ddot{q}^5 }{ \dot{q}^{13}} \big) + \cdots    \notag \\ 
&- \int \Big( m\dot{q} \ddot{q} + \hbar ( c_{110} \frac{ \ddot{q}^2}{ \dot{q}^2} + c_{21} \frac{ \dot{\ddot{q}}}{\dot{q}} ) + \hbar ^2 ( \frac{ c_{120}}{m} \frac{\ddot{q}^3}{\dot{q}^5} + \frac{ c_{121} + c_{22}}{m} \frac{ \ddot{q} \dot{\ddot{q}}}{\dot{q}^4} )  \notag \\ 
& \ \ \ \ \ \ \ \ \ \ \ \ \ \ \ \ \ \   + \hbar ^3 ( \frac{c_{130}}{m^2} \frac{ \ddot{q}^4}{\dot{q}^8} + \frac{c_{131} + c_{23}}{m^2} \frac{ \ddot{q}^2 \dot{\ddot{q}}}{ \dot{q}^7}) 
+ \hbar ^4 ( \frac{c_{140}}{m^3} \frac{ \ddot{q}^5}{\dot{q}^{11}} + \frac{ c_{141} + c_{24}}{m^3} \frac{ \ddot{q}^3\dot{\ddot{q}}}{\dot{q}^{10}} + \frac{c_{142}}{m^3} \frac{ \ddot{q} \dot{\ddot{q}} ^2}{\dot{q}^9} ) \notag \\  
& \ \ \ \ \ \ \ \ \ \ \ \ \ \ \ \ \ \ + \hbar ^5 ( \frac{c_{150}}{m^4} \frac{\ddot{q} ^6}{\dot{q}^{14}} + \frac{ c_{151} + c_{25}}{m^4} \frac{ \ddot{q}^4 \dot{\ddot{q}}}{\dot{q}^{13}} + 
\frac{c_{152}}{m^4} \frac{ \ddot{q}^2 \dot{\ddot{q}}^2}{\dot{q}^{12}})  + \cdots \Big) dt   \ .           \label{nov0714-1} 
\end{align}

The $p^{NW}$ (\ref{nov0314-1}) has more adjustable parameters than $p^{EL}$ (\ref{jun1211-3}) has. 
For example, the $\hbar ^4$-term of $p^{EL}$ has one adjustable parameter $c_4$, while 
$p^{NW}$ has four ones ($c_{140}$, $c_{141}$, $c_{142}$ and $c_{24}$).

\newpage

\section{Behavior of Solutions of (\ref{feb1115-5}) and (\ref{feb1115-6})} \label{apr1615-1}

We examine behavior of solutions of (\ref{feb1115-6}) and (\ref{feb1115-5}).

First, we show that solutions $r \textsf{R} ^C$ and $r \textsf{R} ^D $ of (\ref{feb1115-6}) which behave like (\ref{may1515-5}) exist only for discrete values of $E$. 
We assume in (\ref{feb1115-6}) that $ r^2 V(r) \to 0$ \cite{Landau(1977)} for $r \to 0$ and $0< const. \leq V(r) -E \leq + \infty $ for large $r$.

\noindent 
\textit{Case 1)} $l(l+1) > 0$: We have $V(r) + \hbar ^2 l(l+1)/2m r^2  -E \geq const. >0$ for both $r \to 0$ and $r \to \infty$. Then, there exists a solution $r \textsf{R}$ which converges like $e^{-p(r) \ln r}$ for $r = +\infty $; if a solution converges for $r = +\infty $, another solution diverges like $e^{q(r) \ln r}$ for $r = +\infty $,  where $p(r) , q(r) >0$ is some function of $r$, which we see from an eq. $ - (d^2 y  / d x ^2) + k^2 y = 0 $ having two linearly independent solutions $y=e^{kx}$ and $y=e^{-kx}$. Likewise for $r \to 0$. 
A solution which converges for $r \to 0$ either diverges or converges for $r \to +\infty $. 
If a solution converges for both $r \to 0$ and $r \to \infty$, the other solution diverges to the opposite direction  in $r \to 0$ and $r \to \infty$ ($r \textsf{R} \to \pm \infty$ for $r \to 0$ and  $r\textsf{R} \to \mp \infty$ for $r \to \infty$ with corresponding double sign) because, in general, $y_2 (x) / y_1(x)$ constructed from two linearly independent solutions $y_1$ and $y_2$ of $d^2 y / dx^2 + Q(x) y =0$ is monotonic according to an eq. $ y_2 (x) = y_1 (x) \int ^x y_1 ^{-2} (q)dq \ \ \cdots (*1)$ obtained from  $ y_1 ^2 (d/dx)( y_2/y_1 ) = y_1 y_2 ' - y_1 ' y_2 = const.$ \cite{Arfken(2005)}  
Because a solution $r \textsf{R}$ of (\ref{feb1115-6}) converging for  $r \to 0$ converges for $r \to \infty$ at discrete values of $E$, \cite{Messiah(1961)}$^,$
\footnote{
\label{jan1016-1}Let $\psi ( x;E)$ be a solution of $\psi '' + (E-V(x)) \psi =0 \ \ \cdots (*1)$, where $\psi '' = \partial ^2\psi / \partial x^2$. 
Then, to an infinitesimal $\delta E$, an infinitesimal $\delta \psi$ corresponds in $\psi (x ; E + \delta E ) = \psi ( x; E ) + \delta \psi  ( x; E )$. 
Multiplying $(*1)$ with $ \psi (x ;E+ \delta E)$, and $\psi ''(x ;E+ \delta E) + (E + \delta E -V) \psi (x ;E+ \delta E) =0$ with $ \psi (x ;E )$, we have 
$ \psi ''(x ;E) \psi (x ;E+ \delta E) -  \psi (x ;E) \psi ''(x ;E+ \delta E) - \delta E \psi (x ;E)  \psi (x ;E+ \delta E) =0$. 
With a symbol $W ( \psi _1 ,\psi _2 ) = \psi _1' \psi _2 - \psi _2 '\psi _1 $, we write the eq. as $W' ( \psi (x ;E) ,  \psi (x ;E+ \delta E) ) - \delta E \psi ^2 ( x ; E ) =0$, which is transformed to a integral form 
$W( \psi (q ;E) \psi (q ;E+ \delta E) ) |_{q=c}^{q=x} = \delta E \int _c^x \psi ^2 (q;E ) dq \ \ \cdots (*2)$. 
While, we have $W( \psi (x ;E) ,  \psi (x ;E+ \delta E) ) = W( \psi (x ;E), \delta \psi (x ;E)) = - \psi ^2 \delta ( \psi ' /\psi ) \ \ \cdots (*3)$. From $(*2)$ and $(*3)$, we have 
$\delta ( \psi ' /\psi ) |_c^x = - \delta E \psi ^ {-2} \int _c^x \psi ^2 ( q; E) dq \ \  \cdots (*4)$. 

Assume $0 < const. \leq V(x) -E < + \infty$ for $x$ satisfying $x<a$ and $b <x$, where $a< 0 < b$. 
Then, two linearly independent solutions in $x<a$ are written as $e^{ p_+(x) x}$  and  $e^{ - p_- (x) x}$ , where $p_ \pm (x)>0$.  
Likewise, in $b<x$, they are written as $e^{ q_+(x) x}$ and $e^{ - q_- (x) x}$, where $q_\pm (x) >0$. 
Let a solution $\psi$ converging ($ \psi \to e^{ p_+ (x) x}$) for $ x \to -\infty$ be $\psi _a$, that converging ($ \psi \to e^{ -q _-(x) x}$) for  $ x \to +\infty$ be $\psi _b$. 
Then, say, at $x =0$, we have $\delta (\psi _a ' /\psi _a) / \delta E < 0$ and $\delta (\psi _b ' /\psi _b) / \delta E >0$  from $(*4)$. The $\psi _a $ and $\psi _b$ therefore cannot be parts of a single solution for continuous values of $E$. 
This footnote is a modified excerpt from chap.3 of Ref. \refcite{Messiah(1961)}. 
} 
the (\ref{may1515-5}) is satisfied for discrete values of $E$.

\noindent 
\textit{Case 2)} $l(l+1) =0 $:  The behavior of two solutions for $r = +\infty $ are the same as the $l(l+1) \neq 0 $ case. 
Near $r =0$, two linearly independent solutions are given as $( r \textsf{R})^C \simeq r$ and $( r \textsf{R})^D \simeq const. \neq 0$ by Frobenius method, \cite{Arfken(2005)} which assumes a solution $r \textsf{R}$ of (\ref{feb1115-6}) in a form: $r \textsf{R} = \sum a_s r^s$, then insert it into (\ref{feb1115-6}) to determine $a_s$. 
If a solution converging  ($r \textsf{R} \to 0$) for $r \to 0$     converges for $r = +\infty $, the (\ref{may1515-5}) is satisfied. Such a solution exists for discrete values of $E$ by the same reason as above. 
Thus, (\ref{may1515-5}) is satisfied for discrete values of $E$.

Second, we show that solutions $\sqrt{ \sin \theta } \ \Theta ^C$ and $\sqrt{ \sin \theta } \ \Theta ^D$ of (\ref{feb1115-5}) which behave like (\ref{may1515-1}) exist only for discrete values of $l$. 
In three cases below, $ \textsf{m} \in \mathbb{Z}$ is assumed.    

\noindent 
\textit{Case 1)} $|\textsf{m}| \leq |l| \in \mathbb{Z} $: 
We first solve (\ref{feb1115-5}). 
We transform (\ref{feb1115-5}) to 
$( 1-x^2) d^2 z / dx^2 - 2x d z / dx + ( l(l+1) - \textsf{m}^2 / ( 1-x^2) ) z =0 \ \ \cdots (*2)$ with $x = \cos \theta$ and $\Theta ( \theta) = z(x)$. 
A solution of $(*2)$ is constructed from a solution $y$ of $( 1-x^2) d^2 y / dx^2 - 2x d y / dx + l(l+1) y =0 \ \ \cdots (*3)$ as 
$ z= (1-x^2) ^{  |\textsf{m}| /2 } d^ {|\textsf{m}|} y  /dx^ { |\textsf{m}| }$.  
Two linearly independent solutions $y_1(x)$ and $y_2(x)$ of $(*3)$ are given by the Frobenius method as \cite{Arfken(2005)} 
$y_1(x) =F( -l/2, l/2+1/2;1/2;x^2)  $ and 
$y_2(x) = x F( 1/2 -l/2 , l/2 +1 ;3/2;x^2 ) $, 
where $F$ is hypergeometric function \cite{Olver(2010)} defined as $F(a,b;c;x ) = \sum _{s=0}^ \infty (a) _s (b)_s x^s/ (c)_s s! $, where $(a)_0 =1$, $(a)_s = a (a+1) \cdots (a+s-1)$, etc. 
Solutions of $(*2)$ therefore are given as 
$z_1 (x) = (1-x^2) ^{ |\textsf{m}| /2 } d^ {|\textsf{m}|} y_1 /dx^ { |\textsf{m}|} $ and 
$z_2 (x) = (1-x^2) ^{  |\textsf{m}| /2 } d^ {|\textsf{m}|} y_2 /dx^  {|\textsf{m}|} $, from which solutions of (\ref{feb1115-5}) are given as $ \Theta ^C ( \theta ) = z_1 ( \cos \theta )$ and $ \Theta ^D ( \theta ) = z_2 ( \cos \theta )$. 
For $|\textsf{m}| \leq |l| \in \mathbb{Z} $, $z_1$ and $z_2$ constitute two linearly independent solutions of $(*2)$---note, if $| \textsf{m} | > |l| \in \mathbb{Z}$, they do not because one of them vanishes. 
If $z_1$ is even: $z_1(x) = z_1 (-x)$, then $z_2$ is odd: $z_2 (x) = -z_2(-x)$, and vice versa, which is seen from forms of $y_1(x)$ and $y_2(x)$. 
For $l \in \mathbb{Z} $, either of  $z_1$ and $z_2$ is a finite series having finite value at $x= \pm 1$, which is seen from the form of the $F$. 
Assume $z_1$ has a finite value at $x= \pm 1$. Then, $\Theta ^C ( \theta )$ has a finite value at $\theta = 0 ,  \, \pi$. 
From $(*1)$, we see 
$\lim _{ \theta \to 0 , \, \, \pi}( \Theta ^D /\Theta ^C ) = \lim _ { \theta \to 0 , \, \, \pi } \ \textsf{k} \int ^\theta \sin ^{-1} \eta \cdot ( \Theta ^C )^{-2} ( \eta ) d \eta = \pm \infty$, where $\textsf{k} = const. \neq 0$. 
 With the eq. together with the even-odd property mentioned above, we see (\ref{may1515-1}) is satisfied. 
Likewise for the case that $z_2$ is a finite series.

\noindent 
\textit{Case 2)} $| \textsf{m} | > | l | \in \mathbb{Z}$: 
In this case, $\frac{ \textsf{m}^2 }{ \sin ^2 \theta }- \frac{ \cos ^2 \theta }{ 4 \sin ^2 \theta } > \frac{1}{2} + l(l+1)$ in (\ref{feb1115-5}). 
Two linearly independent solutions of (\ref{feb1115-5}) therefore is written as $ \sqrt{ \sin \theta } \ \Theta ^C= e^{ p( \theta ) \tan ( \theta -\pi /2)  }$ and $ \sqrt{ \sin \theta } \ \Theta ^D= e^{- q ( \theta ) \tan ( \theta -\pi /2) }$ for $0 < \theta < \pi$, where $p( \theta ) ,  q( \theta )   >0$ is some function of $\theta$. 
Because $ \Theta ^D /\Theta ^C = e^{  - ( q(\theta ) + p(\theta )) \tan ( \theta -\pi /2) }$, (\ref{may1515-1}) is not satisfied.

\noindent 
\textit{Case 3)} $l \not\in \mathbb{Z}$: We use representations: \cite{Olver(2010)} $z_1(x) = \textsf{P}_l^ \textsf{m} (x)$ and $z_2(x) = \textsf{P}_l^ \textsf{m} (-x)$ as two linearly independent solutions $z_1 (x) $ and $z_2 (x)$  of $(*2)$, where $ \textsf{P}_l^ \textsf{m} (x) = ( (1+x)/ (1-x) )^{  \textsf{m} /2}  F (l+1 , -l ; 1- \textsf{m} ; (1-x)/2  ) /\Gamma ( 1- \textsf{m})   $ in which $\Gamma$ is gamma function \cite{Olver(2010)} and $F$ is the hypergeometric function defined above. The $\textsf{P}_l^ \textsf{m} (x)$ stays within $\mathbb{R}$ in $-1 < x < 1$ even if $0 \geq 1- \textsf{m} \in \mathbb{Z}$ thanks to a formula $\Gamma (z) \Gamma (1-z ) = \pi / \sin \pi z \, $. \cite{Olver(2010)}
We have
\begin{align}
 &\lim _{\theta \to 0} \frac{ \Theta ^D  }{   \Theta ^C  }
= \lim _{x \to 1} \frac{z_2}{z_1} 
 = \lim _{x \to 1} \frac{ (1-x)^\textsf{m}   F (l+1 , -l ; 1- \textsf{m} ;(1+x)/2 )}{(1+x)^\textsf{m}  F (l+1 , -l ; 1- \textsf{m} ; (1-x)/2 )}   \notag \\ 
  &= \lim _{x \to 1} \frac{     F (l+1 , -l ; 1- \textsf{m} ;(1+x)/2 )}{2^ \textsf{m}(1-x)^{-\textsf{m}}}    \ , \ \ \ \ \ \ \ \ \ \ \ \ 
 \lim _{\theta \to \pi} \frac{ \Theta ^D  }{   \Theta ^C  } 
= \lim _{x \to -1} \frac{z_2}{z_1} 
= \Big( \lim _{x \to 1} \frac{z_2}{z_1} \Big)^{-1}      .  \notag 
\end{align}
Using formulas: \cite{Olver(2010)} 
\begin{align}
&\textrm{if} \   c>a+b  ,  F( a,b;c;1) = \frac{ \Gamma (c) \Gamma (c-a-b)}{ \Gamma ( c-a) \Gamma (c-b)}  ; \ 
\textrm{if} \ c=a+b ,  \lim _{x \to 1} \frac{ F ( a,b;c;x )}{ - \ln (1-x) } 
= \frac{ \Gamma (a+b)}{\Gamma (a) \Gamma (b) } ;                \notag \\ 
&\textrm{if} \ c<a+b  , \lim  _{x \to 1}    \frac{F(a,b;c;x )}{ (1-x)^ {c-a-b}} = \frac{ \Gamma (c) \Gamma ( a+b-c)}{ \Gamma (a) \Gamma (b) }  \ \ \ \ \ \ \ \ \ \ \ \ \ \ \  (a,b,c \in \mathbb{R})    \ ,        \notag 
\end{align}
\sloppy
which read for the present case as 
$F (l+1 , -l ; 1- \textsf{m} ; 1 ) \in \mathbb{R}$ for $0> \textsf{m} \in \mathbb{Z} $,   
 $\lim _{x \to 1} F( l+1 , -l ; 1- \textsf{m}; (1+x)/2  )/ \ln ((1-x)/2) \in \mathbb{R}$ for $\textsf{m} =0$, and  $\lim _{x \to 1} F (l+1 , -l ; 1- \textsf{m} ; (1+x)/2  ) / ((1 -x)/2 )^{ - \textsf{m} } = \{ \pm \infty \}$ for $0< \textsf{m} \in \mathbb{Z}$, 
we see $\lim _{x \to 1} (z_2/z_1) = \{ \pm \infty \}$ for $ \textsf{m} \in \mathbb{Z} $.  The (\ref{may1515-1}) therefore is not satisfied. 
Thus, the (\ref{may1515-1}) is satisfied only for $l$ satisfying $| \textsf{m} | \leq |l|  \in \mathbb{Z}$.

\fussy

\newpage

\section{
Determination of $ \Psi _{\textrm{diff}}$ and $\nabla \mathcal{S}_0^{HJ}$ behind the electron biprism} \label{nov1608-4}

We determine, in the experimental setting of \S \ref{jan2107-6},  $ \Psi _{\textrm{diff}}$ and $ \textbf{p}^{HJ} = \nabla \mathcal{S}_0^{HJ}$ (see \S \ref{jan2107-6}) behind the electron biprism caused by the incoming Gaussian beam $\Psi _G$ of (\ref{may0809-e5}).

  We use three coordinate systems $(x,y,z) $,  $(\tilde{x} , \tilde{y}, \tilde{z})$, and $( \tilde{\tilde{x}} , \tilde{\tilde{y}} , \tilde{\tilde{z}} )$ related with (see Fig.\ref{fig-coordinates})
\begin{equation}
\tilde{x} = \frac{ k_z x - k_x z}{ \sqrt{ k_x ^2 + k_z ^2}}  , \   \tilde{y} =y  , \  
\tilde{z} = \frac{ k_x x + k_z z}{ \sqrt{ k_x ^2 + k_z ^2}}  ,  \ \ \ \  
\tilde{\tilde{x}} = \frac{ k_z x + k_x z}{ \sqrt{ k_x ^2 + k_z ^2}}  , \    \tilde{\tilde{y}} =y  , \   
\tilde{\tilde{z}} = \frac{ -k_x x + k_z z}{ \sqrt{ k_x ^2 + k_z ^2}} \ .  \label{may1807-6}
\end{equation}
We consider two Gaussian beams $ \tilde{\Psi}_G $ and   $\tilde{\tilde{\Psi}}_G $. The $ \tilde{\Psi}_G $ ($\tilde{\tilde{\Psi}}_G $)  propagates parallel to the $\tilde{z}$ ($\tilde{\tilde{z}} $)  axis toward the $\tilde{z}+$ ($\tilde{\tilde{z}} + $) direction.
 The $ \tilde{\Psi}_G $ is 
diffracted by  the knife-edge in the $(\tilde{x}, \tilde{y})$ plane  at $\tilde{z}=0$  open downward from  $\tilde{x} =- \textsf{d} /2 $ to $ -\infty$, where $\textsf{d} $ is the diameter of the filament. The $\tilde{\tilde{\Psi}}_G$ is diffracted by  the knife-edge 
 open upward from   $ \tilde{\tilde{x}} =  + \textsf{d} /2 $ to $+\infty$.

The disturbance $\tilde{U}(P)$ of  $ \tilde{\Psi}_G $ at the point $P = (\tilde{X},\tilde{Y},\tilde{Z})$ behind the downward open knife-edge, where $(\tilde{X},\tilde{Y},\tilde{Z})$ are coordinate values on the $(\tilde{x},\tilde{y},\tilde{z})$-system, is given as \cite{Born(1999)}
\begin{equation}
\tilde{U}(P) \simeq - \frac{iA  }{ \lambda \tilde{s} \ '} \int_{- \infty} ^{- \textsf{d} /2} d \tilde{\xi} \int _ {- \infty}^{+ \infty}  d \tilde{\eta}      \exp ( -\frac{(\tilde{\xi} - \tilde{x}_0)^2 +(\tilde{\eta} - \tilde{y}_0)^2}{w_0^2} +ik \tilde{s} ) \ ,  \label{may1507-5}
\end{equation}
where $A$: amplitude of the incoming beam,  $\lambda (= 2 \pi /k )$: wave length, $(\tilde{\xi} , \tilde{\eta} )$: coordinates on the $(\tilde{x},\tilde{y})$ plane at $\tilde{z} =0$, 
$\tilde{s} = \surd ( ( \tilde{X} -\tilde{\xi} )^2 + ( \tilde{Y}- \tilde{\eta} )^2 + \tilde{Z}^2 )$, and 
 $\tilde{s   } \ ' = \surd ( \tilde{X} ^2 +  \tilde{Y}^2 + \tilde{Z}^2 ) $.

\sloppy 

We rewrite (\ref{may1507-5}) with Fresnel approximation. \cite{Born(1999)} 
We expand $\tilde{s}$ with power series of $\tilde{\xi}$ and $\tilde{\eta} $ (remind $\surd ( 1+x ) = 1 + x/2 -x^2/8 + \cdots $). 
Then, we take terms of the expansion up to second order
\footnote{
Because (tildes omitted)
\begin{equation*}
 \sqrt{ Z^2 + (X -\xi) ^2  } = s' - \xi l + \frac{ \xi ^2 ( 1-l^2)}{ 2s' } + \frac{ \xi ^3 l( 1 - l^2) }{ 2s'^2} + \frac{\xi ^4}{ s'^3} ( - \frac{1}{8} + \frac{3 l^2}{4} - \frac{5l^4}{8} ) + \cdots \ , 
\end{equation*}
where  $s'  = \surd ( X^2 + Z^2 )$, if 
$ k| \, \xi ^3 l(1 - l^2) / 2s'^2 + \xi ^4 ( \cdots )/ s'^3 + \cdots | \ll 2 \pi$  is satisfied at $P$, the Fresnel approximation is  effective at the $P$. 
For $k= 1.45 \times 10^9 mm ^{-1} $, $ | \xi | \leq 2.2 \sqrt{3} \, \mu m $, $X/Z \leq 0.05$, and a criterion $k | \cdots | \leq 2 \pi \times 0.05$, the Fresnel approximation is  effective for $2.5 mm \lesssim Z$.
}
 to have 
\begin{equation*}
\tilde{s} \simeq \tilde{s} \ ' - (\tilde{l} \tilde{\xi} + \tilde{m} \tilde{\eta} ) + \frac{ (1- \tilde{l}^{ \ 2} ) \tilde{\xi} ^2 + (1-\tilde{m}^2 ) \tilde{\eta} ^2 - 2 \tilde{l} \tilde{m} \tilde{\xi}  \tilde{\eta}  }{ 2\tilde{s} \ ' }  \ , 
\end{equation*}
where $\tilde{l}  = \tilde{X}/\tilde{s} \ ' , \ \tilde{m} = \tilde{Y} /\tilde{s} \ '$. 
We consider only the case $\tilde{m} = \tilde{Y}/ \tilde{s} \ ' = 0$. 
In this case, 
(\ref{may1507-5}) is written as
\begin{multline}
\tilde{U}(P)  
\simeq  - \frac{iA}{ \lambda  \tilde{s} \ '}  \exp ik \tilde{s} \ ' \int   _{-\infty} ^{-\textsf{d} /2}    d \tilde{\xi}  \int  _{- \infty }^{ + \infty }  d \tilde{\eta}       
 \exp ( -\frac{ ( \tilde{\xi} -\tilde{x}_0)^2}{ w_0^2} + ik( -\tilde{l} \tilde{\xi}+ \frac{ (1-\tilde{l} \ ^2) \tilde{\xi} ^2 }{ 2 \tilde{s} \ '} ))    \\ 
  \times \exp ( - \frac{ ( \tilde{\eta} -\tilde{y}_0) ^2}{ w_0^2} + ik \frac{\tilde{ \eta} ^2}{ 2 \tilde{s} \ '} )  \ .   \label{may1807-3}
\end{multline}
Setting  $ \tilde{\xi} = \tilde{ \zeta } / \sqrt{\tilde{\alpha}} + k \tilde{l}  / (\pi \tilde{\alpha}) $, where $\tilde{\alpha} =  k ( 1-\tilde{l}^2) /(\pi \tilde{ s} \ ')$, we rewrite the real and imaginary parts of the $\tilde{\xi}$-direction integral as 
\begin{align}
& \textrm{Re part:} \ \ \ \  \int   _{- \infty } ^{ - \textsf{d} /2 }     d \tilde{\xi}  \exp ( -\frac{ ( \tilde{\xi} -\tilde{x}_0)^2}{ w_0^2} ) \cos  k( - \tilde{l} \tilde{\xi} + \frac{ (1-\tilde{l} ^2) \tilde{\xi} ^2 }{ 2 \tilde{s} \ '}  )     \notag  \\  
&= \frac{1}{ \sqrt{\tilde{\alpha}}} \int  _{- \infty }^ { -\frac{ \textsf{d}  \sqrt{\tilde{\alpha}}}{2} - \frac{ k \tilde{l} }{ \pi \sqrt{   \tilde{\alpha}   }} } d \tilde{ \zeta }    \exp ( - \frac{ ( \tilde{ \zeta } +  \frac{ k \tilde{l} }{\pi  \sqrt{\tilde{\alpha}}} - \tilde{x}_0 \sqrt{\alpha})^2}{ \tilde{\alpha} w_0^2 }) \cos ( \frac{\pi \tilde{ \zeta } ^2 }{2} - \frac{ k^2 \tilde{l}^{ \ 2}  }{2\pi \tilde{\alpha}} )           \notag \\ 
&=  \cos (  \frac{ k^2 \tilde{l}^{ \ 2} }{2\pi \tilde{\alpha}} ) \cdot \frac{1}{ \sqrt{\tilde{\alpha}}} \int  _{- \infty }^ { -\frac{ \textsf{d}  \sqrt{\tilde{\alpha}}}{2} - \frac{ k \tilde{l} }{ \pi \sqrt{   \tilde{\alpha}   }} } d \tilde{ \zeta }   \exp ( - \frac{ ( \tilde{ \zeta } +  \frac{k \tilde{l} }{\pi \sqrt{\tilde{\alpha}}} - \tilde{x}_0 \sqrt{\tilde{\alpha}})^2}{ \tilde{\alpha} w_0^2 }) \cos   \frac{\pi \tilde{ \zeta } ^2  }{2}          \notag \\ 
&+ \sin (  \frac{ k^2 \tilde{l}^{ \ 2}  }{2 \pi \tilde{\alpha}} ) \cdot \frac{1}{ \sqrt{\tilde{\alpha}}} \int  _{- \infty }^ { -\frac{ \textsf{d}  \sqrt{\tilde{\alpha}}}{2} - \frac{k\tilde{l} }{ \pi \sqrt{   \tilde{\alpha}   }} }  d \tilde{ \zeta }   \exp ( - \frac{ (\tilde{ \zeta } +  \frac{k \tilde{l} }{\pi \sqrt{\tilde{\alpha}}} - \tilde{x}_0 \sqrt{\tilde{\alpha}})^2}{ \tilde{\alpha} w_0^2 }) \sin \frac{\pi \tilde{ \zeta } ^2  }{2}         \notag  \\ 
&= \cos (  \frac{ k^2 \tilde{l}^{ \ 2}  }{2\pi \tilde{\alpha}} )   \cdot \frac{1}{ \sqrt{\tilde{\alpha}}}   \textsf{M}_1 +  \sin (  \frac{k^2\tilde{l}^{ \ 2} }{2\pi \tilde{\alpha}} )  \cdot \frac{1}{ \sqrt{\tilde{\alpha}}} \textsf{N}_1  \ ,  \notag 
\end{align} 
\begin{align}
& \textrm{Im part:} \ \ \ \ \int  _{- \infty } ^{ -\textsf{d} /2 }  d  \tilde{\xi}  \exp ( -\frac{ ( \tilde{\xi} -\tilde{x}_0)^2}{ w_0^2} ) \sin  k( -\tilde{l} \tilde{\xi} + \frac{ (1-\tilde{l}^2) \tilde{\xi} ^2 }{ 2 \tilde{s} \ '} )   \notag \\ 
  &= \frac{1}{ \sqrt{\tilde{\alpha}}} \int _{- \infty }^ { -\frac{ \textsf{d} \sqrt{\tilde{\alpha}}}{2} - \frac{ k \tilde{l} }{ \pi \sqrt{   \tilde{\alpha}   }} }  d \tilde{ \zeta }   \exp ( - \frac{ ( \tilde{ \zeta } +  \frac{k \tilde{l}}{2 \sqrt{\tilde{\alpha}}} - \tilde{x}_0 \sqrt{\tilde{\alpha}})^2}{ \tilde{\alpha} w_0^2 }) \sin (   \frac{\pi \tilde{ \zeta } ^2  }{2}   - \frac{ k^2\tilde{l}^{ \ 2} }{2\pi \tilde{\alpha}} )           \notag \\ 
 &= \cos (  \frac{k^2\tilde{l}^ { \ 2}  }{2\pi \tilde{\alpha}} )   \cdot \frac{1}{ \sqrt{\tilde{\alpha}}  }  \textsf{N} _1 - \sin (  \frac{k^2\tilde{l}^{ \ 2}  }{2\pi \tilde{\alpha}} )  \cdot \frac{1}{ \sqrt{\tilde{\alpha}} } \textsf{M}_1  \ , \notag 
\end{align} 
where $\textsf{M}_1$ and $ \textsf{N}_1  $ are
\begin{align}
\textsf{M}_1 &=  \int _{- \infty }^ { -\frac{ \textsf{d}  \sqrt{\tilde{\alpha}}}{2} - \frac{ k \tilde{l} }{ \pi \sqrt{   \tilde{\alpha}   }} }  d \tilde{ \zeta}    \exp ( - \frac{ ( \tilde{ \zeta} +  \frac{k \tilde{l} }{\pi \sqrt{\tilde{\alpha}}} - \tilde{x}_0 \sqrt{\tilde{\alpha}})^2}{ \tilde{\alpha} w_0^2 }) \cos   \frac{\pi \tilde{ \zeta } ^2  }{2}  \notag  \\  
& \ \ \ \  \ \ \ \  \ \ \ \  \ \ \ \ \ 
\simeq \int _{- \infty }^ { -\sqrt{ \frac{k}{\pi \tilde{Z}}} ( \frac{ \textsf{d}}{2} + \tilde{X}) } 
 d \tilde{ \zeta}    \exp ( - \frac{ ( \tilde{ \zeta} + \sqrt{ \frac{k}{\pi \tilde{Z}}}(\tilde{X} - \tilde{x}_0)   )   ^2}{ \frac{k w_0^2}{\pi \tilde{Z} }}) \cos   \frac{\pi \tilde{ \zeta } ^2  }{2}  \ , \notag  
\end{align}
\begin{align} 
\textsf{N}_1 &=  \int  _{- \infty }^ { -\frac{ \textsf{d}  \sqrt{\tilde{\alpha}}}{2} - \frac{ k \tilde{l} }{ \pi \sqrt{   \tilde{\alpha}   }} }  d \tilde{ \zeta }    \exp ( - \frac{ ( \tilde{ \zeta} +  \frac{k \tilde{l} }{ \pi \sqrt{\tilde{\alpha}}} - \tilde{x}_0 \sqrt{\tilde{\alpha}})^2}{ \tilde{\alpha} w_0^2 }) \sin  \frac{\pi \tilde{ \zeta } ^2  }{2} \notag     \\  
& \ \ \ \  \ \ \ \  \ \ \ \  \ \ \ \ \ 
\simeq  \int  _{- \infty }^  { -\sqrt{ \frac{k}{\pi \tilde{Z}}} ( \frac{ \textsf{d}}{2} + \tilde{X}) }  d \tilde{ \zeta }    \exp ( - \frac{ ( \tilde{ \zeta} + \sqrt{ \frac{k}{\pi \tilde{Z}}}(\tilde{X} - \tilde{x}_0)   )   ^2}{ \frac{k w_0^2}{\pi \tilde{Z} }}) \sin  \frac{\pi \tilde{ \zeta } ^2  }{2}  \ . \notag 
 \end{align}
Likewise, setting  $ \tilde{\eta }= \tilde{\rho } \sqrt{ \pi  \tilde{s} \ ' /k }  $, we write the real and imaginary parts of the $\tilde{\eta} $-direction integral as
\begin{align}
\textrm{Re part:} \ \ \ \  \int  _{- \infty }^{ + \infty }  d \tilde{\eta}  & \exp ( - \frac{ ( \tilde{\eta}  -\tilde{y}_0) ^2}{ w_0^2}) \cos  \frac{ k \tilde{\eta}  ^2}{ 2\tilde{s} \ '}   \notag   \\
&= \sqrt{\frac{\pi \tilde{s} \ '}{k}} \int  _{- \infty }^{ + \infty }  d \tilde{\rho } \exp ( - \frac{( \tilde{\rho } -  \frac{\tilde{y}_0}{ \sqrt{ \pi \tilde{s} \ '/k} }  )^2}{ \frac{ kw_0^2}{ \pi \tilde{s} \ '}}) \cos \frac{ \pi \tilde{ \rho }^2 }{2} = \sqrt{\frac{ \pi \tilde{s} \ '}{k}} \ \textsf{Q} \ ,  \notag  
\end{align} 
\begin{align}
\textrm{Im part:} \ \ \ \   \int  _{- \infty }^{ + \infty }  d \tilde{\eta}  & \exp ( - \frac{ ( \tilde{\eta}  -\tilde{y}_0) ^2}{ w_0^2}) \sin  \frac{k \tilde{\eta}  ^2}{ 2 \tilde{s} \ '}   \notag  \\  
&= \sqrt{\frac{\pi \tilde{s} \ '}{k}} \int  _{- \infty }^{ + \infty }  d \tilde{\rho } \exp ( - \frac{( \tilde{\rho } - \frac{\tilde{y}_0}{ \sqrt{\pi   \tilde{s} \ '/k} } )^2}{ \frac{ kw_0^2}{ \pi \tilde{s} \ '}}) \sin \frac{\pi \tilde{ \rho }^2 }{2} = \sqrt{\frac{ \pi \tilde{s} \ '}{k}} \ \textsf{R} \ ,    \notag 
\end{align}
in which $\textsf{Q}$ and $\textsf{R} $ represent the integrals of the left sides. 
Thus, noting $- \frac{ i A}{ \lambda \tilde{s} \ '} \frac{1}{\sqrt{\tilde{\alpha}}} \sqrt{\frac{ \pi \tilde{s} \ '}{k}} (\textsf{Q} +i \textsf{R} ) = \frac{- i A  (\textsf{Q} +i \textsf{R} ) }{ 2 \sqrt{ 1-\tilde{l}^2}}  $, we write (\ref{may1807-3})  as
\begin{align}
\tilde{U}(P) &=    \frac{- i A  (\textsf{Q} +i \textsf{R} ) }{ 2 \surd ( 1-\tilde{l}^2 )} e^{ik\tilde{s} \ '}       \notag  \\  
\times 
&\Big( (\cos (  \frac{k^2\tilde{l}^{ \ 2}  }{ 2 \pi \tilde{\alpha} } )    \cdot   \textsf{M}_1 +  \sin (  \frac{k^2\tilde{l}^{ \ 2}  }{ 2 \pi \tilde{\alpha} } )  \cdot  \textsf{N}_1 ) +i( \cos (  \frac{k^2\tilde{l}^{ \ 2}  }{ 2 \pi \tilde{\alpha} } )    \cdot   \textsf{N} _1- \sin (  \frac{k^2\tilde{l}^{ \ 2}  }{ 2 \pi \tilde{\alpha} } )   \cdot \textsf{M} _1 ) \Big)  \ .  \notag    
\end{align}

\fussy

In a same manner, $\tilde{ \tilde{U}}(P) $ is obtained as 
\begin{align}
\tilde{ \tilde{U}}(P) &=  \frac{- i A  (\textsf{Q} +i \textsf{R} ) }{ 2 \surd ( 1-\tilde{\tilde{l}}^2 )} e^{ik \tilde{\tilde{s}} \ '}  \notag   \\  
\times 
&\Big( (\cos (  \frac{ k ^2 \tilde{\tilde{l}}^{ \ 2}  }{ 2 \pi \tilde{\tilde{\alpha}} } )   \cdot   \textsf{M}_2 +  \sin (  \frac{ k  ^2 \tilde{\tilde{l}}^{ \ 2}  }{ 2 \pi \tilde{\tilde{\alpha}} } )   \cdot  \textsf{N}_2 ) +i( \cos (  \frac{ k ^2 \tilde{\tilde{l}}^{ \ 2}  }{ 2 \pi \tilde{\tilde{\alpha}} } )    \cdot   \textsf{N} _2- \sin (  \frac{ k ^2 \tilde{\tilde{l}}^{ \ 2}  }{ 2 \pi \tilde{\tilde{\alpha}} } )   \cdot \textsf{M} _2 ) \Big)  \ , \notag 
\end{align}
where
\begin{align}
\textsf{M}_2 =  \int _{ \frac{ d \sqrt{ \tilde{\tilde{\alpha}} }}{2} - \frac{   k \tilde{\tilde{l}}    }{ \pi \sqrt{ \tilde{\tilde{\alpha}} }} } ^ {+\infty }  d \tilde{\tilde{ \zeta }}  &  \exp ( - \frac{ ( \tilde{\tilde{ \zeta }}  +  \frac{ k\tilde{\tilde{l}} }{\pi \sqrt{ \tilde{\tilde{\alpha}} }} - \tilde{\tilde{ x}} _0 \sqrt{ \tilde{\tilde{\alpha}} })^2}{  \tilde{\tilde{\alpha}}  w_0^2 }) \cos  \frac{  \pi \tilde{\tilde{ \zeta }} ^2 }{2} \notag   \\ 
 & \simeq \int _{ \sqrt{ \frac{k}{\pi \tilde{\tilde{Z}}}} ( \frac{ \textsf{d}}{2} - \tilde{\tilde{X}}) } ^ {+\infty }  d \tilde{\tilde{ \zeta }}    \exp ( - \frac{ ( \tilde{  \tilde{ \zeta}} + \sqrt{ \frac{k}{\pi \tilde{\tilde{Z}}}}(\tilde{\tilde{X}} - \tilde{ \tilde{x}}_0)   )   ^2}{ \frac{k w_0^2}{\pi \tilde{\tilde{Z} } }})  \cos  \frac{  \pi \tilde{\tilde{ \zeta }} ^2 }{2} \ , \notag 
\end{align}
\begin{align}
\textsf{N}_2  =  \int  _{ \frac{ d \sqrt{ \tilde{\tilde{\alpha}} }}{2} - \frac{  k\tilde{\tilde{l}}    }{ \pi \sqrt{ \tilde{\tilde{\alpha}} }} } ^ {+\infty }  d \tilde{\tilde{ \zeta }}   &  \exp ( - \frac{ ( \tilde{\tilde{ \zeta }}  +  \frac{ k \tilde{\tilde{l}} }{\pi \sqrt{ \tilde{\tilde{\alpha}} }} -  \tilde{\tilde{ x}} _0 \sqrt{ \tilde{\tilde{\alpha}} })^2}{  \tilde{\tilde{\alpha}}  w_0^2 }) \sin  \frac{  \pi \tilde{\tilde{ \zeta }} ^2 }{2}   \notag  \\  
& \simeq    \int  _{ \sqrt{ \frac{k}{\pi \tilde{\tilde{Z}}}} ( \frac{ \textsf{d}}{2} - \tilde{\tilde{X}}) }^ {+\infty }  d \tilde{\tilde{ \zeta }}     \exp ( - \frac{ ( \tilde{  \tilde{ \zeta}} + \sqrt{ \frac{k}{\pi \tilde{\tilde{Z}}}}(\tilde{\tilde{X}} - \tilde{ \tilde{x}}_0)   )   ^2}{ \frac{k w_0^2}{\pi \tilde{\tilde{Z} } }}) \sin  \frac{  \pi \tilde{\tilde{ \zeta }} ^2 }{2} \ . \notag  
\end{align}

The diffracted field $ \Psi _{\textrm{diff}}$  at $P$ behind the biprism is the superposition of $\tilde{U}$ and $\tilde{ \tilde{U}}$: 
\begin{multline}
\Psi _{\textrm{diff}} = \tilde{U}(P)  + \tilde{ \tilde{U}}(P) \\ 
= -  \frac{i A  (\textsf{Q} +i \textsf{R} ) }{ 2 \surd ( 1-\tilde{l}^2 )}  (\textsf{M}_1 + i \textsf{N}_1) e^{i(  k \tilde{s} \,\, ' -\tilde{ \beta}) } 
-  \frac{ i A  (\textsf{Q} +i \textsf{R} ) }{ 2 \surd ( 1-\tilde{\tilde{l}}^2)}(\textsf{M}_2 + i \textsf{N}_2) e^{i (  k \tilde{\tilde{s}} \,\, ' -  \tilde{ \tilde{ \beta}})}  \ , \label{jul0508-1}
\end{multline}
where $ \tilde{\beta }  =  (  k    ^2 \tilde{l}^{ \ 2} )/( 2 \pi \tilde{\alpha} ) $ and $\tilde{  \tilde{\beta }  }=  (  k ^2 \tilde{\tilde{l}}^{ \ 2}  )/(2 \pi \tilde{\tilde{\alpha}} ) $.

We obtain the momentum field $\textbf{p}^{HJ} = (p_x ,p_y , p_z ) $ at the $P$ inserting (\ref{jul0508-1}) into $\psi$ of (\ref{may0609-1}). 
We set $y$-coordinate of the incoming beam center to zero ($y_0 =0$). 
Accordingly, $p_y=0$ on $y=0$ plane only on which we consider the $\textbf{p}^{HJ}$.  
The $z$-direction momentum $\hbar k_z$ of the incoming beam is kept almost intact through the biprism  because $k_x \ll k_z$. 
Accordingly, $p_z \simeq \hbar k_z^{\textrm{in}}$, where $k_z^{\textrm{in}}$ is the wave number of the incoming beam. 
The $p_x$ is given as 
$p_x =    \hbar ( \bar{ \Psi } _{\textrm{diff}} \partial _x   \Psi _{\textrm{diff}} - \Psi _{\textrm{diff}}  \partial _x  \bar{ \Psi } _{\textrm{diff}}) /   2 i \, \Psi _{\textrm{diff}} \bar{\Psi} _{\textrm{diff}} $. 
Exploiting 
 $  \partial_ x ( k \tilde{s} \, ' - \tilde{ \beta} ) \simeq k_x $,   
$ \partial_ x ( k \tilde{\tilde{s}} \, ' - \tilde{\tilde{\beta}} ) \simeq - k_x $ 
\footnote{
The $ \partial_x ( k \tilde{s} \, ' - \tilde{ \beta} ) \simeq k_x $ follows from $k = \surd ( k_x^2 + k_z ^2 )$, 
$ \tilde{s} \, ' = \surd ( \tilde{x} ^2 + \tilde{z}^2 )$, 
$\partial / \partial x  = ( \partial \tilde{x} / \partial x ) ( \partial  / \partial \tilde{x} ) + ( \partial \tilde{z} / \partial x ) ( \partial / \partial \tilde{z} )$, and 
$ k \tilde{s} \, ' - \tilde{ \beta} =  k \tilde{s} \, ' ( 1- \tilde{x} ^2 / 2\tilde{z} ^2) )$. 
Likewise for $ \partial_ x ( k \tilde{\tilde{s}} \, ' - \tilde{\tilde{\beta}} ) \simeq - k_x $. 
} 
and 
$\tilde{ \beta } - \tilde{ \tilde{ \beta }  } \simeq  - 2k_x x $, 
we express the denominator and numerator of the $p_x$-formula as 
\begin{subequations}
\begin{align}
&\Psi _{\textrm{diff}} \bar{\Psi} _{\textrm{diff}}    
= \textsf{M}_1^2 + \textsf{N}_1 ^2 + \textsf{M}_2^2 + \textsf{N}_2^2 \notag \\ 
& \ \ \ \  \ \ \ \  \ \ \ \  \ \ \ \  
 + 2( \textsf{M}_1 \textsf{M}_2 +  \textsf{N}_1 \textsf{N}_2) \cos  2k_x x  
 + 2(  \textsf{M}_1 \textsf{N}_2 - \textsf{N}_1 \textsf{M}_2 ) \sin  2k_x x \ ,    \label{may2608-e4}     \\ 
&\frac{ \bar{ \Psi }_{\textrm{diff}}  \partial _x  \Psi _{\textrm{diff}} - \Psi _{\textrm{diff}}  \partial _x \bar{ \Psi }_{\textrm{diff}} }{2i} =  \textrm{Im} (  \bar{ \Psi } _{\textrm{diff}} \partial _x  \Psi _{\textrm{diff}} )      \notag \\  
&=   k_x  (\textsf{M}_1^2 + \textsf{N}_1^2)  -  k_x(\textsf{M}_2 ^2 +  \textsf{N}_2^2 ) +  \textsf{M}_1 \partial _x \textsf{N}_1 -  \textsf{N}_1 \partial _x \textsf{M}_1 + \textsf{M}_2  \partial _x \textsf{N}_2 -  \textsf{N}_2  \partial _x \textsf{M}_2    \notag \\  
&  + \big(- \textsf{N}_1 \partial _x \textsf{M}_2  +\textsf{M}_1 \partial _x \textsf{N}_2      -\textsf{N}_2 \partial _x \textsf{M}_1  +\textsf{M}_2 \partial _x \textsf{N}_1  \big) \cos 2 k_x x  \notag \\  
&  + \big( - \textsf{M}_1 \partial _x \textsf{M}_2  - \textsf{N}_1 \partial _x \textsf{N}_2   
+  \textsf{M}_2 \partial _x \textsf{M}_1  +\textsf{N}_2 \partial _x \textsf{N}_1 \big) \sin  2 k_x x     \ , \label{may2608-e5}  
\end{align} \label{aug1411-1}
\end{subequations}
where we ignored $ - iA (  \textsf{Q} + i \textsf{R}  ) / 2 \surd (  1 -\tilde{l} ^2 ) $ and  $ - iA (  \textsf{Q} + i \textsf{R}  ) / 2 \surd ( 1 -\tilde{\tilde{l}} ^2 ) $ in $\Psi _{\textrm{diff}}$ because $\textsf{Q} , \textsf{R} \simeq const.$ for $z= const.$
\footnote{
Influence of transverse profile of the incoming beam to values of $\textsf{Q}$ and $\textsf{R}$ however is unignorable. 
Indeed, we have $\textsf{Q} = 1.13$ and $\textsf{R} = 0.76$ for $ kw_0^2 / \pi  \tilde{s} \ ' = 1.58$ obtained from $k=1.45 \times 10^6 \mu m ^{-1}$,  $\tilde{s} \ ' = 33770 \mu m$, and $w_0 = 0.34 \mu m$---these values are used in \S \ref{jan2107-6}---, while we have $\textsf{Q}= \textsf{R} = 1.00$ for $w_0 = + \infty$. 
} 
and $\surd ( 1 -\tilde{l} ^2 ) \simeq \surd ( 1 -\tilde{\tilde{l}} ^2 ) \simeq 1$; 
we express the $\textsf{M}_1,   \textsf{N}_1, \textsf{M}_2$, and  $\textsf{N}_2$ at $P = ( X,0,Z)$ on the coordinate system $(x,y,z)$, ignoring terms of second and higher powers of $X/Z$, as
\begin{subequations}
\begin{align}
&\textsf{M}_1 |_{(X,0,Z)}    
=  \int _{- \infty }^ { \sqrt{ \frac{k}{ \pi Z }} ( -\frac{ \textsf{d}  }{2} - X + \frac{k_x }{k_z} Z)  }    \exp \big( - \frac{ \big(  \tau + \sqrt{ \frac{k}{ \pi Z }} (X -  \frac{k_x }{k_z} Z - x_0 ) \big)^2}{ \frac{ k w_0^2}{ \pi Z} } \big) \cos   \frac{\pi  \tau  ^2  }{2}     d  \tau   \label{nov2517-1}  \\ 
&\textsf{N}_1 |_{(X,0,Z)}    
=  \int _{- \infty }^ { \sqrt{ \frac{k}{ \pi Z }} ( -\frac{ \textsf{d}  }{2} - X + \frac{k_x }{k_z} Z)  }     \exp \big( - \frac{ \big( \tau + \sqrt{ \frac{k}{ \pi Z }} (X -  \frac{k_x }{k_z} Z - x_0 ) \big)^2}{ \frac{ k w_0^2}{ \pi Z} } \big) \sin  \frac{\pi  \tau  ^2  }{2}    d  \tau    \ .   \\ 
&\textsf{M}_2  |_{(X,0,Z)}  
=  \int _{ \sqrt{ \frac{k}{ \pi Z }} ( + \frac{ \textsf{d}  }{2} - X -\frac{k_x }{k_z} Z)  } ^ {+\infty }   \exp  \big( - \frac{ \big(  \tau + \sqrt{ \frac{k}{ \pi Z }} (X +  \frac{k_x }{k_z} Z - x_0 ) \big)^2}{ \frac{ k w_0^2}{ \pi Z} } \big)  \cos  \frac{  \pi  \tau  ^2 }{2}   d \tau        \\   
&\textsf{N}_2  |_{(X,0,Z)}   
=  \int  _{ \sqrt{ \frac{k}{ \pi Z }} ( + \frac{ \textsf{d}  }{2} - X - \frac{k_x }{k_z} Z)  }^ {+\infty }    \exp  \big( - \frac{ \big( \tau + \sqrt{ \frac{k}{ \pi Z }} (X +  \frac{k_x }{k_z} Z -  x_0 ) \big)^2}{ \frac{ k w_0^2}{ \pi Z} } \big)  \sin  \frac{  \pi  \tau  ^2 }{2}   d  \tau   \ .       
\end{align} \label{may1409-rr1} 
\end{subequations}

\newpage

\section{Electron density on the screen when the one in the incoming beam is  $| \Psi _s |^2$} \label{sep0811-1}

We show that, if electron ($e^-$) density in the incoming beam (\ref{jul0808-1}) is given as $| \Psi _s |^2$, 
the largest Fresnel peaks of the $e^-$ density on the screen are located more inwardly than those of the field intensity.

We construct the $e^-$ density from trajectories calculated  with (\ref{may1309-31}). 
We construct  the field intensity not from (\ref{may2608-e4}) but from trajectories calculated  with (\ref{may1009-f3}) to avoid  misjudgment in the comparison which may arise from the limited number of trajectory calculations.

For the calculation, we assume 
i) incoming beam flux is comprised of  $\Psi _s$'s  of which center coordinates  $x_0$'s are $x_0 = \pm 0.1 \mu m ,  \pm 0.3 \mu m,  \pm 0.5 \mu m ,  \cdots $,  
ii) each $\Psi _s$ accompanies $e^-$'s at $x_0, \ x_0 \pm 0.01 , \ x_0 \pm 0.02 , \ \cdots , x_0 \pm 4.00 \mu m$, and 
iii) the density of the lower-slot-passed $e^-$'s at  $z=4mm$ is given as $|  0.8 \textsf{M}_1 + 0.2\textsf{M}_{11} |^2 + |  0.8 \textsf{N}_1 + 0.2\textsf{N}_{11} |^2  $ for each $\Psi _s$; see footnote \ref{sep1411-1}.


We calculate trajectories of lower-slot-passed $e^-$'s of which locations at $z= 4mm$ range from $-0.07 $ to $-2.00 \mu m$. Trajectories of upper-slot-passed $e^-$'s are obtained by  symmetry. 
We obtain the  $e^-$ density on the screen caused by a $\Psi _s$ from (\ref{jul-0108-1}). 
They add  up to  the one caused by the incoming beam flux.

\begin{wrapfigure}{r}{52 ex}
\vspace*{-1.5 em} 
\centering
\includegraphics[width = 52ex ,  height = 22ex , clip]{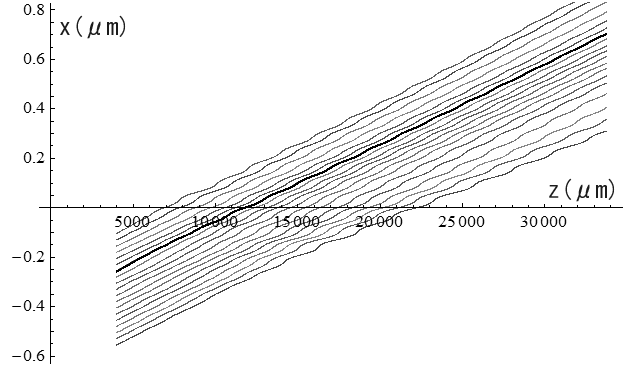}
\caption{Trajectories of electrons accompanying the incoming beam $\Psi _s$ of which center $x$-coordinate is $ -0.4 \mu m$. The thick  trajectory is of the electron which came along the center of the incoming beam.}
\label{fig-skirt}
\end{wrapfigure}

Fig.\ref{fig-psi-square} (p.\pageref{sep2212-19}) is the result. We see i) the field intensity calculated with (\ref{may1009-f3}) well reproduces the one  (Fig.\ref{nov2317-1})       calculated with (\ref{may2608-e4}), and ii) the largest Fresnel peaks of the dens. distr. around $x = \pm 0.6 \mu m$ are located more inwardly than    those of the field intensity. 
The ii) results because $e^-$'s  on the skirts of the incoming beams are more slowed down in $x$-direction than those at centers ($(|\tilde{U}|^2 - |  \tilde{\tilde{U}}|^2) |_{\textrm{center}} >(|\tilde{U}|^2 - |\tilde{\tilde{U}}  |^2 )|_{\textrm{skirt}}$) as is typically seen in Fig.\ref{fig-skirt}.

\sloppy 

Spikes, which are artifacts, on the upper graph appear because, when a long and narrow  $p_x^{HJ} = \partial \mathcal{S}_0 / \partial x \simeq 0$ region is parallel to $z$-axis, trajectories accumulate there because of prohibited velocity reversal. 
Most spikes are generated by lower-slot-passed $e^-$'s in beams of which $x_0$'s  are $-0.1$, $0.1$  and $ 0.3 \mu m$, and by their symmetric counterparts because they give $p_x^{HJ} \simeq 0$ regions suitable for the accumulation. 
They are expected to become ignorable if $e^-$ density on the screen is

\noindent 
constructed from a larger number of incoming beams because the spike position is  different if $x_0$ is different. 
Indeed, the spike  at $x=  -0.0673 \mu m $ spawned by $x_0 =  -0.10 \mu m$-beam moves to $x =  -0.0659 $, $-0.0669 $, $-0.0656  $, 
$-0.0637 \mu m $ if $x_0 = -0.05 $, $-0.075 $, $-0.125 $, \\ 
 $-0.15  \mu m$ respectively.

\fussy

\end{document}